\documentclass[reprint,pra,longbibliography,superscriptaddress,nofootinbib]{revtex4-2}
\usepackage{lipsum}
\usepackage{fancyhdr}
\usepackage{graphicx}
\usepackage[caption=false]{subfig}
\usepackage{braket}
\usepackage[dvipsnames]{xcolor}
\usepackage{mathtools}
\usepackage{amsmath}
\usepackage{amssymb}
\usepackage{hyperref}
\usepackage{esvect}
\usepackage{float}
\usepackage[normalem]{ulem}
\graphicspath{{./PaperFigures/}}

\newcommand{\edit}[1]{{\color{black} #1}}
\newcommand{\reedit}[1]{{\color{black} #1}}

\allowdisplaybreaks

\begin{document}

\title{Quantum trajectory theory and simulations of nonlinear spectra and multi-photon effects in waveguide-QED systems with a time-delayed coherent feedback}

\author{Gavin Crowder}
\email{gcrow088@uottawa.ca}
 \affiliation{Department of Physics, Queen's University, Kingston, Ontario, Canada, K7L 3N6}
\affiliation{Department of Physics, University of Ottawa, Ottawa, ON K1N 6N5, Canada}
\author{Lora Ramunno}
\affiliation{Department of Physics, University of Ottawa, Ottawa, ON K1N 6N5, Canada}
%\author{\sh{Dan, Robin - welcome to join ...}}
\author{Stephen Hughes}
\affiliation{Department of Physics, Queen's University, Kingston, Ontario, Canada, K7L 3N6}

\begin{abstract}
We study the nonlinear spectra and multi-photon correlation functions for the waveguide output of a two-level system (including realistic dissipation channels) with a time-delayed coherent feedback. We compute these observables by extending a recent quantum trajectory discretized-waveguide (QTDW) approach which exploits quantum trajectory simulations and a collisional model for the waveguide to tractably simulate the dynamics. Following a description of the general technique, we show how to calculate the first and second order quantum correlation functions, in the presence of a coherent pumping field. With a short delay time, we show how feedback can be used to filter out the central peak of the Mollow triplet or switch the output between bunched and anti-bunched photons by proper choice of round trip phase. We further show how the loop length and round trip phase effects the zero-time second order quantum correlation function, an indicator of bunching or anti-bunching. New resonances introduced through the feedback loop are also shown through their appearance in the incoherent output spectrum from the waveguide. We explain these results in the context of the waiting time distributions of the system output and individual trajectories, uniquely stochastic observables that are easily accessible with the QTDW model.
\end{abstract}
\maketitle

\section{Introduction}
\label{sec:Intro}

Feedback has been well used as a stabilizing and control mechanism both in photonics and other areas of cutting-edge technology \cite{Dorner2002,Tufarelli2013,Carmele2013,Naumann2016,Lu2017,Pichler2017,Wiseman2006,Arias2020,Grigoletto2020,Hjelme1991,Biology,Franklin2014}. Most commonly, measurement-based feedback has been employed where the output of the system is used to act back on the system to achieve better stability, state generation and error suppression \cite{Wiseman2006,Kubanek2009,Gillett2010,Arias2020,Rafiee2020,Grigoletto2020,Tan2020,Giovanni2021}. This type of  feedback control is common practice in laser design and has also been shown to improve quantum systems as well \cite{Hjelme1991,Shi2021}.
However, for use in many quantum technology applications, it is important to preserve the system {\it coherence}. Thus, a time-delayed coherent feedback has been studied as an avenue of increasing the coherent lifetime in such systems. In contrast to measurement-based feedback, coherent feedback is integrated at the system level and no measurements are taken to avoid introducing further decoherence in the system \cite{Dorner2002,Koshino2012,Tufarelli2013,Carmele2013,Hoi2015,Naumann2016,Lu2017,Pichler2017,Grimsmo2015,Kabuss2016,Pichler2016,Nemet2016,Hein2016,Guimond2016,Whalen2017,Naumann2017,Guimond2017,Forn2017,Whalen2019,Nemet2019,Calajo2019,Crowder2020,Harwood2021,Barkemeyer2021,Shi2021,Regidor2021a}.

The regime of waveguide quantum electrodynamics (QED), where quantum systems are coupled together via waveguide modes, is especially sensitive to loss of coherence \cite{Hughes2004,Shen2005,Zhou2008,Zheng2010,Longo2011,Roy2011,Yan2014,Sanchez2014,Gonzalez2014,Kornovan2017,Mahmoodian2018,Foster2019,Mukhopadhyay2019,Roman2020,Mukhopadhyay2020,Mahmoodian2020,Jeannic2021,Regidor2021b,Sheremet2021,Solano2021}. These systems have many applications in quantum information technology, where they can act as sources for single photons or pairs of photons, photon frequency converters, and single photon detectors which can be further integrated into circuit QED architectures \cite{Blais2007}. Previous work has shown that including coherent feedback in waveguide-QED systems can significantly improve the coherent lifetime of the system or enhance the system emission beyond the typical spontaneous emission rate \cite{Carmele2013,Hoi2015,Lu2017,Whalen2017,Naumann2017,Nemet2019} \edit{as well as enable the generation of shaped single photon sources \cite{Forn2017}}. Not only has it been shown to improve these systems, but also to introduce new system behavior beyond well-known waveguide-QED results, such as new resonances produced by the waveguide modes set up by the feedback \cite{Pichler2016,Crowder2020}.

By including a time-delayed coherent feedback in the waveguide-QED system, a non-Markovian dynamic is introduced in the evolution which adds an additional complexity to the simulation of such systems. Typical models (e.g., Lindblad master equations) make the Markovian approximation, that the evolution of the system only depends on the state of the system at the present, which is no longer valid when feedback is included. Instead, a ``collisional model'' of the waveguide can be used where the waveguide is accounted for at the Hamiltonian level as a series of interactions with a localized coupled quantum system \cite{Kretschmer2016,Calajo2019,Whalen2019,Cilluffo2020,Ciccarello2021}. By expanding the Hamiltonian in this way, Markovian numerical solutions are again possible which are less restricting than the specialized non-Markovian mathematical solutions which have also been used \cite{Carmele2013,Crowder2020}. Note that we refer to the unique dynamics that arise from the introduction of the feedback using the common nomenclature of non-Markovian, since the round trip memory effects are still included in this model. 
%\sh{Still suggest now quantifying the earlier statement then of ``introduces a non-Markovian dynamic to the system'' so as not to create any confusion - the dynamics as seen by a local system operator depends on its past round trip dynamics, and so quantum effects are included, but now in a way that uses a Markovian mathematical solution ... the round trip memory effects are still included}

When the waveguide is included in the Hamiltonian, the Hilbert space can quickly become very large and so specialized methods are used to model its evolution. When limited to the linear regime, results can be computed analytically but this significantly restricts the phenomena that can be investigated \cite{Dorner2002,Carmele2013,Zhang2020,Sommer2020}. A popular method for modelling coherent feedback is matrix product states (MPSs) \cite{Schollwoeck2011,Pichler2016,Naumann2017,Droenner2019,Finsterholzl2020,Regidor2021a,Regidor2021b,Kaestle2021}, a powerful technique where tensor networks are used to limit the entanglement within the Hilbert space. Quantum trajectory (QT) theory is a less popular technique which has also recently been used to investigate the effects of a time-delayed coherent feedback \cite{Whalen2019,Crowder2020,Regidor2021a}. This technique uses stochastic individual realizations of the system to obtain the ensemble average behavior of the system \cite{Dum1992,Tian1992,Dalibard1992,Molmer1993,Brun2002}. The advantage of this technique is that it can give unique insights into the underlying stochastic phenomena of the system behavior and numerically, it scales linearly with the Hilbert space and is completely parallelizable.

Previous QT approaches have been limited to the observables of the coupled quantum system rather than the waveguide output, important for experimental investigations of feedback. In this paper, we extend a previous QT discretized waveguide (QTDW) model to investigate the quantum correlation functions and waiting time distribution function, a uniquely accessible observable from QT theory, of the waveguide output. These observables give intuitive and powerful insight into the multi-quanta effects present when a time-delayed coherent feedback is introduced to the system. Additionally, they are also readily available to experimental realizations of these systems, and thus give fresh insight into the experimental observables.
 
The rest of our paper is organized as follows: In Sec.~\ref{sec:Model}, we present the waveguide-QED model of interest and introduce the QT formalism used in our approach \edit{which accounts for the full non-Markovian dynamics introduced by the feedback loop.} Subsequently, in Sec.~\ref{sec:PopDynamics}, we define the waveguide population parameters of interest and in Sec.~\ref{sec:Correlations}, we explain how to compute the first and second order quantum auto-correlation functions in the QT picture. In Sec.~\ref{sec:Results}, we show our results, including  how a time-delayed feedback can act to increase the coherence of the system output and introduce multi-quanta resonances. We also show how these results are affected by the inclusion of Markovian output channels such as pure dephasing, which is important for modelling realistic qubits. These results are explained through examples of individual trajectories and the waiting time distribution of the waveguide output, an experimentally accessible observable. Lastly, in Sec.~\ref{sec:Conc} we conclude.

\section{Theory}
\label{sec:Theory}

\subsection{Model and Hamiltonian}
\label{sec:Model}

\begin{figure}[tbp]
\subfloat{%
  \includegraphics[width=\columnwidth]{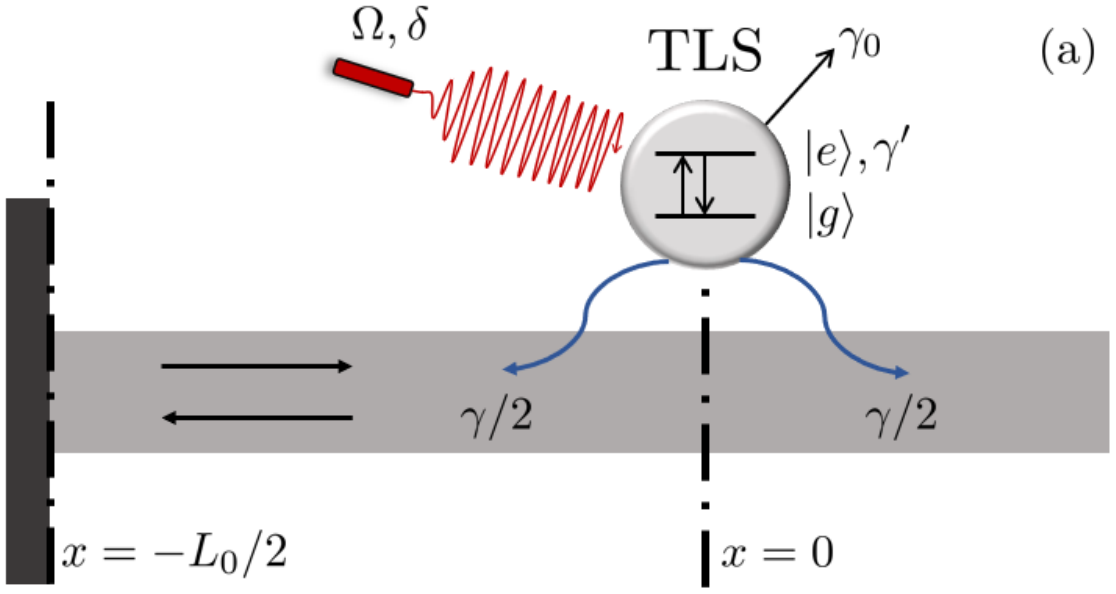}%
}

\subfloat{%
  \includegraphics[width=\columnwidth]{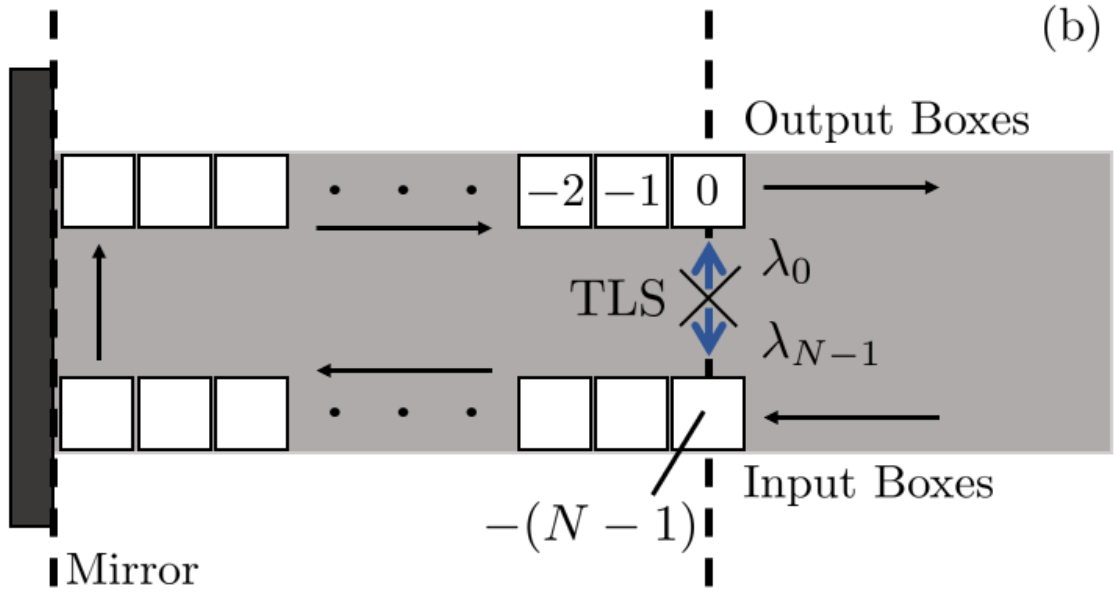}%
}
  \caption{(a) Model of the waveguide-QED system of interest, which includes a TLS with ground state $\ket{g}$ and excited state $\ket{e}$ embedded in a terminated waveguide of length $L_0$. The TLS couples symmetrically to the waveguide with a total decay rate of $\gamma$. The TLS is driven by a CW laser of strength $\Omega$ and detuning $\delta=\omega_{0}-\omega_L$. The TLS also undergoes off-chip decay with rate $\gamma_0$ and pure dephasing with rate $\gamma'$. (b) Representation in the QTDW picture where the modes of the waveguide have been discretized into spatial bins. There are $N$ bins created by the discretization and bin $n$ spans a spatial length from $-n \Delta t$ to $-(n+1) \Delta t$ relative to the output bin. In the QTDW model, the TLS couples to bins $N-1$ and $0$ with coupling constants $\lambda_{N-1}$ and $\lambda_{0}$ respectively.}
  \label{fig:ModelFig}
\end{figure}

We investigate a typical setup for including feedback in a waveguide-QED system, which includes a two level system (TLS), such as a single atom or quantum dot (QD) or flux qubit, coupled to a truncated waveguide depicted in Fig.~\ref{fig:ModelFig}(a). The TLS has ground (excited) state represented by $\ket{g}$ ($\ket{e}$) and couples \reedit{bidirectionally} to the waveguide with \reedit{total radiative decay rate $\gamma$ \cite{Pichler2016,Regidor2021a}}.

The TLS is driven with a continuous wave (CW) laser of Rabi frequency $\Omega$ and the detuning between the laser and TLS is $\delta = \omega_0 - \omega_{\rm L}$, where $\omega_{\rm L}$ is the frequency of the laser and $\omega_{0}$ is the frequency of the TLS. The location of the TLS is chosen as a reference to be $x = 0$, and the waveguide is truncated with a mirror (which we assume is lossless and introduces a phase change of $\phi_{\rm M}$) at $x = -L_0 / 2$. Then the round trip length of the feedback loop is $L_0$ which introduces a delay time of $\tau = L_0 / c(\omega_0)$, where $c(\omega_0)$ is the group velocity of the waveguide mode of reference frequency $\omega_0$ in the waveguide.

The main Hamiltonian for this system, $H = H_{0} + H_{\rm pump} + H_{\rm W} + H_{\rm I}$, includes four important components: the non-interacting Hamiltonian for the TLS, $H_{\rm S}$, the Hamiltonian of the CW-laser drive, $H_{\rm pump}$, the non-interacting Hamiltonian for the waveguide, $H_{\rm W}$, and the interaction Hamiltonian, $H_{\rm I}$, between the TLS and waveguide.
 
In the interaction picture of the laser drive frequency, $\omega_{\rm L}$, the Hamiltonian of the TLS and the pump is
\begin{equation}
    H_{0} + H_{\rm pump} = \delta \sigma^+ \sigma^- + \frac{\Omega}{2} (\sigma^+ + \sigma^-),
\end{equation}
where $\sigma^+$ ($\sigma^-$) is the Pauli raising (lowering) operator, and we have applied a rotating wave approximation. \edit{We have also adopted natural units with $\hbar = 1$.} Next, in the (continuous) frequency domain, the waveguide Hamiltonian is
\begin{equation}
     H_{\rm W} = \int_{-\infty}^{\infty} d\omega  \reedit{\left(\omega - \omega_{\rm L} \right)} b^\dagger (\omega) b (\omega),
     \label{WG_Ham}
\end{equation}
where we choose $b^{\dagger} (\omega)$ ($b(\omega)$) to be the raising (lowering) operator for the propagating photon modes in the waveguide. \edit{These field operators have the usual commutator $[ b (\omega), b^{\dagger} (\omega') ] = \delta(\omega-\omega')$.} 

Finally, the influence of the feedback loop is seen in the interaction Hamiltonian through a modification of the typical coupling rates:
\begin{align}
\label{Hint}
    H_{\rm I} ={} & \int_{-\infty}^{\infty} d\omega \left[ \left( \reedit{\sqrt{\frac{\gamma}{4\pi}}} \sigma^+ b(\omega) \right. \right. \\
    & {}+ \left. \left. \reedit{\sqrt{\frac{\gamma}{4\pi}}} e^{i(\reedit{\phi_M} + \omega \tau)} \sigma^+ b(\omega) \right) + \rm H.c. \right], \nonumber
\end{align}
where the coupling to the right propagating mode picks up the round trip phase change. \edit{The form of this interaction is derived in Appendix~\ref{sec:App1}.}

A full derivation of the QTDW model is given in Ref.~\cite{Regidor2021a}, \edit{including the component Hamiltonians in the discrete frequency picture}, thus we only highlight the important points of the model below. In that paper the formalism is also derived to allow for an open waveguide where the left and right moving fields are treated independently, important for systems such as two QDs spatially separated in the waveguide \reedit{where chiral coupling can also play a significant role in the system behavior \cite{Yao2009,Ballestero2015,LeFeber2015,Young2015,Sollner2015,Coles2016,Scheucher2016,Lodahl2017,Barik2018,Martin-Cano2019,Mehrabad2020,Hauff2021,Regidor2021b}}. \edit{To represent the waveguide with a collisional model,} we transform Eq.~\eqref{WG_Ham} to the discrete time domain with operators $B_{n}$, using a discrete Fourier transform on the waveguide mode operators in the discrete frequency picture \edit{($b(\omega) = \sqrt{L_0/2\pi} b_k$)}. This gives the explicit relationship
\begin{equation}
    \begin{aligned}
    B_{n} ={} & \frac{1}{\sqrt{N}} \sum_{k=0}^{N-1} b_{k} e^{i \reedit{(\omega_k - \omega_0)} n \Delta t}, \\
    b_{k} ={} & \frac{1}{\sqrt{N}} \sum_{n=0}^{N-1} B_{n} e^{\edit{-}i \reedit{(\omega_k - \omega_0)} n \Delta t},
    \end{aligned}
\end{equation}
where $N$ is the total number of discrete bins used to model the waveguide, $\Delta t$ is the corresponding time discretization, and $\omega_k = 2 \pi k / L_0$, which assumes linear dispersion for the waveguide mode. \edit{In this domain, the delay time from the round trip can be expressed as $\tau = N \Delta t$.} Note, the commutator for the time domain operators is $[B_{n} , B^{\dagger}_{n'}] = \delta_{n,n'}$, and $B$ and $B^\dagger$ have dimensionless units. We make the approximation that $N$ is chosen sufficiently large (equivalently $\Delta t$ sufficiently small) that the probability to have more than one photon in any individual bin is negligible, which is valid when $\Delta t \ll 1/\reedit{\gamma}$. \edit{Formally, $B_n$ represents the slice of the waveguide field that interacts with the TLS over a certain time interval, $[n \Delta t, (n+1) \Delta t)$. We can equivalently represent these $B_n$ as a spatial section of the waveguide through the intrinsic relationship between the space and time domains related by the group velocity.}

By transforming into this discrete spatial bin model, the waveguide is effectively a sequence of bins which pass the feedback (and output) field forward one bin each time step. Mathematically, this is seen through the evolution of the time domain operator, $B_n$, under the waveguide Hamiltonian:
\begin{equation}
    U_{\rm W}^{\dagger}(\Delta t) B_{n} U_{\rm W} (\Delta t) = \reedit{e^{-i \delta \Delta t}} B_{n-1},
\end{equation}
where $U_{\rm W} (\Delta t) = e^{-i H_{\rm W} \Delta t}$. \edit{The derivation for this relation is presented in Appendix~\ref{sec:App2}.} Note that only the two spatial bins located at the position of the TLS interact with the TLS at a single time. This is shown schematically in Fig.~\ref{fig:ModelFig}(b).

The TLS and pump Hamiltonians, $H_{0} + H_{\rm pump}$, are unmodified by our transformation into the discrete time domain. The interaction Hamiltonian becomes
\begin{equation}
    H_{\rm I} = \left ( \reedit{\sqrt{\frac{\gamma} {2\Delta t}}} \sigma^+ B_{N-1}  + e^{i \phi} \reedit{\sqrt{\frac{\gamma} {2\Delta t}}} \sigma^+ B_{0} \right) + \rm{H.c.},
\end{equation}
\reedit{where $\phi = \phi_{\rm M} + \omega_0 \tau$ is the round trip phase change.} The TLS first interacts with the $(N-1)^{\rm th}$ bin which then passes the field from bin to bin around the loop and returns to the TLS after $N$ time steps to interact with the TLS again in bin $0$. Thus, the TLS couples to bins $N-1$ and $0$ with coupling constants $\lambda_{N-1} = \reedit{\sqrt{\gamma / 2 \Delta t}}$ and $\lambda_0 = e^{i\phi} \reedit{\sqrt{\gamma / 2\Delta t}}$. \edit{The form of this interaction Hamiltonian was fully derived in Sec. IV A (b) of Ref.~\cite{Regidor2021a} for reference.}

The ket vector, \edit{in factorized form}, for the complete TLS and waveguide system is
\begin{equation}
    \ket{\psi (t)} = \ket{\psi_{\rm S} (t)} \ket{\psi_{\rm W} (t)},
    \label{ketVec}
\end{equation}
where $\ket{\psi_{\rm S} (t)}$ is the ket vector for the TLS and $\ket{\psi_{\rm W} (t)}$ is the ket vector for the waveguide from $x = 0$ to $x = -L_0 /2$, which we limit to a maximum of two photons in this section of the waveguide.\footnote{These ket vectors are $\ket{\psi_{\rm S} (t)} = a(t) \ket{g} + b(t) \ket{e}$ and $\ket{\psi_{\rm W} (t)}= c^{(0)}(t) \ket{ \{0\} } + \sum_{j=1}^N c^{(1)}_j(t) \ket{1_j} + \sum_{j=1}^{N-1} \sum_{k=j+1}^{N} c^{(2)}_{j,k}(t) \ket{1_j} \ket{1_k}$, which can be multiplied together to get the form in Eq.~\eqref{Expanded_ket}.} We justify this approximation later in our results where we report the population for two photons in the feedback loop.

\edit{The waveguide field in the $x>0$ region is modeled through simulated measurements of the output field in the $0$'th bin. The ket vector can be split as 
\begin{equation}
    \ket{\psi (t)} = \ket{\psi_0 (t)} \ket{0_0} + \ket{\psi_1 (t)} \ket{1_0},
\end{equation}
where $\ket{0_0}$ is an empty bin 0 and $\ket{1_0}$ represents a photon in bin $0$. Before moving the final bin forward and out of the waveguide section of interest, its information is retrieved by projecting the full ket vector into one of the two states $\ket{\psi_0 (t)}$ or $\ket{\psi_1 (t)}$ with respective probabilities $\braket{\psi_i (t) | \psi_i (t)}$. This process is not norm conserving so renormalization must occur each time step.}

Two dissipation channels are also included in our model as quantum jump operators: off-chip decay from the TLS, $C_0 = \sqrt{\gamma_0} \sigma^-$, with rate $\gamma_0$ and pure dephasing in the TLS, $C_1 = \sqrt{\gamma'} \sigma^+ \sigma^-$, with rate $\gamma'$. Following standard QT theory, these are included in the system evolution through stochastic application of the jump operator with probabilities $\braket{C^{\dagger}_0 C_0}$ and $\braket{C^{\dagger}_1 C_1}$ for off-chip decay and pure dephasing respectively. The Hamiltonian is also modified to a non-Hermitian effective Hamiltonian
\begin{equation}
    H_{\rm eff} = H_{\rm S} + H_{\rm pump} + H_{\rm I} - \frac{i}{2} \sum_{j = 0}^1 C^{\dagger}_j C_j,
\end{equation}
which is used to evolve the \edit{combined TLS and waveguide ket vector} whenever no quantum jump occurs. The inclusion of the final term (which is non-Hermitian) moves the model beyond that of the semi-classical weak excitation approximation approaches. Both quantum jumps are Lindblad channels which have the superoperator form
\begin{equation}
    \mathcal{L}(\rho) = -\frac{1}{2} \sum_{j = 0}^{1} \left( C^{\dagger}_j C_j \rho + \rho C^{\dagger}_j C_j \right) + \sum_{j = 0}^{1} C_j \rho C_j^{\dagger},
\end{equation}
where $\rho$ is the density matrix for the TLS. \edit{These dissipation channels could be dealt with through additional collisional interactions, however, this is not necessary as we take them to be Markovian, and doing so would add to the size of the Hilbert space and increase the computational complexity of the problem.}

A single time step for a simulation with this model follows a four step process:
\begin{enumerate}
    \item Evolve $\ket{\psi(t)}$ using a regular QT step under the effective Hamiltonian, $H_{\rm eff}$, with the two quantum jump operators $C_0$ and $C_1$.
    \item Take a {\it simulated measurement} on the final bin of the waveguide by calculating the population of the bin, $N_{B_0}$ (defined in Eq.~\eqref{BinPop}), and comparing it to a uniformly distributed random number, $x \in (0,1)$. The ket vector is then projected according to whether a photon is detected or not.
    \item Evolve the waveguide bins under $H_{\rm W}$ which acts to step each bin forward one space. Note that the final bin is emptied and contains no information so it can be dropped while the incoming bin is empty for this setup, although this is not a strict limitation in the model.
    \item Renormalize the complete system ket vector since the operation of measuring and moving the bins along does not conserve the norm.
\end{enumerate}

\subsection{Waveguide Population Observables in the QTDW Approach}
\label{sec:PopDynamics}

To investigate the photon population dynamics in the waveguide, 
%one can calculate 
we begin with the probability of finding a photon in a single waveguide bin $j$ within the feedback loop,
\begin{equation}
    N_{B_j} (t) = \braket{ \psi (t) | B^{\dagger}_j B_j | \psi (t) },
    \label{BinPop}
\end{equation}
where bin $j$ is travelling towards the mirror if $j \geq N/2$, or travelling away from the mirror if $j < N/2$. As the spatial length of the bins depends on $\Delta t$, then so does $ N_{B_j}$ so that the total population around the whole loop does not. Thus, a more useful quantity to use is the flux through a particular bin,
\begin{equation}
     n_{B_j} (t) =  N_{B_j} (t) / \Delta t,
\end{equation}
which is in units of $1/{\rm s}$. The flux of the outgoing bin (at $j=0$) then gives an observable for the photon flux leaving the system, independent of our choice of $\Delta t$, which will prove very useful.

Another useful quantum observable is the photon number distribution function in the feedback loop, namely the probability to have zero, one, or two photons in the feedback loop. With this metric, we can evaluate the applicability of our ``two-photons-in-the-loop'' approximation. Since the QTDW model directly evolves the ket vector for the system (Eq.~\eqref{ketVec}), this becomes a straight-forward quantity to compute if desired. However, with other techniques such as MPS, these quantities have been calculated via the correlation functions, which can be computationally demanding and does not scale well with the total simulation run time \cite{Kabuss2011,Droenner2019}.

Explicitly, the ket vector is
\begin{equation}
    \begin{aligned}
    \ket{\psi(t)} ={} & \left[ \alpha^{(0)}(t) \ket{g} + \beta^{(0)}(t) \ket{e} \right] \ket{ \{ 0 \} } \\ 
    & {}+ \sum_{j=1}^N \left[ \alpha_j^{(1)}(t) \ket{g} + \beta_j^{(1)}(t) \ket{e}  \right] \ket{1_j} \\
    & {}+ \sum_{j=1}^{N-1} \sum_{k=j+1}^{N} \left[ \alpha_{j,k}^{(2)} (t) \ket{g} + \beta_{j,k}^{(2)} (t) \ket{e} \right] \ket{1_j} \ket{1_k},
    \label{Expanded_ket}
\end{aligned}
\end{equation}
and thus the probabilities can be read off almost immediately as
\begin{subequations}
\begin{align}
    p(0,t) ={} & |\alpha^{(0)}(t)|^2 + |\beta^{(0)}(t)|^2, \\
    p(1,t) ={} & \sum_{j=1}^N |\alpha_j^{(1)}(t)|^2 + |\beta_j^{(1)}(t)|^2,  \\
    p(2,t) ={} &  \sum_{j=1}^{N-1} \sum_{k=j+1}^{N} |\alpha_{j,k}^{(2)}(t)|^2 + |\beta_{j,k}^{(2)}(t)|^2, 
\end{align}
\end{subequations}
where $p(n,t)$ represents the probability of having $n$ photons in the loop at time $t$.

Lastly, an important experimental observable is the waiting time distribution (WTD) \cite{Carmichael2}, $W(t')$, for photons detected via the waveguide. This is defined as the distribution of delay times, $t'$, between detection events via the waveguide output channel,
\edit{\begin{equation}
    W(t') = \frac{N\left( \{ A \} | t(A_j) - t(A_{j-1}) = t' \right)}{N_{\rm tot}},
\end{equation}
where $N_{\rm tot}$ is the total number of photons detected via the waveguide and $\sum_j N\left( \{ A \} | t(A_j) - t(A_{j-1}) = t' \right)$ is the number of detections in the detection record $\{ A \}$ with delay time $t',$ between the $A_j$'th event and the previous event, $A_{j-1}$.} Since simulating these detection events are a direct part of the QTDW approach to modelling the response, we can record a jump record for each trajectory and then easily calculate the waiting times for each trajectory. This is uniquely accessible by using a QT based approach as the WTD requires individual realizations of the system rather than the ensemble average. This observable can be very useful in explaining the phenomena seen in the ensemble average response 
%\lr{ what do you mean by average system response exactly? you mean, ensemble average? but with the full distribution function you can get so much more. presumably other methods can also get the average system response?} 
as we will show in Sec.~\ref{sec:Results}.

\subsection{Output Quantum Correlation Functions}
\label{sec:Correlations}

In previous work with QT based approaches to modelling feedback \cite{Crowder2020,Regidor2021a}, the results were restricted to how the feedback affected the population of the TLS. Although many useful insights can be gained from this population dynamic, in this paper we focus our observables around the output photons leaving through the waveguide, which is more practical 
%\lr{ what do you mean by more practical}
and experimentally relevant. Below, we also show how to calculate the first- and second-order quantum auto-correlation functions for waveguide output photons where we use the first order correlation function to calculate the incoherent spectra. This is usually an extremely difficult theoretical 
%\lr{ theoretical? problem or task?}
problem for the usual master equation approach because the quantum regression theorem cannot be used for a non-Markovian dynamic \cite{Carmichael1}. By including the waveguide at the Hamiltonian level, the QTDW model side steps this problem and we can directly calculate the correlation functions at any photon bin, including the output bin.

We begin by describing how to obtain the second-order quantum auto-correlation function as it is the more straightforward of the two. Since we are working with a CW pump field, we are interested in the steady state behavior of the following two-time correlation function:
\begin{equation}
    [g^{(2)}_{\rm out} (t_2)]_{\rm ss} = \frac{\braket{B_0^{\dagger}(0) B_0^{\dagger}(t_2) B_0 (t_2) B_0  (0)}_{\rm ss}}{\braket{B_0^{\dagger} B_0}_{\rm ss}^2},
    \label{g2Def}
\end{equation}
where $t_2$ represents the time after steady state is reached (defined here at $(0)$ in the correlation function). Due to the stochastic nature of the QT technique, we calculate this quantity for individual trajectories and then average over a large number of realizations to arrive at the ensemble average. For good convergence, typically thousands of trajectories are required, which in total take on the order of tens of minutes to run on a single computer.

In the numerical algorithm, an initial set of regular trajectories must first be run to identify when steady state is reached, denoted by $t_{\rm ss}$ which is defined as 0 in Eq.~\eqref{g2Def}. Then, the trajectory, $\ket{\psi(t)}$, is run following the prescribed algorithm until $t_{\rm ss}$ is reached. On the final time step, the process is stopped before the simulated measurement is done on the final bin of the system in step 2 of the algorithm (to avoid losing the information in the $0^{\rm th}$ bin). At this point, the operator $B_0$ is applied to the ket vector, $\ket{\psi' (t_{\rm ss})} = B_0 \ket{\psi (t_{\rm ss})}$, i.e., we force a simulated detection of a photon in the bin leaving the system. Then the final two steps of the algorithm are finished by stepping all of the bins forward and renormalizing. The new $\ket{\psi'}$ is then evolved forward in time following the four-step algorithm, with the observable
\begin{equation}
    G^{(2)}_{\rm out} (t_2) = \braket{\psi' (t_2) | B_0^{\dagger} B_0 | \psi' (t_2)},
\end{equation}
being calculated at each time step. The final normalized second order correlation function is
\begin{equation}
    [g^{(2)}_{\rm out} (t_2)]_{\rm ss} = \frac{G^{(2)}_{\rm out} (t_2)}{\braket{B_0^{\dagger} B_0}_{\rm ss}},
\end{equation}
where, due to the renormalization that already occurs each time step when evolving $\ket{\psi' (t_2)}$, $\braket{B_0^{\dagger} B_0}_{\rm ss}$ is used rather than the square as appears in Eq.~\eqref{g2Def}.

As a consequence of the approximation that there is only one photon in any single bin\footnote{Note that we can still have two photons in different spatial bins.} (which is not a serious restriction as the bin populations are typically very small), this formally sets
\begin{equation}
    [g^{(2)}_{\rm out} (0)]_{\rm ss} = 0.
\end{equation}
However, the quantity $[g^{(2)}_{\rm out} (\Delta t)]_{\rm ss}$ can be used as an analogous value to $[g^{(2)}_{\rm out} (0)]_{\rm ss}$ because it gives the smallest time step between photon emissions in the QTDW model and can still indicate whether sub- or super-Poissonian light is being emitted. 
%\lr{i dont' understand why these are the same}

The {\it incoherent} spectrum from the waveguide output is also a good observable to investigate multi-quanta effects, as it will contain resonances beyond those from the TLS, and shows signatures of coherent bath control through feedback. This spectrum is obtained from
\begin{align}
    S_{\rm incoh}^{\rm out}(\omega) ={} & \int_0^{\infty} d t_2 e^{i(\omega-\omega_{\rm L}) t_2} \left[ \braket{B_0^{\dagger}(t_2) B_0 (0)}_{\rm ss} \right. \\
    & \left. {}- \braket{B_0^{\dagger}(0)}_{\rm ss} \braket{B_0(0)}_{\rm ss} \right], \nonumber
\end{align}
where the first term in the integrand is the unnormalized first order auto-correlation function, $G_{\rm out}^{(1)} (t_2)$, and the second term is composed of steady-state expectation values that can be readily calculated with the QTDW model.

Calculating $G_{\rm out}^{(1)} (t_2)$ with the QTDW model is less intuitive than the second order correlation function \cite{Dum1992,Dalibard1992,Molmer1993}. Similar to our approach for the second order correlation function, the first step is to identify when steady state is reached, $t_{\rm ss}$, from the ensemble average system dynamics. Then the trajectory is evolved until $t_{\rm ss}$, but on the final time step we stop before step 2 of the algorithm, as we did before when calculating $[g^{(2)}_{\rm out} (t_2)]_{\rm ss}$. Here, the process differs, two copies of the system ket vector are made, a ``lead trajectory'' which is unmodified, $\ket{ \psi_{\rm lead}' (0) } = \ket{ \psi(t_{\rm ss}) }$, and a ``follower trajectory'' to which we apply $B_0$, $\ket{ \psi_{\rm fol}' (0) } = B_0 \ket{ \psi(t_{\rm ss}) }$ 
%\lr{hwat do the two trajectories represent physically?}. 
The time step is then completed for the lead trajectory by completing steps 3 and 4 of the algorithm. For the follower trajectory, the bins are stepped forward in step 3, but in step 4 the trajectory is renormalized as
\begin{equation}
    \ket{ \psi_{\rm fol} (0) } = \frac{ \ket{ \psi_{\rm fol}' (0) } }{ \sqrt{ \braket{  \psi_{\rm lead}' (0) | \psi_{\rm lead}' (0) } } },
    \label{FollowRenorm}
\end{equation}
so that it follows the normalization of the lead trajectory.

The lead trajectory then evolves following the four-step algorithm, during which it continues to have stochastic jump events from $C_0$, $C_1$, and (simulated) detections from the final bin.
%\lr{i dont' get this}. 
The {\it follower} trajectory evolves by following the same sequence of jump events that the lead trajectory underwent and at the end of each time step is then projected onto the length of the lead trajectory following Eq.~\ref{FollowRenorm}. %\lr{don't get this} 
To calculate the correlation function, we compute
\begin{equation}
    G_{\rm out}^{(1)} (t_2) = \braket{\psi_{\rm lead} (t_2) | B_0^{\dagger} | \psi_{\rm fol} (t_2)}.
\end{equation}
However, rather than calculating this at the end of each time step, it needs to be calculated after step one of the algorithm is completed but before the final bin is measured in step two.

If the normalized correlation function is desired, then one can calculate this through
\begin{equation}
    [g_{\rm out}^{(1)} (t_2)]_{\rm ss} = \frac{ G_{\rm out}^{(1)} (t_2) }{ \sqrt{\braket{ B_0^{\dagger} (t_2) B_0 (t_2)}_{\rm ss}\braket{B_0^{\dagger}(0) B_0 (0)}_{\rm ss}} },
\end{equation}
where the first expectation value in the denominator is the bin population of $B_0$ at time $t_2$ from the lead trajectory and the second expectation value is the population of $B_0$ from the lead trajectory when $t_2 = 0$.

\section{Results}
\label{sec:Results}

In this section, we apply our extended QTDW approach to several systems. First we show that it correctly produces feedback induced changes to the Mollow triplet and explain these changes in the context of $[g^{(2)}_{\rm out} (t_2)]_{\rm ss}$ and $W(t')$. Then we characterize the effect of the delay time, $\tau$, and the round trip phase change, $\phi$, on the waveguide output and the photon number distribution in the feedback loop. Finally, we drive the system with a strong CW-pump to excite the additional resonances from the feedback loop in the output spectrum.

\begin{figure}[tbp]
    \centering
    \includegraphics[width=\columnwidth]{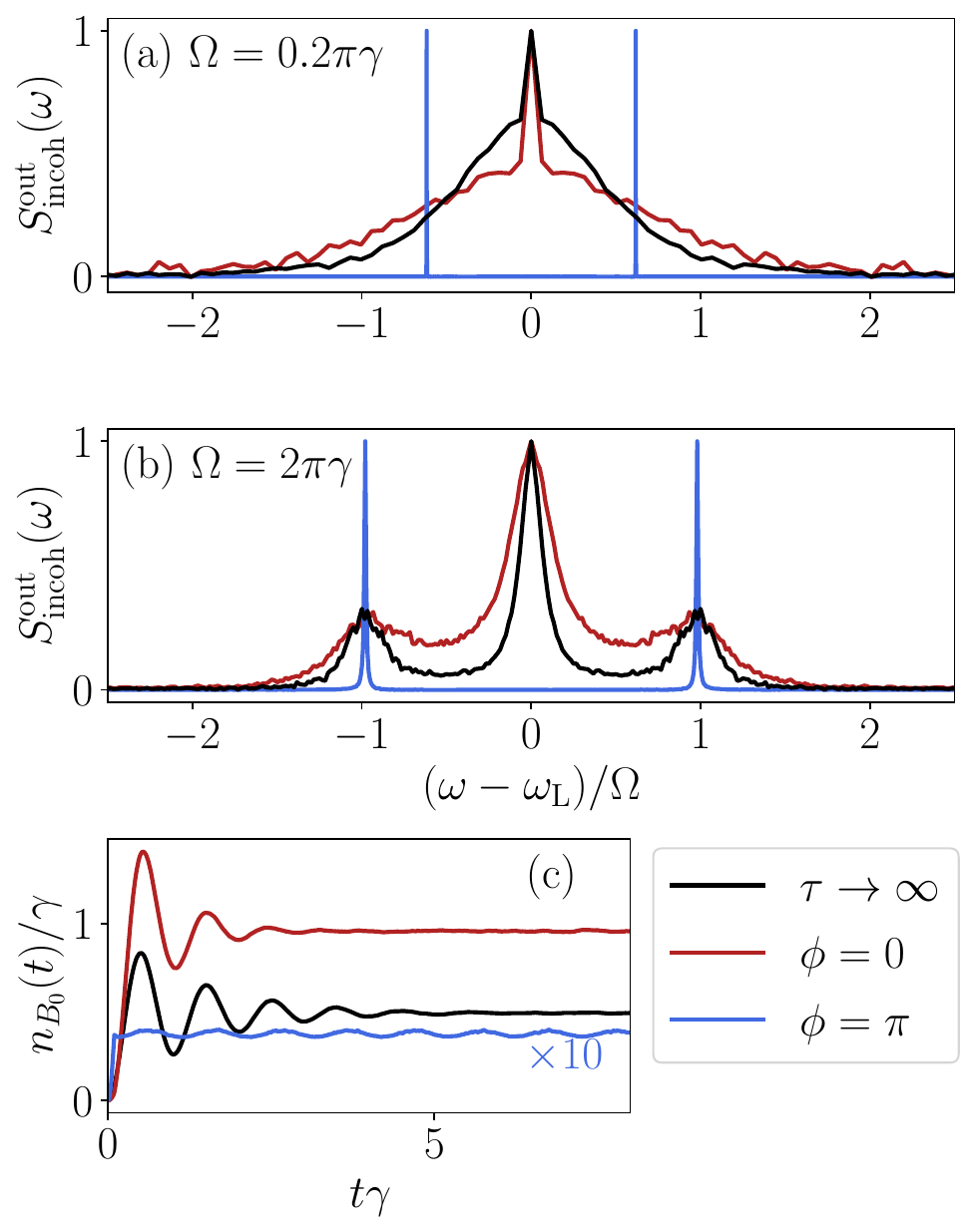}
    \vspace{-0.6cm}
    \caption{Incoherent spectra of the waveguide output for (a) $\Omega = 0.2 \pi \gamma$ and (b) $\Omega = 2 \pi \gamma$. For the spectrum in (b) the output photon flux is in (c). For a TLS driven on resonance without feedback, $\tau \rightarrow \infty$ (black) and with a short feedback loop, $\tau = 0.1 \gamma^{-1}$, for $\phi = 0$ (red) and $\phi = \pi$ (blue); 20,000 trajectories were used for $\tau \rightarrow \infty$ and $\phi = \pi$, and 50,000 trajectories for $\phi = 0$.}
    \label{fig:ShortLoop_Spectras}
\end{figure}

A single TLS excited by an on-resonance CW laser, with Rabi frequency $\Omega = 2 \pi \gamma$, will emit a spectrum with the distinctive Mollow triplet, including a central resonance at the frequency of the laser $\omega_{\rm L}$ and two side peaks at $\omega - \omega_{\rm L} = \pm \Omega$. By introducing a relatively short feedback loop, the central peak can disappear, such that with the proper choice of the round trip phase only the side peaks remain \cite{Pichler2016}. In Fig.~\ref{fig:ShortLoop_Spectras}(b) the incoherent output spectra for the TLS without feedback (with round trip time $\tau \rightarrow \infty$), and the TLS coupled to a short feedback loop (with round trip time $\tau = 0.1 \gamma$) are shown, where in the short loop case two values of round trip phase are considered, $\phi = 0$ and $\phi = \pi$. When $\phi = 0$, the effect of feedback is such that the spectral peaks have the same heights as the no feedback case, but are all broadened. For $\phi =\pi$, the side peaks are significantly sharpened and the central peak almost completely removed. 

\edit{In the Heitler regime, when the Rabi frequency is small, the spectrum becomes single peaked and the Mollow Triplet is no longer seen \cite{Heitler1954,Carmichael1,Matthiesen2012}. For a Rabi frequency of $\Omega = 0.2 \pi \gamma$, Fig.~\ref{fig:ShortLoop_Spectras}(a) shows the output spectrum both with and without feedback. With this weak pump and short delay time, the field within the waveguide remains quite weak. A choice of $\phi = 0$ again leads to broadening of the central peak as the weak field constructively interferes with itself. When $\phi = \pi$, the single photon in the loop field destructively interferes with the TLS output and only a very small output flux remains. This causes the central peak to be suppressed and only extremely sharp sidepeaks remain. For a Mollow triplet, the side peaks are at $\pm \Omega$, however the feedback modifies this as these two peaks are no longer purely the dressed states of the TLS and laser. Instead, these states are dressed by the additional field modes that are set up in the waveguide causing the peaks to shift.}

\edit{To achieve maximal constructive and destructive interference, the condition for the round trip phase change is $\Omega \tau / 2 - \phi = (2k-1) \pi$ for destructive interference and $\Omega \tau / 2 - \phi = 2k\pi$ for constructive interference with $k \in \mathbb{Z}$. For the interference to be complete (i.e. no output for the case of destructive interference) there is a further requirement of $\Omega \tau = 2 n \pi$ for $n \in \mathbb{Z}$ to match the phase of the Rabi oscillations. For a short loop and weak drive, this interference occurs maximally when $\phi \approx 0$ for constructive interference and $\phi \approx \pi$ for destructive interference.}

This is readily seen by comparing the output flux from the loop, $n_{B_0} (t)$, in Fig.~\ref{fig:ShortLoop_Spectras}(c), for the three setups. In the {\it one-photon-in-the-loop} approximation, the output from the waveguide at $\omega_{\rm L}$ will be suppressed when $\phi = \pi$ is consistent with the findings in Refs.~\cite{Grimsmo2015,Whalen2019,Crowder2020}. \edit{The approximation is good in the limit when $\gamma \tau \rightarrow 0$, i.e. when the field in the feedback loop is weak ($\gamma \rightarrow 0$) or the dynamics can be treated in the Markovian regime ($\tau \rightarrow 0$), but fails when the non-Markovian dynamics are required.} Allowing for two photons in the loop, as we do here, means that the feedback {\it cannot} match the phase requirements for destructive interference of the two different photon frequencies simultaneously when $\phi = \pi$ \cite{Regidor2021a}. 
%\lr{don't get this}. 
Thus, the loop acts to meet the phase requirements of the central $\omega = \omega_{\rm L}$ peak (and thus suppress the peak) while the side peaks remain. When the feedback interferes constructively when $\phi = 0$, the waveguide emission from the system is much faster 
%\lr{don't get this} 
and this acts to broaden the emission while retaining the general Mollow triplet shape. \edit{For intermediate choices of $\phi$ between $0$ and $\pi$, the interference is no longer maximal and as the phase is varied, the observables vary continuously between the behavior of the two maximal interference cases.}

\begin{figure}[tbp]
    \hspace{-0.6cm}
    \includegraphics[width=\columnwidth]{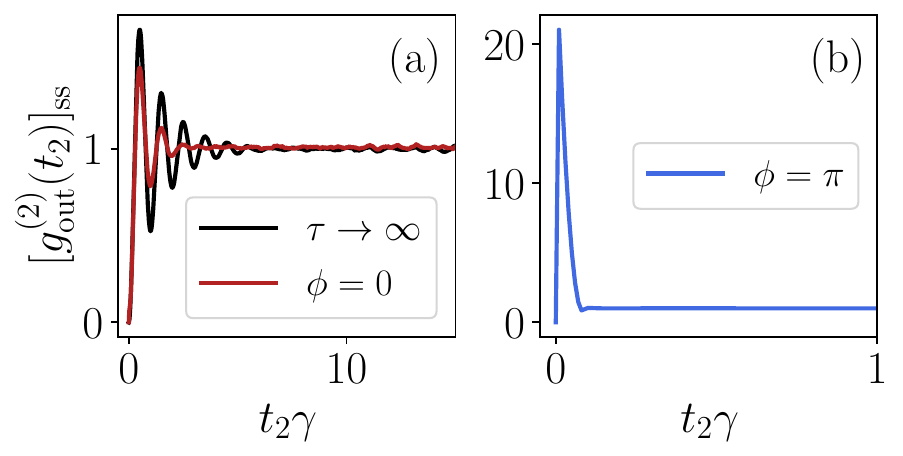}
    \vspace{-0.4cm}
    \caption{The second-order quantum correlation function for the three setups described in Fig.~\ref{fig:ShortLoop_Spectras}(b). For both with and without feedback in (a) the output is sub-Poissonian while in (b) it is super-Poissonian. Note that in (b) the correlation function begins at 0 but then immediately jumps to $>$1 on the next time step since there is a maximum of one photon in each bin. These results are an average of 20,000 trajectories each.}
    \label{fig:ShortLoop_g2}
\end{figure}

The photon counting statistics of the waveguide output are also strikingly effected by the introduction of a time-delayed coherent feedback. By virtue of having only two energy levels, a TLS in an infinite waveguide will nominally emit sub-Poissonian light with $[g^{(2)}_{\rm out} (0)]_{\rm ss} = 0$. The response is similar when a short feedback loop, $\tau = 0.1 \gamma^{-1}$, is included with $\phi = 0$ and these are overlaid in Fig.~\ref{fig:ShortLoop_g2}(a). When feedback is included, the response remains sub-Poissonian but loses the second-order correlations in time much faster (the red line in Fig.~\ref{fig:ShortLoop_g2}(a) decaying to steady state more quickly than the black) due to the increased output from the system. However, when $\phi = \pi$, as in Fig.~\ref{fig:ShortLoop_g2}(b), the statistics of the outgoing light switches to highly super-Poissonian over a period of time equal to the delay time of the feedback. These two starkly different regimes can allow feedback to be used to tune the characteristics of the outgoing light depending on the intended use.

In Fig.~\ref{fig:ShortLoop_Spectras}, a relatively short feedback loop is used, $\tau = 0.1 \gamma^{-1}$, in order to maintain coherence between the returning field in the feedback loop and the emitted field from the TLS. \edit{As the delay time increases, the feedback becomes increasingly sensitive to meeting the exact phase matching conditions for interference as the coherence between the two fields decreases.} However, other channels for decoherence, such as pure dephasing, also have an effect on how well the feedback works. 

\begin{figure}[tbp]
    \centering
    \includegraphics[width=\columnwidth]{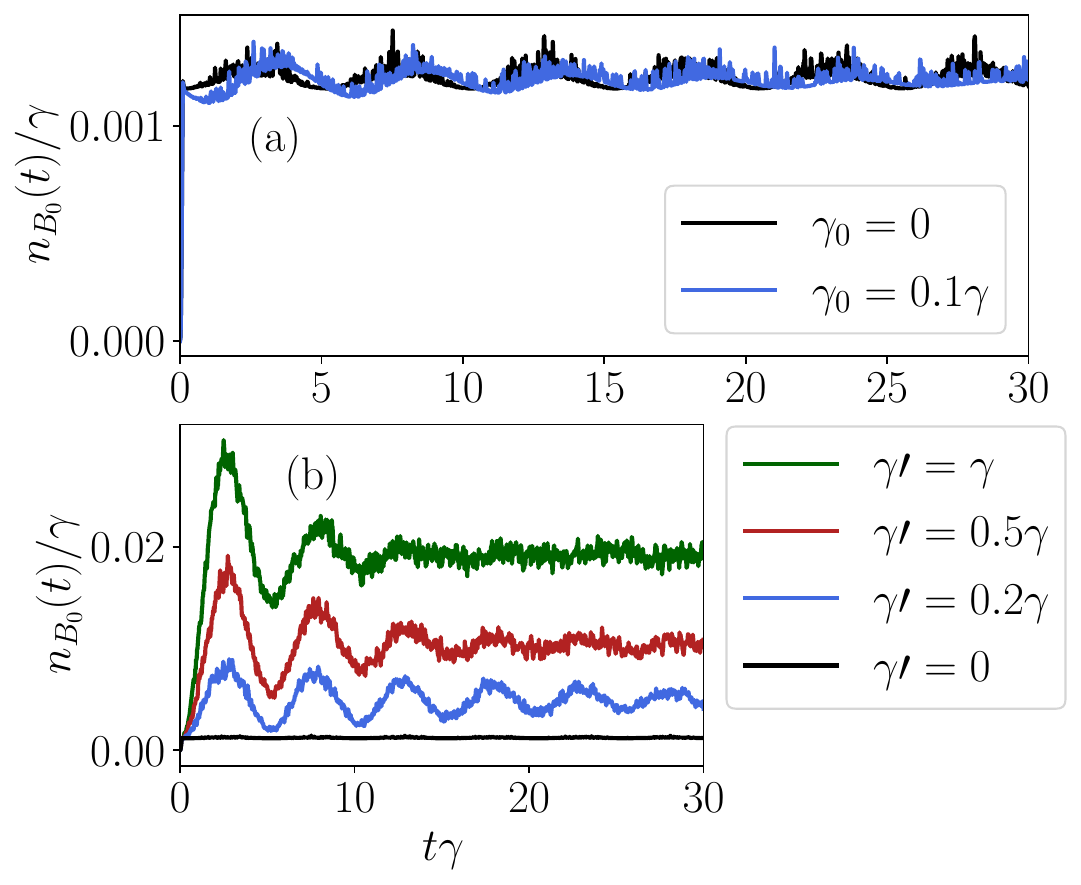}
    \vspace{-0.6cm}
    \caption{Effect on the output photon flux from the inclusion of the Lindblad output channels; (a) off chip decay and (b) pure dephasing. The TLS is driven on resonance with $\Omega = 0.4 \pi \gamma$ with a feedback loop of delay time $\tau = 0.1 \gamma^{-1}$ and $\phi = \pi$. Each result is an average of 20,000 trajectories.}
    \label{fig:LindbladOuts_Comp}
\end{figure}

In Fig.~\ref{fig:LindbladOuts_Comp} we have introduced the Lindblad output channels (off chip decay and pure dephasing),  to see to what degree they disrupt the effect of feedback when $\phi = \pi$ leading to destructive interference. Off chip decay from the TLS is included in Fig.~\ref{fig:LindbladOuts_Comp}(a) at a rate of $\gamma_0 = 0.1\gamma$ which is typical of state of the art TLS sources such as QDs \cite{Manga2007,Reimer2016}. This output channel does not greatly effect the response of the output flux because whenever a quantum jump occurs, the TLS is emptied and so any returning feedback is met with no field; thus, no interference occurs and the system quickly returns to its steady state behavior. However, when pure dephasing is included in Fig.~\ref{fig:LindbladOuts_Comp}(b), we see a much more pronounced effect as the output flux out of the system is increased approximately twentyfold when $\gamma' = \gamma$. In this case, when a jump occurs, the phase of the TLS flips. The returning feedback is thus met with constructive interference and the output from the system is enhanced leading to the increased flux out of the system. Note though, that the steady state output flux when $\gamma' = \gamma$ is still much reduced from the case without feedback which is shown as the black line in Fig.~\ref{fig:ShortLoop_Spectras}(c). \edit{In the case of $\phi = 0$, similar physics is seen when including these output channels. Off-chip decay continues to have little effect on the steady state output flux while dephasing now works to spoil the constructive interference in the output fields. This causes the steady state output flux to decrease as the dephasing rate increases, from $[n_{B_0} / \gamma]_{\rm ss} \approx 0.45$ without any output channels to $[n_{B_0} / \gamma]_{\rm ss} \approx 0.34$ when $\gamma' = \gamma$.}
To limit the amount of pure dephasing in
QD systems, usually one works
at lower temperatures, and gated QDs can also reduce the charge noise significantly~\cite{War2013,Som2016}.
% \sh{here we want to say teh importance then or working at low T for QD systems, and possibly using contacts to reduce charge noise - Warburton (I can add the ref), which is also how they improve current single photon sources.} 

\begin{figure}[tbp]
    \centering
    \includegraphics[width=\columnwidth]{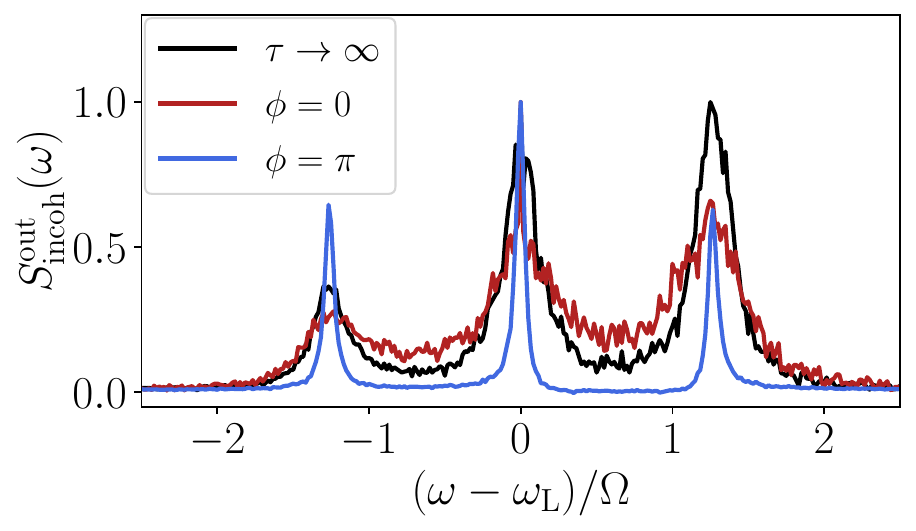}
    \vspace{-0.6cm}
    \caption{\edit{Asymmetric Mollow triplet from the TLS driven off resonance ($\delta = 5\gamma$, $\Omega = 2 \pi \gamma$) with pure dephasing ($\gamma' = 0.5 \gamma$) with and without feedback ($\tau = 0.1 \gamma^{-1}$). Each result is an average of 20,000 trajectories.}}
    \label{fig:Detune_Dephase_Spec}
\end{figure}

Pure dephasing can also act to qualitatively change the emitted spectrum. Under off-resonant pumping, the Mollow triplet becomes asymmetrical when pure dephasing is included~\cite{Edwards1983,Ulhaq2013,Gustin2018}. We show how the inclusion of feedback affects this result in Fig.~\ref{fig:Detune_Dephase_Spec}. The detuning between the laser and TLS is $\delta = 5 \gamma$ and the Rabi frequency if $\Omega = 2\pi$. The rate of pure dephasing is $\gamma' = 0.5\gamma$. When a short feedback loop is included ($\tau = 0.1 \gamma^{-1}$), \reedit{the spectral asymmetry is not as pronounced. When $\phi = 0$, the asymmetry remains, however now the largest peak is the central peak. When $\phi = \pi$, only a small amount of asymmetry in the spectrum shape remains and the peak heights are now symmetrical.} For the remaining results we will neglect the two dissipation channels to present optimal results as we remain on resonance.

\begin{figure}[tbp]
    \centering
    \includegraphics[width=\columnwidth]{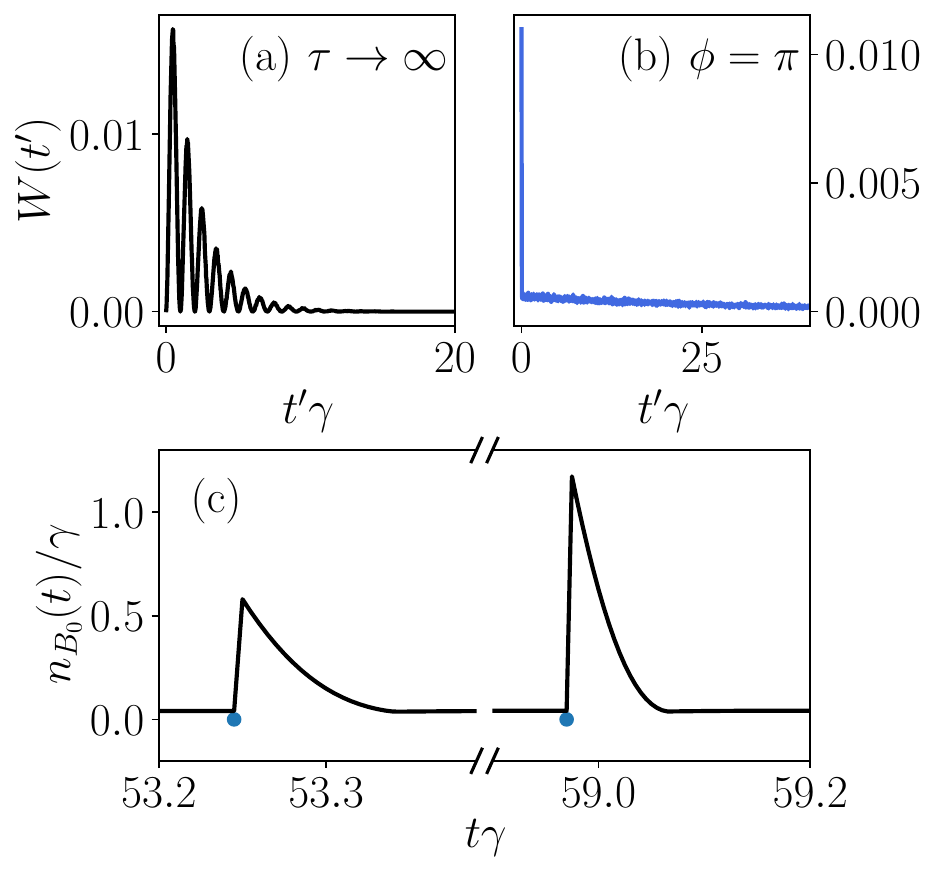}
    \vspace{-0.6cm}
    \caption{The WTDs for a TLS driven on resonance with $\Omega = 2 \pi \gamma$ and with (a) no feedback or (b) feedback from a short loop with $\tau = 0.1 \gamma^{-1}$ and $\phi = \pi$. In (c) two jump events during a trajectory are shown for the setup in (b) to illustrate the increase in output flux after a jump occurs.}
    \label{fig:ShortLoop_WTDs}
\end{figure}

The WTDs of the waveguide output give us additional insight into how feedback is changing the system dynamics. Without feedback, in Fig.~\ref{fig:ShortLoop_WTDs}(a), the system goes through periodic peaks of photon emission which quickly die off as the waiting time gets much longer than the lifetime of the TLS. As expected, there is a very small probability of back to back jumps since the output light is sub-Poissonian. 

When feedback is introduced, as shown in Fig.~\ref{fig:ShortLoop_WTDs}(b), there is a sharp jump at the beginning of the WTD, over the length of the delay time, indicating that there is a significant population of back-to-back emissions occurring. After this initial spike, the distribution is almost flat, out to very long waiting times; indeed the graph in Fig.~\ref{fig:ShortLoop_WTDs}(b) is truncated for readability and the distribution tapers out to $t' = 60 \gamma^{-1}$. This indicates that if a photon pair emission does not occur, the next emission is uncorrelated with the previous emission, which agrees with the correlation function of Fig.~\ref{fig:ShortLoop_g2}(b). This dynamic also plays out in the output flux from the system after a jump event occurs. 

Figure~\ref{fig:ShortLoop_WTDs}(c) shows a snapshot of the output flux in a trajectory immediately after a jump occurs for two separate jump events. The flux immediately increases and becomes much larger over the timescale of a single round trip. This sharp increase is due to the suppressed probability of a single photon in the loop from the destructive interference of the output bin. If a photon is detected, than it is more likely to have a second photon in the loop coming around than it is to have an empty loop. If a second emission event does not occur before the round trip ends then the flux settles back into the near constant value before the jump occurred. There are slight fluctuations in the between jump output flux, which is what determines the peak output flux after a jump occurs.

\begin{figure}[tbp]
    \centering
    \includegraphics[width=\columnwidth]{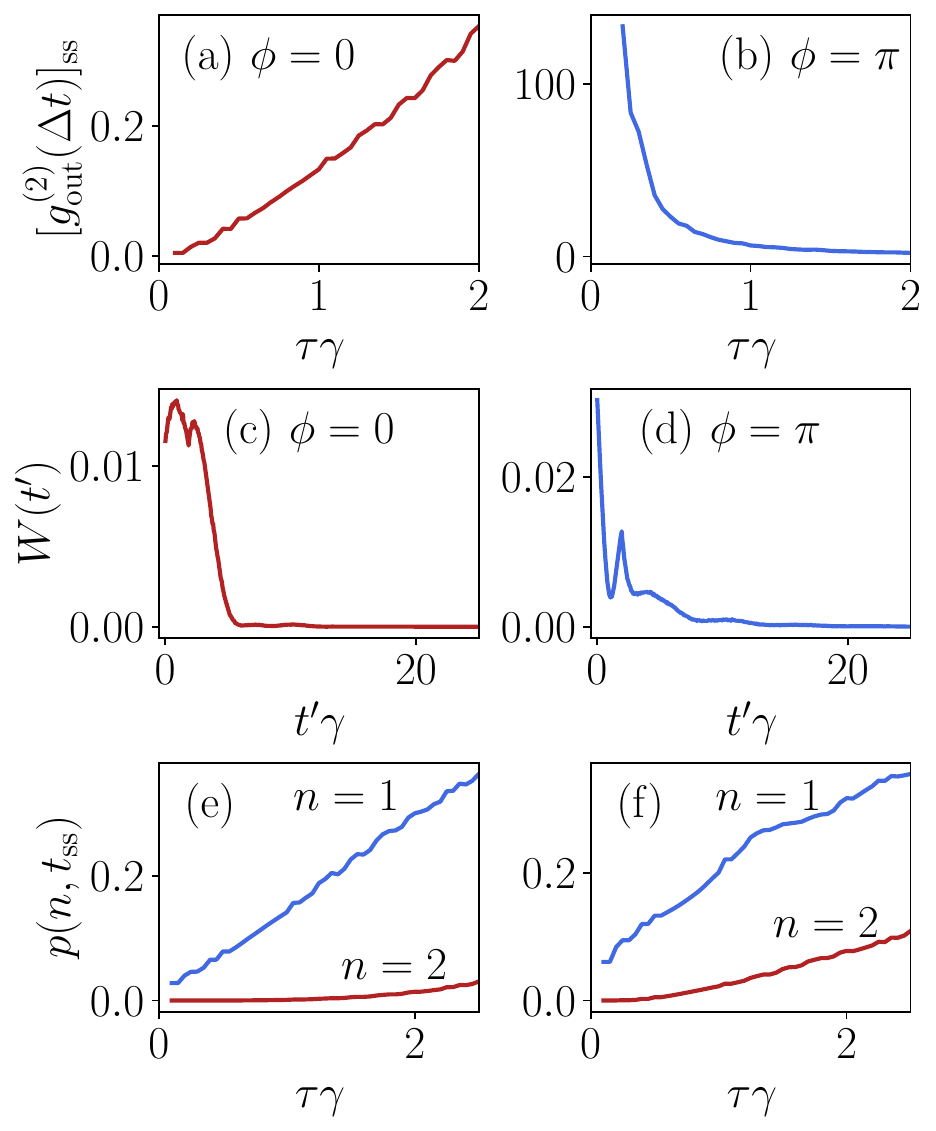}
    \vspace{-0.6cm}
    \caption{Comparison of the near-equal time second-order correlation function as a function of delay time for a TLS driven on resonance with $\Omega = 0.4 \pi \gamma$ and round trip phase change of (a) $\phi = 0$ and (b) $\phi = \pi$. For the setup with the longest delay time in (b) ($\tau = 2.0 \gamma^{-1}$), we show the WTD in (c) for $\phi = 0$ and in (d) for $\phi = \pi$. We show the probability to have $n = 1$ and $2$ photons in the loop (the blue and red lines respectively) as a function of delay time when $\phi = 0$ for drive strengths of (e) $\Omega = 0.4 \pi \gamma$ and (f) $\Omega = 2 \pi \gamma$. The results are an average of 10,000 trajectories for each delay time.}
    \label{fig:LoopLengthVary}
\end{figure}

We now turn to investigating the effect of increasing the length of the feedback loop (and thus the delay time).  Figs.~\ref{fig:LoopLengthVary}(a) and (b) show the effect of increasing the delay time on the anti-bunching and bunching for both $\phi = 0$ and $\phi = \pi$, respectively; here, the pump strength is decreased to $\Omega = 0.4 \pi \gamma$ to avoid driving too many of the system resonances, which increase the complexity of the photon counting statistics. As the loop length increases, both setups show worse statistics as they become less anti-bunched and less bunched. This is again due to the loss of coherence between the feedback and TLS. When $\phi = 0$, increasing the delay time acts to increase probability of having two photons in the loop. Therefore, when a jump occurs the remaining field in the feedback loop is non-zero. Conversely, when $\phi = \pi$, the loss of coherence causes the destructive interference between the one photon in the loop field and TLS to no longer be perfect. Somewhat counter intuitively, this means when a jump occurs, it becomes less probable that there are two photons in the loop as the loop increases in length.

It is interesting to note that the larger population in the loop actually increases the likelihood of a photon pair emission when $\phi = \pi$. When $\tau = 2 \gamma^{-1}$, the WTD for the waveguide output is shown in Fig.~\ref{fig:LoopLengthVary}(c) for $\phi = 0$ and (d) for $\phi = \pi$. The lifetime of the WTD when $\phi = \pi$ is much longer showing the increased waiting time between individual jumps, but also the initial peak is larger than the $\phi = 0$ case showing an increase in photon pair emission. Figure~\ref{fig:LoopLengthVary}(e) and (f) show the probability for one and two photons in the loop in the steady state as a function of delay time for $\Omega = 0.4 \pi \gamma$ and $\Omega = 2 \pi \gamma$, respectively, when $\phi = 0$ (note that the probabilities for one or two photons remain about same in the steady state when $\phi = \pi$). The probability to have two photons in the loop remains small past a delay time of $\tau = 2 \gamma^{-1}$. For example at $\tau = 2.5 \gamma^{-1}$, the probability is $p(2,t_{\rm ss}) \approx 0.03$ for $\Omega = 0.4 \pi \gamma$ and the probability is $p(2,t_{\rm ss}) \approx 0.11$ for $\Omega = 2 \pi \gamma$, small enough for our approximation of a maximum of two photons in the loop to be valid.

\begin{figure}[tbp]
    \centering
    \includegraphics[width=\columnwidth]{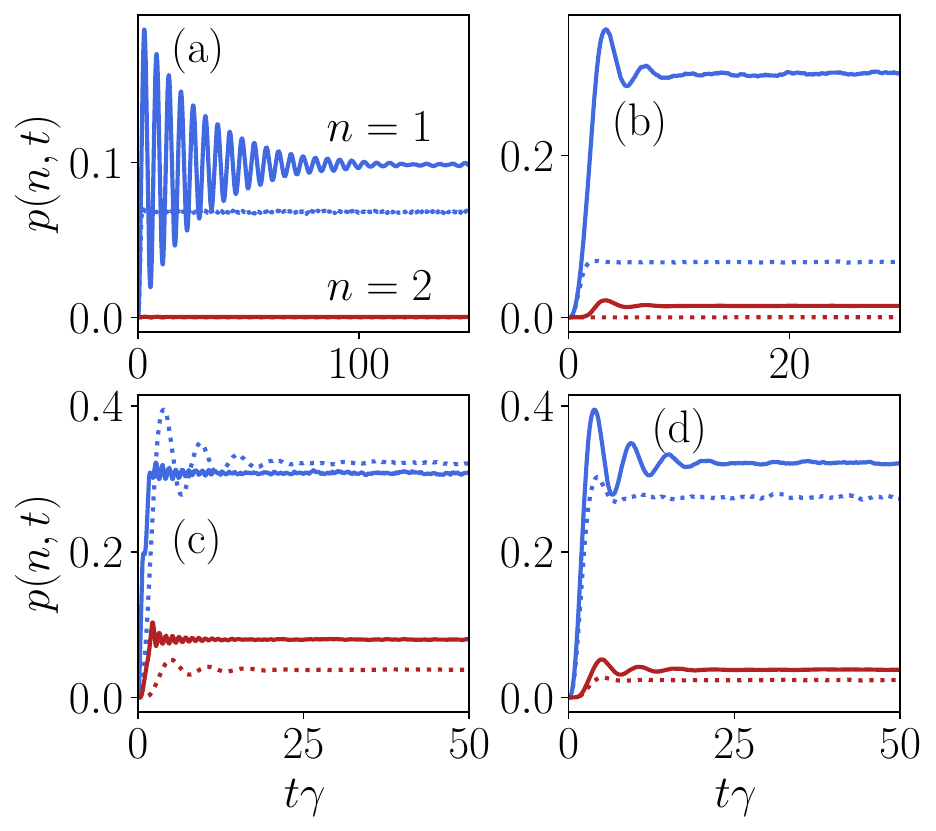}
    \vspace{-0.6cm}
    \caption{Time evolution of the probability of having $n = 1$ and $2$ photons in the loop (the blue and red lines, respectively). In (a), we set $\Omega = 0.4 \pi \gamma$ and $\tau = 0.5 \gamma^{-1}$ for $\phi = 0$ (dashed) and $\phi = \pi$ (solid). In (b), $\Omega = 0.4 \pi \gamma$ and $\phi = 0$ for $\tau = 0.5 \gamma^{-1}$ (dashed) and $\tau = 2 \gamma^{-1}$ (solid). In (c), $\tau = 2 \gamma^{-1}$ and $\phi = \pi$ for $\Omega = 0.4 \pi \gamma$ (dashed) and $\Omega = 2 \pi \gamma$ (solid). Lastly, in (d) when $\Omega = 0.4 \pi \gamma$, $\tau = 2 \gamma^{-1}$, and $\phi = \pi$, we compare the case with no outputs (solid, $\gamma_0 = \gamma' = 0$) to that where both Lindblad output channels are included (dashed, $\gamma_0 = 0.1\gamma$ and $\gamma' = 0.5\gamma$). Each result is an average of 20,000 trajectories.}
    \label{fig:LoopProbs_Time}
\end{figure}

In Fig.~\ref{fig:LoopProbs_Time}, we compare the time dynamics of the $p(n,t)$ as we vary the round trip phase change, delay time, pump strength and introduce off-chip decay and pure dephasing. In Figs.~\ref{fig:LoopProbs_Time}(a) and (b) we plot $p(n,t)$ as a function of the round trip phase change and the delay time, respectively, for $n=1$ and $2$. As expected, we see that for $\phi = \pi$ the long-lived system Rabi oscillations are passed on to the loop probabilities, while for $\phi = 0$ we do not see these oscillations. The increasing delay time also acts as expected to increase the probabilities of having both one and two photons in the loop since the feedback loop itself is longer and holds more of the emitted field from the TLS. In Fig.~\ref{fig:LoopProbs_Time}(c) the pump strength is increased to $\Omega = 2 \pi \gamma$. Although the probability to have one photon in the loop remains approximately the same, the probability for two photons almost doubles as the TLS is pumped harder. We also see the Rabi oscillations reflected in the loop probabilities die out faster for the larger pump strength. Lastly, we introduce both Lindblad output channels with $\gamma_0 = 0.1\gamma$ and $\gamma' = 0.5\gamma$ in Fig.~\ref{fig:LoopProbs_Time}(d), which act to decrease the probability to find either one photon or two photons in the loop and reduce the Rabi oscillations.

\begin{figure}[tbp]
    \centering
    \includegraphics[width=0.9\columnwidth]{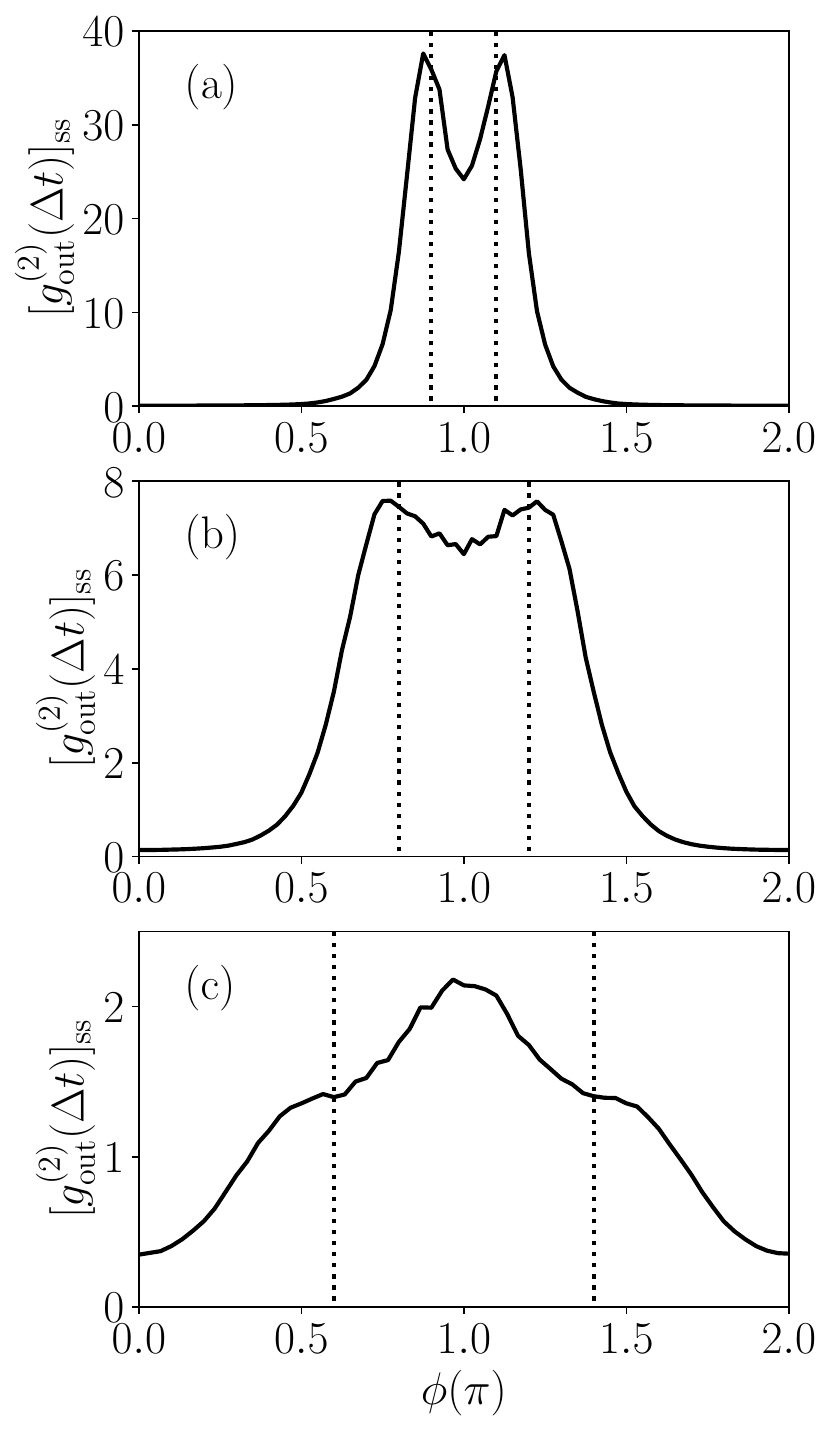}
    \vspace{-0.6cm}
    \caption{Comparison of the near-equal time second-order correlation function as a function of the round trip phase change, for three delay times: (a) $\tau = 0.5 \gamma^{-1}$, (b) $\tau = 1.0 \gamma^{-1}$, and (c) $\tau = 2.0 \gamma^{-1}$. The vertical lines denote the phase matching condition $\Omega \tau / 2 - \phi = (2k-1) \pi, k \in \mathbb{Z}$. The TLS is driven on resonance with $\Omega = 0.4 \pi \gamma$. The result for each round trip phase change is an average of 10,000 trajectories.}
    \label{fig:PhaseChangeVary}
\end{figure}

\begin{figure}[tbp]
    \centering
    \includegraphics[width=\columnwidth]{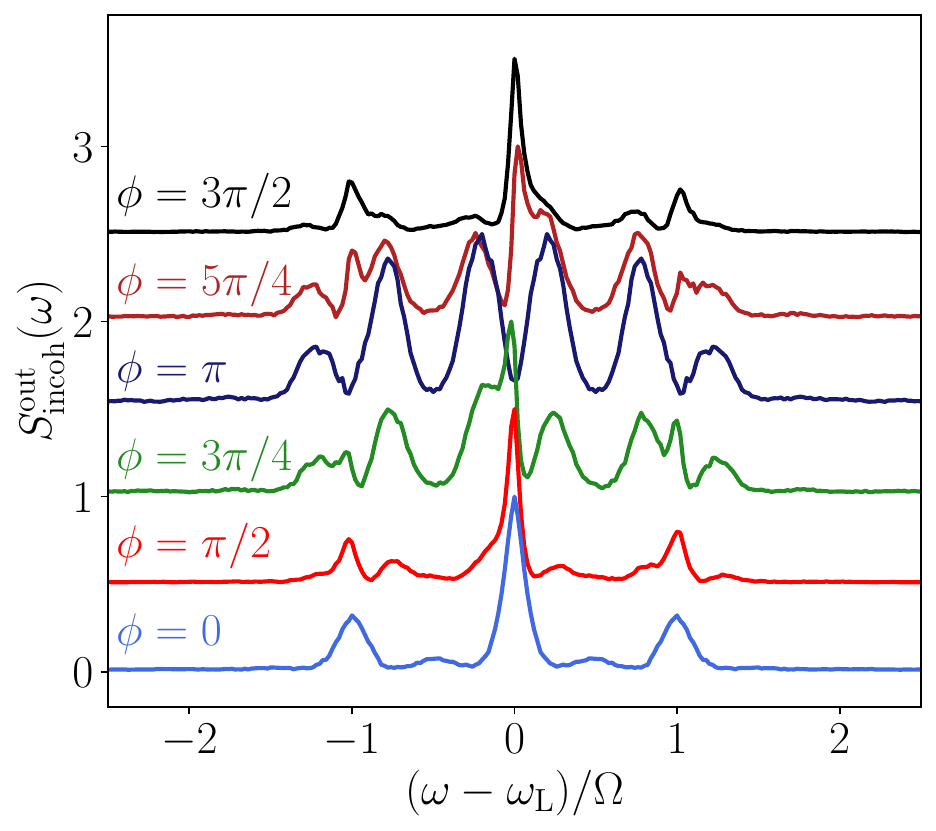}
    \vspace{-0.6cm}
    \caption{Four incoherent output spectra when the system is driven on resonance with $\Omega = 2 \pi \gamma$ and the feedback is introduced with a delay time of $\tau = 2.0 \gamma^{-1}$ for round trip phase changes of $\phi = 0$, $\phi = \pi/2$, $\phi = \pi$, and $\phi = 3\pi/2$. Each spectra is an average of 20,000 trajectories.}
    \label{fig:MultiPhotonSpectra}
\end{figure}

As the delay time from the feedback loop becomes non-negligible, the peak bunching does not occur when $\phi = \pi$ as for the short loop but rather at $\Omega \tau / 2 - \phi = (2k-1) \pi, k \in \mathbb{Z}$ as shown in Fig.~\ref{fig:PhaseChangeVary} for three different choices of loop length. This condition arises due to phase matching of the Rabi oscillations in the TLS with the phase of the returning field. Matching this condition would always give the peak bunching in the one photon in the loop limit; however, when two photons are present this is not the case as the peaks shift slightly and are not always the dominant phase choice for bunching as shown in Fig.~\ref{fig:PhaseChangeVary}(c). Here, $\tau = 2.0 \gamma^{-1}$ and $\Omega = 0.4 \pi \gamma$, so we would expect the peaks at $\phi = 0.6 \pi, 1.4 \pi$; instead these are present but the central peak is at $\phi = \pi$. This likely comes back to the fact that when there is a probability for two photons in the loop, they both cannot be phase matched at the same time for complete interference. Thus, at the longer loop lengths when the loop population increases, a maximal $[g_{\rm out}^{(2)}(\Delta t)]_{\rm ss}$ occurs when $\phi = \pi$ and the returning field interferes at the center frequency rather than trying to match the one photon in the loop field.

Furthermore, as the population in the loop increases this gives rise to additional resonances in the output spectrum arising from the feedback loop. These additional resonances are at $\omega - \omega_{\rm L} = \left( 2 \pi k + \phi \right) / 2 \pi \tau$ and we show example spectra in Fig.~\ref{fig:MultiPhotonSpectra} for $\Omega = 2 \pi \gamma$, $\tau = 2.0\gamma^{-1}$, and \edit{six} choices of round trip phase change. These resonances are a result of the multi-mode cavity resonances set up via multiple round trips around the feedback loop, where the TLS acts as a second mirror, \edit{analogous to the Fabry-P\'{e}rot resonances set up in a closed cavity coupling to the Mollow triplet.} The cavity QED physics that begins to be introduced is fully picked up by the QTDW model at the system level where truncating to two photons in the loop is analogous to truncating the cavity to a ground state and two excited state levels.

\section{Discussion on Potential Experimental Systems to Exploit Coherent Feedback}

%\sh{Potential discussion of experimental systems with real numbers -  add in some real numbers for quantum dots first (we have numbers from before), and I can look up some superconductor systems with so-called giant atoms and the recent work of Oskar Painter - this sort of discussion is important for experimental groups ref:
%\url{https://www.nature.com/articles/s41586-019-1196-1} and anything you can find with controlled quantum dot separations, including any from NRC (Chelsea's paper PRB)}

For optical frequencies, state of the art qubits
in typical photonic waveguides typically exhibit decay rates on the order of $\gamma = 1 \,\, \rm{ns}^{-1}$ \cite{Reimer2016}. For example,  if we embed such emitters into a 
low loss $\rm{SiN}$ waveguide (assuming a group index of 2), then the required waveguide length of $L_0/2 = 0.75$ cm for a ``short" feedback loop, with delay time $\tau = 0.1 \gamma^{\edit{-1}}$. Such integration schemes have now been demonstrated experimentally, using QD nanowires and SiN waveguides~\cite{2019Dan}. For the more extreme delay time of $\tau=2 \gamma^{-1}$, as used in Fig.~\ref{fig:MultiPhotonSpectra}, then $L_0/2 = 15$ cm. So low loss waveguides are required, though the lengths could be reduced by designing slow light waveguides modes with a larger group index. To control the tuning of the round trip phase, this would require moving the QD a distance of half a wavelength to change the phase from $\phi = 0$ and $\phi = \pi$, or a means to tune the mirror phase or group velocity. 

In order to shorten the required feedback length, the QD can also be placed in a cavity (such as a Fabry-P\'erot cavity or ring resonator) to include a Purcell factor, $P_F$, to the decay rate. This would decrease the required waveguide length to be $L_0/2P_F$. For example, a modest Purcell factor of 10, would decrease the range of the waveguide length to be $L_0/2 \in [0.75,15]$ mm, where waveguide loss is less of a problem. 

At microwave frequencies, one can also exploit highly developed circuit QED systems~\cite{2020nori,2020blais}, where multiple qubits can be controlled in position with great precision~\cite{Mirhosseini2019} (even multiple qubits can be spatially controlled to fractions of a wavelength). Here the physical length scales are of course much larger, but all the results equally apply in scaled units.

\section{Conclusions}
\label{sec:Conc}

We have introduced an open-system quantum optics theory to study the nonlinear behavior of an optically pumped TLS in a waveguide system, with a time-delayed coherent feedback. To do this, we described an extension to a previous QT model of coherent feedback \cite{Regidor2021a} to explicitly characterize the waveguide output from the system. We have shown how one can obtain the first- and second-order correlation functions using the QTDW model and in turn the incoherent output spectra. We then showed a number of results focusing on how the multi quanta effects from the feedback loop effect the output observables. Notably, with the proper phase matching conditions, the feedback can be used to filter out the central laser peak, switch the system through anti-bunched and bunched light output, and enhance the photon pair emission from the system. We also show how the feedback loop can introduce new resonances into the output spectra.

By focusing on incorporating practical experimental observables, along with the unique stochastic insights from QT theory, this approach is useful to connect to experimental realizations of time-delayed coherent feedback. The inclusion of pure dephasing, an important experimental consideration, is a natural addition to the QTDW model but can be added into the MPS formalism with some added complexity \cite{Kaestle2021}. To continue expanding on the practical usefulness of this approach, the addition of incoming photon wavepackets from the waveguide input rather than a CW-drive is necessary to investigate the effect of feedback on the single photon source properties of the TLS. Also, an expansion of the TLS to multi-exciton or biexciton systems will more accurately represent the experimental reality of embedded QDs \cite{Reimer2010}.

\vspace{0.5cm}

\acknowledgements
This work was supported by the Natural Sciences and Engineering Research Council of Canada, the Canadian Foundation for Innovation, the National Research Council of Canada, Queen's University and the University of Ottawa. We thank Dan Dalacu and Robin Williams for support and useful discussions.

\bibliography{PaperDraft1}

\begin{thebibliography}{101}%
\makeatletter
\providecommand \@ifxundefined [1]{%
 \@ifx{#1\undefined}
}%
\providecommand \@ifnum [1]{%
 \ifnum #1\expandafter \@firstoftwo
 \else \expandafter \@secondoftwo
 \fi
}%
\providecommand \@ifx [1]{%
 \ifx #1\expandafter \@firstoftwo
 \else \expandafter \@secondoftwo
 \fi
}%
\providecommand \natexlab [1]{#1}%
\providecommand \enquote  [1]{``#1''}%
\providecommand \bibnamefont  [1]{#1}%
\providecommand \bibfnamefont [1]{#1}%
\providecommand \citenamefont [1]{#1}%
\providecommand \href@noop [0]{\@secondoftwo}%
\providecommand \href [0]{\begingroup \@sanitize@url \@href}%
\providecommand \@href[1]{\@@startlink{#1}\@@href}%
\providecommand \@@href[1]{\endgroup#1\@@endlink}%
\providecommand \@sanitize@url [0]{\catcode `\\12\catcode `\$12\catcode
  `\&12\catcode `\#12\catcode `\^12\catcode `\_12\catcode `\%12\relax}%
\providecommand \@@startlink[1]{}%
\providecommand \@@endlink[0]{}%
\providecommand \url  [0]{\begingroup\@sanitize@url \@url }%
\providecommand \@url [1]{\endgroup\@href {#1}{\urlprefix }}%
\providecommand \urlprefix  [0]{URL }%
\providecommand \Eprint [0]{\href }%
\providecommand \doibase [0]{https://doi.org/}%
\providecommand \selectlanguage [0]{\@gobble}%
\providecommand \bibinfo  [0]{\@secondoftwo}%
\providecommand \bibfield  [0]{\@secondoftwo}%
\providecommand \translation [1]{[#1]}%
\providecommand \BibitemOpen [0]{}%
\providecommand \bibitemStop [0]{}%
\providecommand \bibitemNoStop [0]{.\EOS\space}%
\providecommand \EOS [0]{\spacefactor3000\relax}%
\providecommand \BibitemShut  [1]{\csname bibitem#1\endcsname}%
\let\auto@bib@innerbib\@empty
%</preamble>
\bibitem [{\citenamefont {Dorner}\ and\ \citenamefont
  {Zoller}(2002)}]{Dorner2002}%
  \BibitemOpen
  \bibfield  {author} {\bibinfo {author} {\bibfnamefont {U.}~\bibnamefont
  {Dorner}}\ and\ \bibinfo {author} {\bibfnamefont {P.}~\bibnamefont
  {Zoller}},\ }\bibfield  {title} {\bibinfo {title} {Laser-driven atoms in
  half-cavities},\ }\href {https://doi.org/10.1103/PhysRevA.66.023816}
  {\bibfield  {journal} {\bibinfo  {journal} {Physical Review A}\ }\textbf
  {\bibinfo {volume} {66}},\ \bibinfo {pages} {023816} (\bibinfo {year}
  {2002})}\BibitemShut {NoStop}%
\bibitem [{\citenamefont {Tufarelli}\ \emph {et~al.}(2013)\citenamefont
  {Tufarelli}, \citenamefont {Ciccarello},\ and\ \citenamefont
  {Kim}}]{Tufarelli2013}%
  \BibitemOpen
  \bibfield  {author} {\bibinfo {author} {\bibfnamefont {T.}~\bibnamefont
  {Tufarelli}}, \bibinfo {author} {\bibfnamefont {F.}~\bibnamefont
  {Ciccarello}},\ and\ \bibinfo {author} {\bibfnamefont {M.~S.}\ \bibnamefont
  {Kim}},\ }\bibfield  {title} {\bibinfo {title} {Dynamics of spontaneous
  emission in a single-end photonic waveguide},\ }\href
  {https://doi.org/10.1103/PhysRevA.87.013820} {\bibfield  {journal} {\bibinfo
  {journal} {Physical Review A}\ }\textbf {\bibinfo {volume} {87}},\ \bibinfo
  {pages} {013820} (\bibinfo {year} {2013})}\BibitemShut {NoStop}%
\bibitem [{\citenamefont {Carmele}\ \emph {et~al.}(2013)\citenamefont
  {Carmele}, \citenamefont {Kabuss}, \citenamefont {Schulze}, \citenamefont
  {Reitzenstein},\ and\ \citenamefont {Knorr}}]{Carmele2013}%
  \BibitemOpen
  \bibfield  {author} {\bibinfo {author} {\bibfnamefont {A.}~\bibnamefont
  {Carmele}}, \bibinfo {author} {\bibfnamefont {J.}~\bibnamefont {Kabuss}},
  \bibinfo {author} {\bibfnamefont {F.}~\bibnamefont {Schulze}}, \bibinfo
  {author} {\bibfnamefont {S.}~\bibnamefont {Reitzenstein}},\ and\ \bibinfo
  {author} {\bibfnamefont {A.}~\bibnamefont {Knorr}},\ }\bibfield  {title}
  {\bibinfo {title} {Single {Photon} {Delayed} {Feedback}: {A} {Way} to
  {Stabilize} {Intrinsic} {Quantum} {Cavity} {Electrodynamics}},\ }\href
  {https://doi.org/10.1103/PhysRevLett.110.013601} {\bibfield  {journal}
  {\bibinfo  {journal} {Physical Review Letters}\ }\textbf {\bibinfo {volume}
  {110}},\ \bibinfo {pages} {013601} (\bibinfo {year} {2013})}\BibitemShut
  {NoStop}%
\bibitem [{\citenamefont {Naumann}\ \emph {et~al.}(2016)\citenamefont
  {Naumann}, \citenamefont {Droenner}, \citenamefont {Hein}, \citenamefont
  {Carmele}, \citenamefont {Knorr},\ and\ \citenamefont
  {Kabuss}}]{Naumann2016}%
  \BibitemOpen
  \bibfield  {author} {\bibinfo {author} {\bibfnamefont {N.~L.}\ \bibnamefont
  {Naumann}}, \bibinfo {author} {\bibfnamefont {L.}~\bibnamefont {Droenner}},
  \bibinfo {author} {\bibfnamefont {S.~M.}\ \bibnamefont {Hein}}, \bibinfo
  {author} {\bibfnamefont {A.}~\bibnamefont {Carmele}}, \bibinfo {author}
  {\bibfnamefont {A.}~\bibnamefont {Knorr}},\ and\ \bibinfo {author}
  {\bibfnamefont {J.}~\bibnamefont {Kabuss}},\ }\bibfield  {title} {\bibinfo
  {title} {Feedback control of optomechanical systems},\ }in\ \href
  {https://doi.org/10.1117/12.2209338} {\emph {\bibinfo {booktitle} {Physics
  and {Simulation} of {Optoelectronic} {Devices} {XXIV}}}},\ Vol.\ \bibinfo
  {volume} {9742}\ (\bibinfo  {publisher} {International Society for Optics and
  Photonics},\ \bibinfo {year} {2016})\ p.\ \bibinfo {pages}
  {974216}\BibitemShut {NoStop}%
\bibitem [{\citenamefont {Lu}\ \emph {et~al.}(2017)\citenamefont {Lu},
  \citenamefont {Naumann}, \citenamefont {Cerrillo}, \citenamefont {Zhao},
  \citenamefont {Knorr},\ and\ \citenamefont {Carmele}}]{Lu2017}%
  \BibitemOpen
  \bibfield  {author} {\bibinfo {author} {\bibfnamefont {Y.}~\bibnamefont
  {Lu}}, \bibinfo {author} {\bibfnamefont {N.~L.}\ \bibnamefont {Naumann}},
  \bibinfo {author} {\bibfnamefont {J.}~\bibnamefont {Cerrillo}}, \bibinfo
  {author} {\bibfnamefont {Q.}~\bibnamefont {Zhao}}, \bibinfo {author}
  {\bibfnamefont {A.}~\bibnamefont {Knorr}},\ and\ \bibinfo {author}
  {\bibfnamefont {A.}~\bibnamefont {Carmele}},\ }\bibfield  {title} {\bibinfo
  {title} {Intensified antibunching via feedback-induced quantum
  interference},\ }\href {https://doi.org/10.1103/PhysRevA.95.063840}
  {\bibfield  {journal} {\bibinfo  {journal} {Physical Review A}\ }\textbf
  {\bibinfo {volume} {95}},\ \bibinfo {pages} {063840} (\bibinfo {year}
  {2017})}\BibitemShut {NoStop}%
\bibitem [{\citenamefont {Pichler}\ \emph {et~al.}(2017)\citenamefont
  {Pichler}, \citenamefont {Choi}, \citenamefont {Zoller},\ and\ \citenamefont
  {Lukin}}]{Pichler2017}%
  \BibitemOpen
  \bibfield  {author} {\bibinfo {author} {\bibfnamefont {H.}~\bibnamefont
  {Pichler}}, \bibinfo {author} {\bibfnamefont {S.}~\bibnamefont {Choi}},
  \bibinfo {author} {\bibfnamefont {P.}~\bibnamefont {Zoller}},\ and\ \bibinfo
  {author} {\bibfnamefont {M.~D.}\ \bibnamefont {Lukin}},\ }\bibfield  {title}
  {\bibinfo {title} {Universal photonic quantum computation via time-delayed
  feedback},\ }\href {https://doi.org/10.1073/pnas.1711003114} {\bibfield
  {journal} {\bibinfo  {journal} {Proceedings of the National Academy of
  Sciences}\ }\textbf {\bibinfo {volume} {114}},\ \bibinfo {pages} {11362}
  (\bibinfo {year} {2017})}\BibitemShut {NoStop}%
\bibitem [{\citenamefont {Wiseman}\ and\ \citenamefont
  {Milburn}(2002)}]{Wiseman2006}%
  \BibitemOpen
  \bibfield  {author} {\bibinfo {author} {\bibfnamefont {H.~M.}\ \bibnamefont
  {Wiseman}}\ and\ \bibinfo {author} {\bibfnamefont {G.~J.}\ \bibnamefont
  {Milburn}},\ }\href@noop {} {\emph {\bibinfo {title} {Quantum {M}easurement
  and {C}ontrol}}}\ (\bibinfo  {publisher} {Cambridge University Press},\
  \bibinfo {address} {Oxford},\ \bibinfo {year} {2002})\ p.\ \bibinfo {pages}
  {231}\BibitemShut {NoStop}%
\bibitem [{\citenamefont {Mu\~{n}oz Arias}\ \emph {et~al.}(2020)\citenamefont
  {Mu\~{n}oz Arias}, \citenamefont {Deutsch}, \citenamefont {Jessen},\ and\
  \citenamefont {Poggi}}]{Arias2020}%
  \BibitemOpen
  \bibfield  {author} {\bibinfo {author} {\bibfnamefont {M.~H.}\ \bibnamefont
  {Mu\~{n}oz Arias}}, \bibinfo {author} {\bibfnamefont {I.~H.}\ \bibnamefont
  {Deutsch}}, \bibinfo {author} {\bibfnamefont {P.~S.}\ \bibnamefont
  {Jessen}},\ and\ \bibinfo {author} {\bibfnamefont {P.~M.}\ \bibnamefont
  {Poggi}},\ }\bibfield  {title} {\bibinfo {title} {Simulation of complex
  dynamics of mean-field $p$-spin models using measurement-based quantum
  feedback control},\ }\href {https://doi.org/10.1103/PhysRevA.102.022610}
  {\bibfield  {journal} {\bibinfo  {journal} {Physical Review A}\ }\textbf
  {\bibinfo {volume} {102}},\ \bibinfo {pages} {022610} (\bibinfo {year}
  {2020})}\BibitemShut {NoStop}%
\bibitem [{\citenamefont {Grigoletto}\ and\ \citenamefont
  {Ticozzi}(2020)}]{Grigoletto2020}%
  \BibitemOpen
  \bibfield  {author} {\bibinfo {author} {\bibfnamefont {T.}~\bibnamefont
  {Grigoletto}}\ and\ \bibinfo {author} {\bibfnamefont {F.}~\bibnamefont
  {Ticozzi}},\ }\bibfield  {title} {\bibinfo {title} {Stabilization via
  feedback switching for quantum stochastic dynamics},\ }\href
  {http://arxiv.org/abs/2012.08712} {\bibfield  {journal} {\bibinfo  {journal}
  {arXiv:2012.08712 [quant-ph]}\ } (\bibinfo {year} {2020})}\BibitemShut
  {NoStop}%
\bibitem [{\citenamefont {{Hjelme}}\ \emph {et~al.}(1991)\citenamefont
  {{Hjelme}}, \citenamefont {{Mickelson}},\ and\ \citenamefont
  {{Beausoleil}}}]{Hjelme1991}%
  \BibitemOpen
  \bibfield  {author} {\bibinfo {author} {\bibfnamefont {D.~R.}\ \bibnamefont
  {{Hjelme}}}, \bibinfo {author} {\bibfnamefont {A.~R.}\ \bibnamefont
  {{Mickelson}}},\ and\ \bibinfo {author} {\bibfnamefont {R.~G.}\ \bibnamefont
  {{Beausoleil}}},\ }\bibfield  {title} {\bibinfo {title} {Semiconductor laser
  stabilization by external optical feedback},\ }\href
  {https://ieeexplore.ieee.org/document/81333} {\bibfield  {journal} {\bibinfo
  {journal} {IEEE Journal of Quantum Electronics}\ }\textbf {\bibinfo {volume}
  {27}},\ \bibinfo {pages} {352} (\bibinfo {year} {1991})}\BibitemShut
  {NoStop}%
\bibitem [{\citenamefont {Cosentino}\ and\ \citenamefont
  {Bates}(2011)}]{Biology}%
  \BibitemOpen
  \bibfield  {author} {\bibinfo {author} {\bibfnamefont {C.}~\bibnamefont
  {Cosentino}}\ and\ \bibinfo {author} {\bibfnamefont {D.}~\bibnamefont
  {Bates}},\ }\href@noop {} {\emph {\bibinfo {title} {Feedback Control in
  Systems Biology}}}\ (\bibinfo  {publisher} {CRC Press},\ \bibinfo {year}
  {2011})\BibitemShut {NoStop}%
\bibitem [{\citenamefont {Franklin}\ \emph {et~al.}(2014)\citenamefont
  {Franklin}, \citenamefont {Powell},\ and\ \citenamefont
  {Emami-Naeini}}]{Franklin2014}%
  \BibitemOpen
  \bibfield  {author} {\bibinfo {author} {\bibfnamefont {G.~F.}\ \bibnamefont
  {Franklin}}, \bibinfo {author} {\bibfnamefont {J.~D.}\ \bibnamefont
  {Powell}},\ and\ \bibinfo {author} {\bibfnamefont {A.}~\bibnamefont
  {Emami-Naeini}},\ }\href@noop {} {\emph {\bibinfo {title} {Feedback Control
  of Dynamic Systems}}}\ (\bibinfo  {publisher} {Pearson},\ \bibinfo {year}
  {2014})\BibitemShut {NoStop}%
\bibitem [{\citenamefont {Kubanek}\ \emph {et~al.}(2009)\citenamefont
  {Kubanek}, \citenamefont {Koch}, \citenamefont {Sames}, \citenamefont
  {Ourjoumtsev}, \citenamefont {Pinkse}, \citenamefont {Murr},\ and\
  \citenamefont {Rempe}}]{Kubanek2009}%
  \BibitemOpen
  \bibfield  {author} {\bibinfo {author} {\bibfnamefont {A.}~\bibnamefont
  {Kubanek}}, \bibinfo {author} {\bibfnamefont {M.}~\bibnamefont {Koch}},
  \bibinfo {author} {\bibfnamefont {C.}~\bibnamefont {Sames}}, \bibinfo
  {author} {\bibfnamefont {A.}~\bibnamefont {Ourjoumtsev}}, \bibinfo {author}
  {\bibfnamefont {P.~W.~H.}\ \bibnamefont {Pinkse}}, \bibinfo {author}
  {\bibfnamefont {K.}~\bibnamefont {Murr}},\ and\ \bibinfo {author}
  {\bibfnamefont {G.}~\bibnamefont {Rempe}},\ }\bibfield  {title} {\bibinfo
  {title} {Photon-by-photon feedback control of a single-atom trajectory},\
  }\href {https://doi.org/10.1038/nature08563} {\bibfield  {journal} {\bibinfo
  {journal} {Nature}\ }\textbf {\bibinfo {volume} {462}},\ \bibinfo {pages}
  {898} (\bibinfo {year} {2009})}\BibitemShut {NoStop}%
\bibitem [{\citenamefont {Gillett}\ \emph {et~al.}(2010)\citenamefont
  {Gillett}, \citenamefont {Dalton}, \citenamefont {Lanyon}, \citenamefont
  {Almeida}, \citenamefont {Barbieri}, \citenamefont {Pryde}, \citenamefont
  {O'Brien}, \citenamefont {Resch}, \citenamefont {Bartlett},\ and\
  \citenamefont {White}}]{Gillett2010}%
  \BibitemOpen
  \bibfield  {author} {\bibinfo {author} {\bibfnamefont {G.~G.}\ \bibnamefont
  {Gillett}}, \bibinfo {author} {\bibfnamefont {R.~B.}\ \bibnamefont {Dalton}},
  \bibinfo {author} {\bibfnamefont {B.~P.}\ \bibnamefont {Lanyon}}, \bibinfo
  {author} {\bibfnamefont {M.~P.}\ \bibnamefont {Almeida}}, \bibinfo {author}
  {\bibfnamefont {M.}~\bibnamefont {Barbieri}}, \bibinfo {author}
  {\bibfnamefont {G.~J.}\ \bibnamefont {Pryde}}, \bibinfo {author}
  {\bibfnamefont {J.~L.}\ \bibnamefont {O'Brien}}, \bibinfo {author}
  {\bibfnamefont {K.~J.}\ \bibnamefont {Resch}}, \bibinfo {author}
  {\bibfnamefont {S.~D.}\ \bibnamefont {Bartlett}},\ and\ \bibinfo {author}
  {\bibfnamefont {A.~G.}\ \bibnamefont {White}},\ }\bibfield  {title} {\bibinfo
  {title} {Experimental feedback control of quantum systems using weak
  measurements},\ }\href {https://doi.org/10.1103/PhysRevLett.104.080503}
  {\bibfield  {journal} {\bibinfo  {journal} {Phys. Rev. Lett.}\ }\textbf
  {\bibinfo {volume} {104}},\ \bibinfo {pages} {080503} (\bibinfo {year}
  {2010})}\BibitemShut {NoStop}%
\bibitem [{\citenamefont {Rafiee}\ \emph {et~al.}(2020)\citenamefont {Rafiee},
  \citenamefont {Nourmandipour},\ and\ \citenamefont {Mancini}}]{Rafiee2020}%
  \BibitemOpen
  \bibfield  {author} {\bibinfo {author} {\bibfnamefont {M.}~\bibnamefont
  {Rafiee}}, \bibinfo {author} {\bibfnamefont {A.}~\bibnamefont
  {Nourmandipour}},\ and\ \bibinfo {author} {\bibfnamefont {S.}~\bibnamefont
  {Mancini}},\ }\bibfield  {title} {\bibinfo {title} {Enforcing dissipative
  entanglement by feedback},\ }\href
  {https://doi.org/10.1016/j.physleta.2020.126748} {\bibfield  {journal}
  {\bibinfo  {journal} {Physics Letters A}\ }\textbf {\bibinfo {volume}
  {384}},\ \bibinfo {pages} {126748} (\bibinfo {year} {2020})}\BibitemShut
  {NoStop}%
\bibitem [{\citenamefont {Tan}\ \emph {et~al.}(2021)\citenamefont {Tan},
  \citenamefont {Camati}, \citenamefont {Cauquil}, \citenamefont {Auff\`eves},\
  and\ \citenamefont {Dotsenko}}]{Tan2020}%
  \BibitemOpen
  \bibfield  {author} {\bibinfo {author} {\bibfnamefont {Z.}~\bibnamefont
  {Tan}}, \bibinfo {author} {\bibfnamefont {P.~A.}\ \bibnamefont {Camati}},
  \bibinfo {author} {\bibfnamefont {G.~C.}\ \bibnamefont {Cauquil}}, \bibinfo
  {author} {\bibfnamefont {A.}~\bibnamefont {Auff\`eves}},\ and\ \bibinfo
  {author} {\bibfnamefont {I.}~\bibnamefont {Dotsenko}},\ }\bibfield  {title}
  {\bibinfo {title} {Alternative experimental ways to access entropy
  production},\ }\href {https://doi.org/10.1103/PhysRevResearch.3.043076}
  {\bibfield  {journal} {\bibinfo  {journal} {Phys. Rev. Research}\ }\textbf
  {\bibinfo {volume} {3}},\ \bibinfo {pages} {043076} (\bibinfo {year}
  {2021})}\BibitemShut {NoStop}%
\bibitem [{\citenamefont {Di~Giovanni}\ \emph {et~al.}(2021)\citenamefont
  {Di~Giovanni}, \citenamefont {Brunelli},\ and\ \citenamefont
  {Genoni}}]{Giovanni2021}%
  \BibitemOpen
  \bibfield  {author} {\bibinfo {author} {\bibfnamefont {A.}~\bibnamefont
  {Di~Giovanni}}, \bibinfo {author} {\bibfnamefont {M.}~\bibnamefont
  {Brunelli}},\ and\ \bibinfo {author} {\bibfnamefont {M.~G.}\ \bibnamefont
  {Genoni}},\ }\bibfield  {title} {\bibinfo {title} {Unconditional mechanical
  squeezing via backaction-evading measurements and nonoptimal feedback
  control},\ }\href {https://doi.org/10.1103/PhysRevA.103.022614} {\bibfield
  {journal} {\bibinfo  {journal} {Physical Review A}\ }\textbf {\bibinfo
  {volume} {103}},\ \bibinfo {pages} {022614} (\bibinfo {year}
  {2021})}\BibitemShut {NoStop}%
\bibitem [{\citenamefont {Shi}\ and\ \citenamefont {Waks}(2021)}]{Shi2021}%
  \BibitemOpen
  \bibfield  {author} {\bibinfo {author} {\bibfnamefont {Y.}~\bibnamefont
  {Shi}}\ and\ \bibinfo {author} {\bibfnamefont {E.}~\bibnamefont {Waks}},\
  }\bibfield  {title} {\bibinfo {title} {Deterministic generation of
  multidimensional photonic cluster states using time-delay feedback},\ }\href
  {https://doi.org/10.1103/PhysRevA.104.013703} {\bibfield  {journal} {\bibinfo
   {journal} {Physical Review A}\ }\textbf {\bibinfo {volume} {104}},\ \bibinfo
  {pages} {013703} (\bibinfo {year} {2021})}\BibitemShut {NoStop}%
\bibitem [{\citenamefont {Koshino}\ and\ \citenamefont
  {Nakamura}(2012)}]{Koshino2012}%
  \BibitemOpen
  \bibfield  {author} {\bibinfo {author} {\bibfnamefont {K.}~\bibnamefont
  {Koshino}}\ and\ \bibinfo {author} {\bibfnamefont {Y.}~\bibnamefont
  {Nakamura}},\ }\bibfield  {title} {\bibinfo {title} {Control of the radiative
  level shift and linewidth of a superconducting artificial atom through a
  variable boundary condition},\ }\href
  {https://doi.org/10.1088/1367-2630/14/4/043005} {\bibfield  {journal}
  {\bibinfo  {journal} {New Journal of Physics}\ }\textbf {\bibinfo {volume}
  {14}},\ \bibinfo {pages} {043005} (\bibinfo {year} {2012})}\BibitemShut
  {NoStop}%
\bibitem [{\citenamefont {Hoi}\ \emph {et~al.}(2015)\citenamefont {Hoi},
  \citenamefont {Kockum}, \citenamefont {Tornberg}, \citenamefont
  {Pourkabirian}, \citenamefont {Johansson}, \citenamefont {Delsing},\ and\
  \citenamefont {Wilson}}]{Hoi2015}%
  \BibitemOpen
  \bibfield  {author} {\bibinfo {author} {\bibfnamefont {I.-C.}\ \bibnamefont
  {Hoi}}, \bibinfo {author} {\bibfnamefont {A.~F.}\ \bibnamefont {Kockum}},
  \bibinfo {author} {\bibfnamefont {L.}~\bibnamefont {Tornberg}}, \bibinfo
  {author} {\bibfnamefont {A.}~\bibnamefont {Pourkabirian}}, \bibinfo {author}
  {\bibfnamefont {G.}~\bibnamefont {Johansson}}, \bibinfo {author}
  {\bibfnamefont {P.}~\bibnamefont {Delsing}},\ and\ \bibinfo {author}
  {\bibfnamefont {C.~M.}\ \bibnamefont {Wilson}},\ }\bibfield  {title}
  {\bibinfo {title} {Probing the quantum vacuum with an artificial atom in
  front of a mirror},\ }\href {https://doi.org/10.1038/nphys3484} {\bibfield
  {journal} {\bibinfo  {journal} {Nature Physics}\ }\textbf {\bibinfo {volume}
  {11}},\ \bibinfo {pages} {1045} (\bibinfo {year} {2015})}\BibitemShut
  {NoStop}%
\bibitem [{\citenamefont {Grimsmo}(2015)}]{Grimsmo2015}%
  \BibitemOpen
  \bibfield  {author} {\bibinfo {author} {\bibfnamefont {A.~L.}\ \bibnamefont
  {Grimsmo}},\ }\bibfield  {title} {\bibinfo {title} {Time-{Delayed} {Quantum}
  {Feedback} {Control}},\ }\href
  {https://doi.org/10.1103/PhysRevLett.115.060402} {\bibfield  {journal}
  {\bibinfo  {journal} {Physical Review Letters}\ }\textbf {\bibinfo {volume}
  {115}},\ \bibinfo {pages} {060402} (\bibinfo {year} {2015})}\BibitemShut
  {NoStop}%
\bibitem [{\citenamefont {Kabuss}\ \emph {et~al.}(2016)\citenamefont {Kabuss},
  \citenamefont {Katsch}, \citenamefont {Knorr},\ and\ \citenamefont
  {Carmele}}]{Kabuss2016}%
  \BibitemOpen
  \bibfield  {author} {\bibinfo {author} {\bibfnamefont {J.}~\bibnamefont
  {Kabuss}}, \bibinfo {author} {\bibfnamefont {F.}~\bibnamefont {Katsch}},
  \bibinfo {author} {\bibfnamefont {A.}~\bibnamefont {Knorr}},\ and\ \bibinfo
  {author} {\bibfnamefont {A.}~\bibnamefont {Carmele}},\ }\bibfield  {title}
  {\bibinfo {title} {Unraveling coherent quantum feedback for {Pyragas}
  control},\ }\href {https://doi.org/10.1364/JOSAB.33.000C10} {\bibfield
  {journal} {\bibinfo  {journal} {Journal of the Optical Society of America B}\
  }\textbf {\bibinfo {volume} {33}},\ \bibinfo {pages} {C10} (\bibinfo {year}
  {2016})}\BibitemShut {NoStop}%
\bibitem [{\citenamefont {Pichler}\ and\ \citenamefont
  {Zoller}(2016)}]{Pichler2016}%
  \BibitemOpen
  \bibfield  {author} {\bibinfo {author} {\bibfnamefont {H.}~\bibnamefont
  {Pichler}}\ and\ \bibinfo {author} {\bibfnamefont {P.}~\bibnamefont
  {Zoller}},\ }\bibfield  {title} {\bibinfo {title} {Photonic {Circuits} with
  {Time} {Delays} and {Quantum} {Feedback}},\ }\href
  {https://doi.org/10.1103/PhysRevLett.116.093601} {\bibfield  {journal}
  {\bibinfo  {journal} {Physical Review Letters}\ }\textbf {\bibinfo {volume}
  {116}},\ \bibinfo {pages} {093601} (\bibinfo {year} {2016})}\BibitemShut
  {NoStop}%
\bibitem [{\citenamefont {N\'{e}met}\ and\ \citenamefont
  {Parkins}(2016)}]{Nemet2016}%
  \BibitemOpen
  \bibfield  {author} {\bibinfo {author} {\bibfnamefont {N.}~\bibnamefont
  {N\'{e}met}}\ and\ \bibinfo {author} {\bibfnamefont {S.}~\bibnamefont
  {Parkins}},\ }\bibfield  {title} {\bibinfo {title} {Enhanced optical
  squeezing from a degenerate parametric amplifier via time-delayed coherent
  feedback},\ }\href {https://doi.org/10.1103/PhysRevA.94.023809} {\bibfield
  {journal} {\bibinfo  {journal} {Physical Review A}\ }\textbf {\bibinfo
  {volume} {94}},\ \bibinfo {pages} {023809} (\bibinfo {year}
  {2016})}\BibitemShut {NoStop}%
\bibitem [{\citenamefont {Hein}\ \emph {et~al.}(2016)\citenamefont {Hein},
  \citenamefont {Carmele},\ and\ \citenamefont {Knorr}}]{Hein2016}%
  \BibitemOpen
  \bibfield  {author} {\bibinfo {author} {\bibfnamefont {S.~M.}\ \bibnamefont
  {Hein}}, \bibinfo {author} {\bibfnamefont {A.}~\bibnamefont {Carmele}},\ and\
  \bibinfo {author} {\bibfnamefont {A.}~\bibnamefont {Knorr}},\ }\bibfield
  {title} {\bibinfo {title} {Creation and control of entanglement by
  time-delayed quantum-coherent feedback},\ }in\ \href
  {https://doi.org/10.1117/12.2207671} {\emph {\bibinfo {booktitle} {Physics
  and {Simulation} of {Optoelectronic} {Devices} {XXIV}}}},\ Vol.\ \bibinfo
  {volume} {9742}\ (\bibinfo  {publisher} {International Society for Optics and
  Photonics},\ \bibinfo {year} {2016})\ p.\ \bibinfo {pages}
  {97420X}\BibitemShut {NoStop}%
\bibitem [{\citenamefont {Guimond}\ \emph {et~al.}(2016)\citenamefont
  {Guimond}, \citenamefont {Pichler}, \citenamefont {Rauschenbeutel},\ and\
  \citenamefont {Zoller}}]{Guimond2016}%
  \BibitemOpen
  \bibfield  {author} {\bibinfo {author} {\bibfnamefont {P.-O.}\ \bibnamefont
  {Guimond}}, \bibinfo {author} {\bibfnamefont {H.}~\bibnamefont {Pichler}},
  \bibinfo {author} {\bibfnamefont {A.}~\bibnamefont {Rauschenbeutel}},\ and\
  \bibinfo {author} {\bibfnamefont {P.}~\bibnamefont {Zoller}},\ }\bibfield
  {title} {\bibinfo {title} {Chiral quantum optics with {V}-level atoms and
  coherent quantum feedback},\ }\href
  {https://doi.org/10.1103/PhysRevA.94.033829} {\bibfield  {journal} {\bibinfo
  {journal} {Physical Review A}\ }\textbf {\bibinfo {volume} {94}},\ \bibinfo
  {pages} {033829} (\bibinfo {year} {2016})}\BibitemShut {NoStop}%
\bibitem [{\citenamefont {Whalen}\ \emph {et~al.}(2017)\citenamefont {Whalen},
  \citenamefont {Grimsmo},\ and\ \citenamefont {Carmichael}}]{Whalen2017}%
  \BibitemOpen
  \bibfield  {author} {\bibinfo {author} {\bibfnamefont {S.~J.}\ \bibnamefont
  {Whalen}}, \bibinfo {author} {\bibfnamefont {A.~L.}\ \bibnamefont
  {Grimsmo}},\ and\ \bibinfo {author} {\bibfnamefont {H.~J.}\ \bibnamefont
  {Carmichael}},\ }\bibfield  {title} {\bibinfo {title} {Open quantum systems
  with delayed coherent feedback},\ }\href
  {https://doi.org/10.1088/2058-9565/aa8331} {\bibfield  {journal} {\bibinfo
  {journal} {Quantum Science and Technology}\ }\textbf {\bibinfo {volume}
  {2}},\ \bibinfo {pages} {044008} (\bibinfo {year} {2017})}\BibitemShut
  {NoStop}%
\bibitem [{\citenamefont {Naumann}\ \emph {et~al.}(2017)\citenamefont
  {Naumann}, \citenamefont {Hein}, \citenamefont {Kraft}, \citenamefont
  {Knorr},\ and\ \citenamefont {Carmele}}]{Naumann2017}%
  \BibitemOpen
  \bibfield  {author} {\bibinfo {author} {\bibfnamefont {N.~L.}\ \bibnamefont
  {Naumann}}, \bibinfo {author} {\bibfnamefont {S.~M.}\ \bibnamefont {Hein}},
  \bibinfo {author} {\bibfnamefont {M.}~\bibnamefont {Kraft}}, \bibinfo
  {author} {\bibfnamefont {A.}~\bibnamefont {Knorr}},\ and\ \bibinfo {author}
  {\bibfnamefont {A.}~\bibnamefont {Carmele}},\ }\bibfield  {title} {\bibinfo
  {title} {Feedback control of photon statistics},\ }in\ \href
  {https://doi.org/10.1117/12.2251952} {\emph {\bibinfo {booktitle} {Physics
  and {Simulation} of {Optoelectronic} {Devices} {XXV}}}},\ Vol.\ \bibinfo
  {volume} {10098}\ (\bibinfo  {publisher} {International Society for Optics
  and Photonics},\ \bibinfo {year} {2017})\ p.\ \bibinfo {pages}
  {100980N}\BibitemShut {NoStop}%
\bibitem [{\citenamefont {Guimond}\ \emph {et~al.}(2017)\citenamefont
  {Guimond}, \citenamefont {Pletyukhov}, \citenamefont {Pichler},\ and\
  \citenamefont {Zoller}}]{Guimond2017}%
  \BibitemOpen
  \bibfield  {author} {\bibinfo {author} {\bibfnamefont {P.-O.}\ \bibnamefont
  {Guimond}}, \bibinfo {author} {\bibfnamefont {M.}~\bibnamefont {Pletyukhov}},
  \bibinfo {author} {\bibfnamefont {H.}~\bibnamefont {Pichler}},\ and\ \bibinfo
  {author} {\bibfnamefont {P.}~\bibnamefont {Zoller}},\ }\bibfield  {title}
  {\bibinfo {title} {Delayed coherent quantum feedback from a scattering theory
  and a matrix product state perspective},\ }\href
  {https://doi.org/10.1088/2058-9565/aa7f03} {\bibfield  {journal} {\bibinfo
  {journal} {Quantum Science and Technology}\ }\textbf {\bibinfo {volume}
  {2}},\ \bibinfo {pages} {044012} (\bibinfo {year} {2017})}\BibitemShut
  {NoStop}%
\bibitem [{\citenamefont {Forn-Díaz}\ \emph {et~al.}(2017)\citenamefont
  {Forn-Díaz}, \citenamefont {Warren}, \citenamefont {Chang}, \citenamefont
  {Vadiraj},\ and\ \citenamefont {Wilson}}]{Forn2017}%
  \BibitemOpen
  \bibfield  {author} {\bibinfo {author} {\bibfnamefont {P.}~\bibnamefont
  {Forn-Díaz}}, \bibinfo {author} {\bibfnamefont {C.}~\bibnamefont {Warren}},
  \bibinfo {author} {\bibfnamefont {C.}~\bibnamefont {Chang}}, \bibinfo
  {author} {\bibfnamefont {A.}~\bibnamefont {Vadiraj}},\ and\ \bibinfo {author}
  {\bibfnamefont {C.}~\bibnamefont {Wilson}},\ }\bibfield  {title} {\bibinfo
  {title} {On-{Demand} {Microwave} {Generator} of {Shaped} {Single}
  {Photons}},\ }\href {https://doi.org/10.1103/PhysRevApplied.8.054015}
  {\bibfield  {journal} {\bibinfo  {journal} {Physical Review Applied}\
  }\textbf {\bibinfo {volume} {8}},\ \bibinfo {pages} {054015} (\bibinfo {year}
  {2017})}\BibitemShut {NoStop}%
\bibitem [{\citenamefont {Whalen}(2019)}]{Whalen2019}%
  \BibitemOpen
  \bibfield  {author} {\bibinfo {author} {\bibfnamefont {S.~J.}\ \bibnamefont
  {Whalen}},\ }\bibfield  {title} {\bibinfo {title} {Collision model for
  non-{Markovian} quantum trajectories},\ }\href
  {https://doi.org/10.1103/PhysRevA.100.052113} {\bibfield  {journal} {\bibinfo
   {journal} {Physical Review A}\ }\textbf {\bibinfo {volume} {100}},\ \bibinfo
  {pages} {052113} (\bibinfo {year} {2019})}\BibitemShut {NoStop}%
\bibitem [{\citenamefont {N\'{e}met}\ \emph {et~al.}(2019)\citenamefont
  {N\'{e}met}, \citenamefont {Parkins}, \citenamefont {Knorr},\ and\
  \citenamefont {Carmele}}]{Nemet2019}%
  \BibitemOpen
  \bibfield  {author} {\bibinfo {author} {\bibfnamefont {N.}~\bibnamefont
  {N\'{e}met}}, \bibinfo {author} {\bibfnamefont {S.}~\bibnamefont {Parkins}},
  \bibinfo {author} {\bibfnamefont {A.}~\bibnamefont {Knorr}},\ and\ \bibinfo
  {author} {\bibfnamefont {A.}~\bibnamefont {Carmele}},\ }\bibfield  {title}
  {\bibinfo {title} {Stabilizing quantum coherence against pure dephasing in
  the presence of time-delayed coherent feedback at finite temperature},\
  }\href {https://doi.org/10.1103/PhysRevA.99.053809} {\bibfield  {journal}
  {\bibinfo  {journal} {Physical Review A}\ }\textbf {\bibinfo {volume} {99}},\
  \bibinfo {pages} {053809} (\bibinfo {year} {2019})}\BibitemShut {NoStop}%
\bibitem [{\citenamefont {Calaj\'{o}}\ \emph {et~al.}(2019)\citenamefont
  {Calaj\'{o}}, \citenamefont {Fang}, \citenamefont {Baranger},\ and\
  \citenamefont {Ciccarello}}]{Calajo2019}%
  \BibitemOpen
  \bibfield  {author} {\bibinfo {author} {\bibfnamefont {G.}~\bibnamefont
  {Calaj\'{o}}}, \bibinfo {author} {\bibfnamefont {Y.-L.~L.}\ \bibnamefont
  {Fang}}, \bibinfo {author} {\bibfnamefont {H.~U.}\ \bibnamefont {Baranger}},\
  and\ \bibinfo {author} {\bibfnamefont {F.}~\bibnamefont {Ciccarello}},\
  }\bibfield  {title} {\bibinfo {title} {Exciting a {Bound} {State} in the
  {Continuum} through {Multiphoton} {Scattering} {Plus} {Delayed} {Quantum}
  {Feedback}},\ }\href {https://doi.org/10.1103/PhysRevLett.122.073601}
  {\bibfield  {journal} {\bibinfo  {journal} {Physical Review Letters}\
  }\textbf {\bibinfo {volume} {122}},\ \bibinfo {pages} {073601} (\bibinfo
  {year} {2019})}\BibitemShut {NoStop}%
\bibitem [{\citenamefont {Crowder}\ \emph {et~al.}(2020)\citenamefont
  {Crowder}, \citenamefont {Carmichael},\ and\ \citenamefont
  {Hughes}}]{Crowder2020}%
  \BibitemOpen
  \bibfield  {author} {\bibinfo {author} {\bibfnamefont {G.}~\bibnamefont
  {Crowder}}, \bibinfo {author} {\bibfnamefont {H.}~\bibnamefont
  {Carmichael}},\ and\ \bibinfo {author} {\bibfnamefont {S.}~\bibnamefont
  {Hughes}},\ }\bibfield  {title} {\bibinfo {title} {Quantum trajectory theory
  of few-photon cavity-{QED} systems with a time-delayed coherent feedback},\
  }\href {https://doi.org/10.1103/PhysRevA.101.023807} {\bibfield  {journal}
  {\bibinfo  {journal} {Physical Review A}\ }\textbf {\bibinfo {volume}
  {101}},\ \bibinfo {pages} {023807} (\bibinfo {year} {2020})}\BibitemShut
  {NoStop}%
\bibitem [{\citenamefont {Harwood}\ \emph {et~al.}(2021)\citenamefont
  {Harwood}, \citenamefont {Brunelli},\ and\ \citenamefont
  {Serafini}}]{Harwood2021}%
  \BibitemOpen
  \bibfield  {author} {\bibinfo {author} {\bibfnamefont {A.}~\bibnamefont
  {Harwood}}, \bibinfo {author} {\bibfnamefont {M.}~\bibnamefont {Brunelli}},\
  and\ \bibinfo {author} {\bibfnamefont {A.}~\bibnamefont {Serafini}},\
  }\bibfield  {title} {\bibinfo {title} {Cavity optomechanics assisted by
  optical coherent feedback},\ }\href
  {https://doi.org/10.1103/PhysRevA.103.023509} {\bibfield  {journal} {\bibinfo
   {journal} {Physical Review A}\ }\textbf {\bibinfo {volume} {103}},\ \bibinfo
  {pages} {023509} (\bibinfo {year} {2021})}\BibitemShut {NoStop}%
\bibitem [{\citenamefont {Barkemeyer}\ \emph {et~al.}(2021)\citenamefont
  {Barkemeyer}, \citenamefont {Hohn}, \citenamefont {Reitzenstein},\ and\
  \citenamefont {Carmele}}]{Barkemeyer2021}%
  \BibitemOpen
  \bibfield  {author} {\bibinfo {author} {\bibfnamefont {K.}~\bibnamefont
  {Barkemeyer}}, \bibinfo {author} {\bibfnamefont {M.}~\bibnamefont {Hohn}},
  \bibinfo {author} {\bibfnamefont {S.}~\bibnamefont {Reitzenstein}},\ and\
  \bibinfo {author} {\bibfnamefont {A.}~\bibnamefont {Carmele}},\ }\bibfield
  {title} {\bibinfo {title} {Boosting energy-time entanglement using coherent
  time-delayed feedback},\ }\href {https://doi.org/10.1103/PhysRevA.103.062423}
  {\bibfield  {journal} {\bibinfo  {journal} {Physical Review A}\ }\textbf
  {\bibinfo {volume} {103}},\ \bibinfo {pages} {062423} (\bibinfo {year}
  {2021})}\BibitemShut {NoStop}%
\bibitem [{\citenamefont {Arranz~Regidor}\ \emph {et~al.}(2021)\citenamefont
  {Arranz~Regidor}, \citenamefont {Crowder}, \citenamefont {Carmichael},\ and\
  \citenamefont {Hughes}}]{Regidor2021a}%
  \BibitemOpen
  \bibfield  {author} {\bibinfo {author} {\bibfnamefont {S.}~\bibnamefont
  {Arranz~Regidor}}, \bibinfo {author} {\bibfnamefont {G.}~\bibnamefont
  {Crowder}}, \bibinfo {author} {\bibfnamefont {H.}~\bibnamefont
  {Carmichael}},\ and\ \bibinfo {author} {\bibfnamefont {S.}~\bibnamefont
  {Hughes}},\ }\bibfield  {title} {\bibinfo {title} {Modeling quantum
  light-matter interactions in waveguide {QED} with retardation, nonlinear
  interactions, and a time-delayed feedback: {Matrix} product states versus a
  space-discretized waveguide model},\ }\href
  {https://doi.org/10.1103/PhysRevResearch.3.023030} {\bibfield  {journal}
  {\bibinfo  {journal} {Physical Review Research}\ }\textbf {\bibinfo {volume}
  {3}},\ \bibinfo {pages} {023030} (\bibinfo {year} {2021})}\BibitemShut
  {NoStop}%
\bibitem [{\citenamefont {Hughes}(2004)}]{Hughes2004}%
  \BibitemOpen
  \bibfield  {author} {\bibinfo {author} {\bibfnamefont {S.}~\bibnamefont
  {Hughes}},\ }\bibfield  {title} {\bibinfo {title} {Enhanced single-photon
  emission from quantum dots in photonic crystal waveguides and nanocavities},\
  }\href {https://doi.org/10.1364/OL.29.002659} {\bibfield  {journal} {\bibinfo
   {journal} {Optics Letters}\ }\textbf {\bibinfo {volume} {29}},\ \bibinfo
  {pages} {2659} (\bibinfo {year} {2004})}\BibitemShut {NoStop}%
\bibitem [{\citenamefont {Shen}\ and\ \citenamefont {Fan}(2005)}]{Shen2005}%
  \BibitemOpen
  \bibfield  {author} {\bibinfo {author} {\bibfnamefont {J.~T.}\ \bibnamefont
  {Shen}}\ and\ \bibinfo {author} {\bibfnamefont {S.}~\bibnamefont {Fan}},\
  }\bibfield  {title} {\bibinfo {title} {Coherent photon transport from
  spontaneous emission in one-dimensional waveguides},\ }\href
  {https://doi.org/10.1364/OL.30.002001} {\bibfield  {journal} {\bibinfo
  {journal} {Optics Letters}\ }\textbf {\bibinfo {volume} {30}},\ \bibinfo
  {pages} {2001} (\bibinfo {year} {2005})}\BibitemShut {NoStop}%
\bibitem [{\citenamefont {Zhou}\ \emph {et~al.}(2008)\citenamefont {Zhou},
  \citenamefont {Gong}, \citenamefont {Liu}, \citenamefont {Sun},\ and\
  \citenamefont {Nori}}]{Zhou2008}%
  \BibitemOpen
  \bibfield  {author} {\bibinfo {author} {\bibfnamefont {L.}~\bibnamefont
  {Zhou}}, \bibinfo {author} {\bibfnamefont {Z.~R.}\ \bibnamefont {Gong}},
  \bibinfo {author} {\bibfnamefont {Y.-x.}\ \bibnamefont {Liu}}, \bibinfo
  {author} {\bibfnamefont {C.~P.}\ \bibnamefont {Sun}},\ and\ \bibinfo {author}
  {\bibfnamefont {F.}~\bibnamefont {Nori}},\ }\bibfield  {title} {\bibinfo
  {title} {Controllable {Scattering} of a {Single} {Photon} inside a
  {One}-{Dimensional} {Resonator} {Waveguide}},\ }\href
  {https://doi.org/10.1103/PhysRevLett.101.100501} {\bibfield  {journal}
  {\bibinfo  {journal} {Physical Review Letters}\ }\textbf {\bibinfo {volume}
  {101}},\ \bibinfo {pages} {100501} (\bibinfo {year} {2008})}\BibitemShut
  {NoStop}%
\bibitem [{\citenamefont {Zheng}\ \emph {et~al.}(2010)\citenamefont {Zheng},
  \citenamefont {Gauthier},\ and\ \citenamefont {Baranger}}]{Zheng2010}%
  \BibitemOpen
  \bibfield  {author} {\bibinfo {author} {\bibfnamefont {H.}~\bibnamefont
  {Zheng}}, \bibinfo {author} {\bibfnamefont {D.~J.}\ \bibnamefont
  {Gauthier}},\ and\ \bibinfo {author} {\bibfnamefont {H.~U.}\ \bibnamefont
  {Baranger}},\ }\bibfield  {title} {\bibinfo {title} {Waveguide {QED}:
  {Many}-body bound-state effects in coherent and {Fock}-state scattering from
  a two-level system},\ }\href {https://doi.org/10.1103/PhysRevA.82.063816}
  {\bibfield  {journal} {\bibinfo  {journal} {Physical Review A}\ }\textbf
  {\bibinfo {volume} {82}},\ \bibinfo {pages} {063816} (\bibinfo {year}
  {2010})}\BibitemShut {NoStop}%
\bibitem [{\citenamefont {Longo}\ \emph {et~al.}(2011)\citenamefont {Longo},
  \citenamefont {Schmitteckert},\ and\ \citenamefont {Busch}}]{Longo2011}%
  \BibitemOpen
  \bibfield  {author} {\bibinfo {author} {\bibfnamefont {P.}~\bibnamefont
  {Longo}}, \bibinfo {author} {\bibfnamefont {P.}~\bibnamefont
  {Schmitteckert}},\ and\ \bibinfo {author} {\bibfnamefont {K.}~\bibnamefont
  {Busch}},\ }\bibfield  {title} {\bibinfo {title} {Few-photon transport in
  low-dimensional systems},\ }\href
  {https://doi.org/10.1103/PhysRevA.83.063828} {\bibfield  {journal} {\bibinfo
  {journal} {Physical Review A}\ }\textbf {\bibinfo {volume} {83}},\ \bibinfo
  {pages} {063828} (\bibinfo {year} {2011})}\BibitemShut {NoStop}%
\bibitem [{\citenamefont {Roy}(2011)}]{Roy2011}%
  \BibitemOpen
  \bibfield  {author} {\bibinfo {author} {\bibfnamefont {D.}~\bibnamefont
  {Roy}},\ }\bibfield  {title} {\bibinfo {title} {Two-{Photon} {Scattering} by
  a {Driven} {Three}-{Level} {Emitter} in a {One}-{Dimensional} {Waveguide} and
  {Electromagnetically} {Induced} {Transparency}},\ }\href
  {https://doi.org/10.1103/PhysRevLett.106.053601} {\bibfield  {journal}
  {\bibinfo  {journal} {Physical Review Letters}\ }\textbf {\bibinfo {volume}
  {106}},\ \bibinfo {pages} {053601} (\bibinfo {year} {2011})}\BibitemShut
  {NoStop}%
\bibitem [{\citenamefont {Yan}\ and\ \citenamefont {Fan}(2014)}]{Yan2014}%
  \BibitemOpen
  \bibfield  {author} {\bibinfo {author} {\bibfnamefont {W.-B.}\ \bibnamefont
  {Yan}}\ and\ \bibinfo {author} {\bibfnamefont {H.}~\bibnamefont {Fan}},\
  }\bibfield  {title} {\bibinfo {title} {Control of single-photon transport in
  a one-dimensional waveguide by a single photon},\ }\href
  {https://doi.org/10.1103/PhysRevA.90.053807} {\bibfield  {journal} {\bibinfo
  {journal} {Physical Review A}\ }\textbf {\bibinfo {volume} {90}},\ \bibinfo
  {pages} {053807} (\bibinfo {year} {2014})}\BibitemShut {NoStop}%
\bibitem [{\citenamefont {S\'{a}nchez-Burillo}\ \emph
  {et~al.}(2014)\citenamefont {S\'{a}nchez-Burillo}, \citenamefont {Zueco},
  \citenamefont {Garcia-Ripoll},\ and\ \citenamefont
  {Martin-Moreno}}]{Sanchez2014}%
  \BibitemOpen
  \bibfield  {author} {\bibinfo {author} {\bibfnamefont {E.}~\bibnamefont
  {S\'{a}nchez-Burillo}}, \bibinfo {author} {\bibfnamefont {D.}~\bibnamefont
  {Zueco}}, \bibinfo {author} {\bibfnamefont {J.~J.}\ \bibnamefont
  {Garcia-Ripoll}},\ and\ \bibinfo {author} {\bibfnamefont {L.}~\bibnamefont
  {Martin-Moreno}},\ }\bibfield  {title} {\bibinfo {title} {Scattering in the
  {Ultrastrong} {Regime}: {Nonlinear} {Optics} with {One} {Photon}},\ }\href
  {https://doi.org/10.1103/PhysRevLett.113.263604} {\bibfield  {journal}
  {\bibinfo  {journal} {Physical Review Letters}\ }\textbf {\bibinfo {volume}
  {113}},\ \bibinfo {pages} {263604} (\bibinfo {year} {2014})}\BibitemShut
  {NoStop}%
\bibitem [{\citenamefont {Gonzalez-Ballestero}\ \emph
  {et~al.}(2014)\citenamefont {Gonzalez-Ballestero}, \citenamefont {Moreno},\
  and\ \citenamefont {Garcia-Vidal}}]{Gonzalez2014}%
  \BibitemOpen
  \bibfield  {author} {\bibinfo {author} {\bibfnamefont {C.}~\bibnamefont
  {Gonzalez-Ballestero}}, \bibinfo {author} {\bibfnamefont {E.}~\bibnamefont
  {Moreno}},\ and\ \bibinfo {author} {\bibfnamefont {F.~J.}\ \bibnamefont
  {Garcia-Vidal}},\ }\bibfield  {title} {\bibinfo {title} {Generation,
  manipulation, and detection of two-qubit entanglement in waveguide {QED}},\
  }\href {https://doi.org/10.1103/PhysRevA.89.042328} {\bibfield  {journal}
  {\bibinfo  {journal} {Physical Review A}\ }\textbf {\bibinfo {volume} {89}},\
  \bibinfo {pages} {042328} (\bibinfo {year} {2014})}\BibitemShut {NoStop}%
\bibitem [{\citenamefont {Kornovan}\ \emph {et~al.}(2017)\citenamefont
  {Kornovan}, \citenamefont {Petrov},\ and\ \citenamefont
  {Iorsh}}]{Kornovan2017}%
  \BibitemOpen
  \bibfield  {author} {\bibinfo {author} {\bibfnamefont {D.~F.}\ \bibnamefont
  {Kornovan}}, \bibinfo {author} {\bibfnamefont {M.~I.}\ \bibnamefont
  {Petrov}},\ and\ \bibinfo {author} {\bibfnamefont {I.~V.}\ \bibnamefont
  {Iorsh}},\ }\bibfield  {title} {\bibinfo {title} {Transport and collective
  radiance in a basic quantum chiral optical model},\ }\href
  {https://doi.org/10.1103/PhysRevB.96.115162} {\bibfield  {journal} {\bibinfo
  {journal} {Physical Review B}\ }\textbf {\bibinfo {volume} {96}},\ \bibinfo
  {pages} {115162} (\bibinfo {year} {2017})}\BibitemShut {NoStop}%
\bibitem [{\citenamefont {Mahmoodian}\ \emph {et~al.}(2018)\citenamefont
  {Mahmoodian}, \citenamefont {Cepulkovskis}, \citenamefont {Das},
  \citenamefont {Lodahl}, \citenamefont {Hammerer},\ and\ \citenamefont
  {S\o{}rensen}}]{Mahmoodian2018}%
  \BibitemOpen
  \bibfield  {author} {\bibinfo {author} {\bibfnamefont {S.}~\bibnamefont
  {Mahmoodian}}, \bibinfo {author} {\bibfnamefont {M.}~\bibnamefont
  {Cepulkovskis}}, \bibinfo {author} {\bibfnamefont {S.}~\bibnamefont {Das}},
  \bibinfo {author} {\bibfnamefont {P.}~\bibnamefont {Lodahl}}, \bibinfo
  {author} {\bibfnamefont {K.}~\bibnamefont {Hammerer}},\ and\ \bibinfo
  {author} {\bibfnamefont {A.~S.}\ \bibnamefont {S\o{}rensen}},\ }\bibfield
  {title} {\bibinfo {title} {Strongly {Correlated} {Photon} {Transport} in
  {Waveguide} {Quantum} {Electrodynamics} with {Weakly} {Coupled} {Emitters}},\
  }\href {https://doi.org/10.1103/PhysRevLett.121.143601} {\bibfield  {journal}
  {\bibinfo  {journal} {Physical Review Letters}\ }\textbf {\bibinfo {volume}
  {121}},\ \bibinfo {pages} {143601} (\bibinfo {year} {2018})}\BibitemShut
  {NoStop}%
\bibitem [{\citenamefont {Foster}\ \emph {et~al.}(2019)\citenamefont {Foster},
  \citenamefont {Hallett}, \citenamefont {Iorsh}, \citenamefont {Sheldon},
  \citenamefont {Godsland}, \citenamefont {Royall}, \citenamefont {Clarke},
  \citenamefont {Shelykh}, \citenamefont {Fox}, \citenamefont {Skolnick},
  \citenamefont {Itskevich},\ and\ \citenamefont {Wilson}}]{Foster2019}%
  \BibitemOpen
  \bibfield  {author} {\bibinfo {author} {\bibfnamefont {A.~P.}\ \bibnamefont
  {Foster}}, \bibinfo {author} {\bibfnamefont {D.}~\bibnamefont {Hallett}},
  \bibinfo {author} {\bibfnamefont {I.~V.}\ \bibnamefont {Iorsh}}, \bibinfo
  {author} {\bibfnamefont {S.~J.}\ \bibnamefont {Sheldon}}, \bibinfo {author}
  {\bibfnamefont {M.~R.}\ \bibnamefont {Godsland}}, \bibinfo {author}
  {\bibfnamefont {B.}~\bibnamefont {Royall}}, \bibinfo {author} {\bibfnamefont
  {E.}~\bibnamefont {Clarke}}, \bibinfo {author} {\bibfnamefont {I.~A.}\
  \bibnamefont {Shelykh}}, \bibinfo {author} {\bibfnamefont {A.~M.}\
  \bibnamefont {Fox}}, \bibinfo {author} {\bibfnamefont {M.~S.}\ \bibnamefont
  {Skolnick}}, \bibinfo {author} {\bibfnamefont {I.~E.}\ \bibnamefont
  {Itskevich}},\ and\ \bibinfo {author} {\bibfnamefont {L.~R.}\ \bibnamefont
  {Wilson}},\ }\bibfield  {title} {\bibinfo {title} {Tunable photon statistics
  exploiting the {Fano} effect in a waveguide},\ }\href
  {https://doi.org/10.1103/PhysRevLett.122.173603} {\bibfield  {journal}
  {\bibinfo  {journal} {Physical Review Letters}\ }\textbf {\bibinfo {volume}
  {122}},\ \bibinfo {pages} {173603} (\bibinfo {year} {2019})}\BibitemShut
  {NoStop}%
\bibitem [{\citenamefont {Mukhopadhyay}\ and\ \citenamefont
  {Agarwal}(2019)}]{Mukhopadhyay2019}%
  \BibitemOpen
  \bibfield  {author} {\bibinfo {author} {\bibfnamefont {D.}~\bibnamefont
  {Mukhopadhyay}}\ and\ \bibinfo {author} {\bibfnamefont {G.~S.}\ \bibnamefont
  {Agarwal}},\ }\bibfield  {title} {\bibinfo {title} {Multiple {Fano}
  interferences due to waveguide-mediated phase coupling between atoms},\
  }\href {https://doi.org/10.1103/PhysRevA.100.013812} {\bibfield  {journal}
  {\bibinfo  {journal} {Physical Review A}\ }\textbf {\bibinfo {volume}
  {100}},\ \bibinfo {pages} {013812} (\bibinfo {year} {2019})}\BibitemShut
  {NoStop}%
\bibitem [{\citenamefont {Rom\'{a}n-Roche}\ \emph {et~al.}(2020)\citenamefont
  {Rom\'{a}n-Roche}, \citenamefont {S\'{a}nchez-Burillo},\ and\ \citenamefont
  {Zueco}}]{Roman2020}%
  \BibitemOpen
  \bibfield  {author} {\bibinfo {author} {\bibfnamefont {J.}~\bibnamefont
  {Rom\'{a}n-Roche}}, \bibinfo {author} {\bibfnamefont {E.}~\bibnamefont
  {S\'{a}nchez-Burillo}},\ and\ \bibinfo {author} {\bibfnamefont
  {D.}~\bibnamefont {Zueco}},\ }\bibfield  {title} {\bibinfo {title} {Bound
  states in ultrastrong waveguide {QED}},\ }\href
  {https://doi.org/10.1103/PhysRevA.102.023702} {\bibfield  {journal} {\bibinfo
   {journal} {Physical Review A}\ }\textbf {\bibinfo {volume} {102}},\ \bibinfo
  {pages} {023702} (\bibinfo {year} {2020})}\BibitemShut {NoStop}%
\bibitem [{\citenamefont {Mukhopadhyay}\ and\ \citenamefont
  {Agarwal}(2020)}]{Mukhopadhyay2020}%
  \BibitemOpen
  \bibfield  {author} {\bibinfo {author} {\bibfnamefont {D.}~\bibnamefont
  {Mukhopadhyay}}\ and\ \bibinfo {author} {\bibfnamefont {G.~S.}\ \bibnamefont
  {Agarwal}},\ }\bibfield  {title} {\bibinfo {title} {Transparency in a chain
  of disparate quantum emitters strongly coupled to a waveguide},\ }\href
  {https://doi.org/10.1103/PhysRevA.101.063814} {\bibfield  {journal} {\bibinfo
   {journal} {Physical Review A}\ }\textbf {\bibinfo {volume} {101}},\ \bibinfo
  {pages} {063814} (\bibinfo {year} {2020})}\BibitemShut {NoStop}%
\bibitem [{\citenamefont {Mahmoodian}\ \emph {et~al.}(2020)\citenamefont
  {Mahmoodian}, \citenamefont {Calaj\'{o}}, \citenamefont {Chang},
  \citenamefont {Hammerer},\ and\ \citenamefont
  {S\o{}rensen}}]{Mahmoodian2020}%
  \BibitemOpen
  \bibfield  {author} {\bibinfo {author} {\bibfnamefont {S.}~\bibnamefont
  {Mahmoodian}}, \bibinfo {author} {\bibfnamefont {G.}~\bibnamefont
  {Calaj\'{o}}}, \bibinfo {author} {\bibfnamefont {D.~E.}\ \bibnamefont
  {Chang}}, \bibinfo {author} {\bibfnamefont {K.}~\bibnamefont {Hammerer}},\
  and\ \bibinfo {author} {\bibfnamefont {A.~S.}\ \bibnamefont {S\o{}rensen}},\
  }\bibfield  {title} {\bibinfo {title} {Dynamics of {Many}-{Body} {Photon}
  {Bound} {States} in {Chiral} {Waveguide} {QED}},\ }\href
  {https://doi.org/10.1103/PhysRevX.10.031011} {\bibfield  {journal} {\bibinfo
  {journal} {Physical Review X}\ }\textbf {\bibinfo {volume} {10}},\ \bibinfo
  {pages} {031011} (\bibinfo {year} {2020})}\BibitemShut {NoStop}%
\bibitem [{\citenamefont {Le~Jeannic}\ \emph {et~al.}(2021)\citenamefont
  {Le~Jeannic}, \citenamefont {Ramos}, \citenamefont {Simonsen}, \citenamefont
  {Pregnolato}, \citenamefont {Liu}, \citenamefont {Schott}, \citenamefont
  {Wieck}, \citenamefont {Ludwig}, \citenamefont {Rotenberg}, \citenamefont
  {Garc\'{i}a-Ripoll},\ and\ \citenamefont {Lodahl}}]{Jeannic2021}%
  \BibitemOpen
  \bibfield  {author} {\bibinfo {author} {\bibfnamefont {H.}~\bibnamefont
  {Le~Jeannic}}, \bibinfo {author} {\bibfnamefont {T.}~\bibnamefont {Ramos}},
  \bibinfo {author} {\bibfnamefont {S.~F.}\ \bibnamefont {Simonsen}}, \bibinfo
  {author} {\bibfnamefont {T.}~\bibnamefont {Pregnolato}}, \bibinfo {author}
  {\bibfnamefont {Z.}~\bibnamefont {Liu}}, \bibinfo {author} {\bibfnamefont
  {R.}~\bibnamefont {Schott}}, \bibinfo {author} {\bibfnamefont {A.~D.}\
  \bibnamefont {Wieck}}, \bibinfo {author} {\bibfnamefont {A.}~\bibnamefont
  {Ludwig}}, \bibinfo {author} {\bibfnamefont {N.}~\bibnamefont {Rotenberg}},
  \bibinfo {author} {\bibfnamefont {J.~J.}\ \bibnamefont {Garc\'{i}a-Ripoll}},\
  and\ \bibinfo {author} {\bibfnamefont {P.}~\bibnamefont {Lodahl}},\
  }\bibfield  {title} {\bibinfo {title} {Experimental {Reconstruction} of the
  {Few}-{Photon} {Nonlinear} {Scattering} {Matrix} from a {Single} {Quantum}
  {Dot} in a {Nanophotonic} {Waveguide}},\ }\href
  {https://doi.org/10.1103/PhysRevLett.126.023603} {\bibfield  {journal}
  {\bibinfo  {journal} {Physical Review Letters}\ }\textbf {\bibinfo {volume}
  {126}},\ \bibinfo {pages} {023603} (\bibinfo {year} {2021})}\BibitemShut
  {NoStop}%
\bibitem [{\citenamefont {Arranz~Regidor}\ and\ \citenamefont
  {Hughes}(2021)}]{Regidor2021b}%
  \BibitemOpen
  \bibfield  {author} {\bibinfo {author} {\bibfnamefont {S.}~\bibnamefont
  {Arranz~Regidor}}\ and\ \bibinfo {author} {\bibfnamefont {S.}~\bibnamefont
  {Hughes}},\ }\bibfield  {title} {\bibinfo {title} {Cavitylike strong coupling
  in macroscopic waveguide {QED} using three coupled qubits in the deep
  non-{Markovian} regime},\ }\href
  {https://doi.org/10.1103/PhysRevA.104.L031701} {\bibfield  {journal}
  {\bibinfo  {journal} {Physical Review A}\ }\textbf {\bibinfo {volume}
  {104}},\ \bibinfo {pages} {L031701} (\bibinfo {year} {2021})}\BibitemShut
  {NoStop}%
\bibitem [{\citenamefont {Sheremet}\ \emph {et~al.}(2021)\citenamefont
  {Sheremet}, \citenamefont {Petrov}, \citenamefont {Iorsh}, \citenamefont
  {Poshakinskiy},\ and\ \citenamefont {Poddubny}}]{Sheremet2021}%
  \BibitemOpen
  \bibfield  {author} {\bibinfo {author} {\bibfnamefont {A.~S.}\ \bibnamefont
  {Sheremet}}, \bibinfo {author} {\bibfnamefont {M.~I.}\ \bibnamefont
  {Petrov}}, \bibinfo {author} {\bibfnamefont {I.~V.}\ \bibnamefont {Iorsh}},
  \bibinfo {author} {\bibfnamefont {A.~V.}\ \bibnamefont {Poshakinskiy}},\ and\
  \bibinfo {author} {\bibfnamefont {A.~N.}\ \bibnamefont {Poddubny}},\
  }\bibfield  {title} {\bibinfo {title} {Waveguide quantum electrodynamics:
  collective radiance and photon-photon correlations},\ }\href
  {http://arxiv.org/abs/2103.06824} {\bibfield  {journal} {\bibinfo  {journal}
  {arXiv:2103.06824 [quant-ph]}\ } (\bibinfo {year} {2021})}\BibitemShut
  {NoStop}%
\bibitem [{\citenamefont {Solano}\ \emph {et~al.}(2021)\citenamefont {Solano},
  \citenamefont {Barberis-Blostein},\ and\ \citenamefont {Sinha}}]{Solano2021}%
  \BibitemOpen
  \bibfield  {author} {\bibinfo {author} {\bibfnamefont {P.}~\bibnamefont
  {Solano}}, \bibinfo {author} {\bibfnamefont {P.}~\bibnamefont
  {Barberis-Blostein}},\ and\ \bibinfo {author} {\bibfnamefont
  {K.}~\bibnamefont {Sinha}},\ }\bibfield  {title} {\bibinfo {title}
  {Collective directional emission from distant emitters in waveguide {QED}},\
  }\href {http://arxiv.org/abs/2108.12951} {\bibfield  {journal} {\bibinfo
  {journal} {arXiv:2108.12951 [quant-ph]}\ } (\bibinfo {year}
  {2021})}\BibitemShut {NoStop}%
\bibitem [{\citenamefont {Blais}\ \emph {et~al.}(2007)\citenamefont {Blais},
  \citenamefont {Gambetta}, \citenamefont {Wallraff}, \citenamefont {Schuster},
  \citenamefont {Girvin}, \citenamefont {Devoret},\ and\ \citenamefont
  {Schoelkopf}}]{Blais2007}%
  \BibitemOpen
  \bibfield  {author} {\bibinfo {author} {\bibfnamefont {A.}~\bibnamefont
  {Blais}}, \bibinfo {author} {\bibfnamefont {J.}~\bibnamefont {Gambetta}},
  \bibinfo {author} {\bibfnamefont {A.}~\bibnamefont {Wallraff}}, \bibinfo
  {author} {\bibfnamefont {D.~I.}\ \bibnamefont {Schuster}}, \bibinfo {author}
  {\bibfnamefont {S.~M.}\ \bibnamefont {Girvin}}, \bibinfo {author}
  {\bibfnamefont {M.~H.}\ \bibnamefont {Devoret}},\ and\ \bibinfo {author}
  {\bibfnamefont {R.~J.}\ \bibnamefont {Schoelkopf}},\ }\bibfield  {title}
  {\bibinfo {title} {Quantum-information processing with circuit quantum
  electrodynamics},\ }\href {https://doi.org/10.1103/PhysRevA.75.032329}
  {\bibfield  {journal} {\bibinfo  {journal} {Physical Review A}\ }\textbf
  {\bibinfo {volume} {75}},\ \bibinfo {pages} {032329} (\bibinfo {year}
  {2007})}\BibitemShut {NoStop}%
\bibitem [{\citenamefont {Kretschmer}\ \emph {et~al.}(2016)\citenamefont
  {Kretschmer}, \citenamefont {Luoma},\ and\ \citenamefont
  {Strunz}}]{Kretschmer2016}%
  \BibitemOpen
  \bibfield  {author} {\bibinfo {author} {\bibfnamefont {S.}~\bibnamefont
  {Kretschmer}}, \bibinfo {author} {\bibfnamefont {K.}~\bibnamefont {Luoma}},\
  and\ \bibinfo {author} {\bibfnamefont {W.~T.}\ \bibnamefont {Strunz}},\
  }\bibfield  {title} {\bibinfo {title} {Collision model for non-{Markovian}
  quantum dynamics},\ }\href {https://doi.org/10.1103/PhysRevA.94.012106}
  {\bibfield  {journal} {\bibinfo  {journal} {Physical Review A}\ }\textbf
  {\bibinfo {volume} {94}},\ \bibinfo {pages} {012106} (\bibinfo {year}
  {2016})}\BibitemShut {NoStop}%
\bibitem [{\citenamefont {Cilluffo}\ \emph {et~al.}(2020)\citenamefont
  {Cilluffo}, \citenamefont {Carollo}, \citenamefont {Lorenzo}, \citenamefont
  {Gross}, \citenamefont {Palma},\ and\ \citenamefont
  {Ciccarello}}]{Cilluffo2020}%
  \BibitemOpen
  \bibfield  {author} {\bibinfo {author} {\bibfnamefont {D.}~\bibnamefont
  {Cilluffo}}, \bibinfo {author} {\bibfnamefont {A.}~\bibnamefont {Carollo}},
  \bibinfo {author} {\bibfnamefont {S.}~\bibnamefont {Lorenzo}}, \bibinfo
  {author} {\bibfnamefont {J.~A.}\ \bibnamefont {Gross}}, \bibinfo {author}
  {\bibfnamefont {G.~M.}\ \bibnamefont {Palma}},\ and\ \bibinfo {author}
  {\bibfnamefont {F.}~\bibnamefont {Ciccarello}},\ }\bibfield  {title}
  {\bibinfo {title} {Collisional picture of quantum optics with giant
  emitters},\ }\href {https://doi.org/10.1103/PhysRevResearch.2.043070}
  {\bibfield  {journal} {\bibinfo  {journal} {Physical Review Research}\
  }\textbf {\bibinfo {volume} {2}},\ \bibinfo {pages} {043070} (\bibinfo {year}
  {2020})}\BibitemShut {NoStop}%
\bibitem [{\citenamefont {Ciccarello}\ \emph {et~al.}(2022)\citenamefont
  {Ciccarello}, \citenamefont {Lorenzo}, \citenamefont {Giovannetti},\ and\
  \citenamefont {Palma}}]{Ciccarello2021}%
  \BibitemOpen
  \bibfield  {author} {\bibinfo {author} {\bibfnamefont {F.}~\bibnamefont
  {Ciccarello}}, \bibinfo {author} {\bibfnamefont {S.}~\bibnamefont {Lorenzo}},
  \bibinfo {author} {\bibfnamefont {V.}~\bibnamefont {Giovannetti}},\ and\
  \bibinfo {author} {\bibfnamefont {G.~M.}\ \bibnamefont {Palma}},\ }\bibfield
  {title} {\bibinfo {title} {Quantum collision models: {Open} system dynamics
  from repeated interactions},\ }\href
  {https://doi.org/10.1016/j.physrep.2022.01.001} {\bibfield  {journal}
  {\bibinfo  {journal} {Physics Reports}\ }\textbf {\bibinfo {volume} {954}},\
  \bibinfo {pages} {1} (\bibinfo {year} {2022})}\BibitemShut {NoStop}%
\bibitem [{\citenamefont {Zhang}\ \emph {et~al.}(2020)\citenamefont {Zhang},
  \citenamefont {You},\ and\ \citenamefont {Lu}}]{Zhang2020}%
  \BibitemOpen
  \bibfield  {author} {\bibinfo {author} {\bibfnamefont {B.}~\bibnamefont
  {Zhang}}, \bibinfo {author} {\bibfnamefont {S.}~\bibnamefont {You}},\ and\
  \bibinfo {author} {\bibfnamefont {M.}~\bibnamefont {Lu}},\ }\bibfield
  {title} {\bibinfo {title} {Enhancement of spontaneous entanglement generation
  via coherent quantum feedback},\ }\href
  {https://doi.org/10.1103/PhysRevA.101.032335} {\bibfield  {journal} {\bibinfo
   {journal} {Physical Review A}\ }\textbf {\bibinfo {volume} {101}},\ \bibinfo
  {pages} {032335} (\bibinfo {year} {2020})}\BibitemShut {NoStop}%
\bibitem [{\citenamefont {Sommer}\ \emph {et~al.}(2020)\citenamefont {Sommer},
  \citenamefont {Ghosh},\ and\ \citenamefont {Genes}}]{Sommer2020}%
  \BibitemOpen
  \bibfield  {author} {\bibinfo {author} {\bibfnamefont {C.}~\bibnamefont
  {Sommer}}, \bibinfo {author} {\bibfnamefont {A.}~\bibnamefont {Ghosh}},\ and\
  \bibinfo {author} {\bibfnamefont {C.}~\bibnamefont {Genes}},\ }\bibfield
  {title} {\bibinfo {title} {Multimode cold-damping optomechanics with delayed
  feedback},\ }\href {https://doi.org/10.1103/PhysRevResearch.2.033299}
  {\bibfield  {journal} {\bibinfo  {journal} {Physical Review Research}\
  }\textbf {\bibinfo {volume} {2}},\ \bibinfo {pages} {033299} (\bibinfo {year}
  {2020})}\BibitemShut {NoStop}%
\bibitem [{\citenamefont {Schollwoeck}(2011)}]{Schollwoeck2011}%
  \BibitemOpen
  \bibfield  {author} {\bibinfo {author} {\bibfnamefont {U.}~\bibnamefont
  {Schollwoeck}},\ }\bibfield  {title} {\bibinfo {title} {The density-matrix
  renormalization group in the age of matrix product states},\ }\href
  {https://doi.org/10.1016/j.aop.2010.09.012} {\bibfield  {journal} {\bibinfo
  {journal} {Annals of Physics}\ }\textbf {\bibinfo {volume} {326}},\ \bibinfo
  {pages} {96} (\bibinfo {year} {2011})}\BibitemShut {NoStop}%
\bibitem [{\citenamefont {Droenner}\ \emph {et~al.}(2019)\citenamefont
  {Droenner}, \citenamefont {Naumann}, \citenamefont {Sch\"{o}ll},
  \citenamefont {Knorr},\ and\ \citenamefont {Carmele}}]{Droenner2019}%
  \BibitemOpen
  \bibfield  {author} {\bibinfo {author} {\bibfnamefont {L.}~\bibnamefont
  {Droenner}}, \bibinfo {author} {\bibfnamefont {N.~L.}\ \bibnamefont
  {Naumann}}, \bibinfo {author} {\bibfnamefont {E.}~\bibnamefont {Sch\"{o}ll}},
  \bibinfo {author} {\bibfnamefont {A.}~\bibnamefont {Knorr}},\ and\ \bibinfo
  {author} {\bibfnamefont {A.}~\bibnamefont {Carmele}},\ }\bibfield  {title}
  {\bibinfo {title} {Quantum {Pyragas} control: {Selective}-control of
  individual photon probabilities},\ }\href
  {https://doi.org/10.1103/PhysRevA.99.023840} {\bibfield  {journal} {\bibinfo
  {journal} {Physical Review A}\ }\textbf {\bibinfo {volume} {99}},\ \bibinfo
  {pages} {023840} (\bibinfo {year} {2019})}\BibitemShut {NoStop}%
\bibitem [{\citenamefont {Finsterh\"{o}lzl}\ \emph {et~al.}(2020)\citenamefont
  {Finsterh\"{o}lzl}, \citenamefont {Katzer}, \citenamefont {Knorr},\ and\
  \citenamefont {Carmele}}]{Finsterholzl2020}%
  \BibitemOpen
  \bibfield  {author} {\bibinfo {author} {\bibfnamefont {R.}~\bibnamefont
  {Finsterh\"{o}lzl}}, \bibinfo {author} {\bibfnamefont {M.}~\bibnamefont
  {Katzer}}, \bibinfo {author} {\bibfnamefont {A.}~\bibnamefont {Knorr}},\ and\
  \bibinfo {author} {\bibfnamefont {A.}~\bibnamefont {Carmele}},\ }\bibfield
  {title} {\bibinfo {title} {Using {Matrix}-{Product} {States} for {Open}
  {Quantum} {Many}-{Body} {Systems}: {Efficient} {Algorithms} for {Markovian}
  and {Non}-{Markovian} {Time}-{Evolution}},\ }\href
  {https://doi.org/10.3390/e22090984} {\bibfield  {journal} {\bibinfo
  {journal} {Entropy}\ }\textbf {\bibinfo {volume} {22}},\ \bibinfo {pages}
  {984} (\bibinfo {year} {2020})}\BibitemShut {NoStop}%
\bibitem [{\citenamefont {Kaestle}\ \emph {et~al.}(2021)\citenamefont
  {Kaestle}, \citenamefont {Finsterh\"{o}lzl}, \citenamefont {Knorr},\ and\
  \citenamefont {Carmele}}]{Kaestle2021}%
  \BibitemOpen
  \bibfield  {author} {\bibinfo {author} {\bibfnamefont {O.}~\bibnamefont
  {Kaestle}}, \bibinfo {author} {\bibfnamefont {R.}~\bibnamefont
  {Finsterh\"{o}lzl}}, \bibinfo {author} {\bibfnamefont {A.}~\bibnamefont
  {Knorr}},\ and\ \bibinfo {author} {\bibfnamefont {A.}~\bibnamefont
  {Carmele}},\ }\bibfield  {title} {\bibinfo {title} {Continuous and
  time-discrete non-{Markovian} system-reservoir interactions: {Dissipative}
  coherent quantum feedback in {Liouville} space},\ }\href
  {https://doi.org/10.1103/PhysRevResearch.3.023168} {\bibfield  {journal}
  {\bibinfo  {journal} {Physical Review Research}\ }\textbf {\bibinfo {volume}
  {3}},\ \bibinfo {pages} {023168} (\bibinfo {year} {2021})}\BibitemShut
  {NoStop}%
\bibitem [{\citenamefont {Dum}\ \emph {et~al.}(1992)\citenamefont {Dum},
  \citenamefont {Parkins}, \citenamefont {Zoller},\ and\ \citenamefont
  {Gardiner}}]{Dum1992}%
  \BibitemOpen
  \bibfield  {author} {\bibinfo {author} {\bibfnamefont {R.}~\bibnamefont
  {Dum}}, \bibinfo {author} {\bibfnamefont {A.~S.}\ \bibnamefont {Parkins}},
  \bibinfo {author} {\bibfnamefont {P.}~\bibnamefont {Zoller}},\ and\ \bibinfo
  {author} {\bibfnamefont {C.~W.}\ \bibnamefont {Gardiner}},\ }\bibfield
  {title} {\bibinfo {title} {Monte {Carlo} simulation of master equations in
  quantum optics for vacuum, thermal, and squeezed reservoirs},\ }\href
  {https://doi.org/10.1103/PhysRevA.46.4382} {\bibfield  {journal} {\bibinfo
  {journal} {Physical Review A}\ }\textbf {\bibinfo {volume} {46}},\ \bibinfo
  {pages} {4382} (\bibinfo {year} {1992})}\BibitemShut {NoStop}%
\bibitem [{\citenamefont {Tian}\ and\ \citenamefont
  {Carmichael}(1992)}]{Tian1992}%
  \BibitemOpen
  \bibfield  {author} {\bibinfo {author} {\bibfnamefont {L.}~\bibnamefont
  {Tian}}\ and\ \bibinfo {author} {\bibfnamefont {H.~J.}\ \bibnamefont
  {Carmichael}},\ }\bibfield  {title} {\bibinfo {title} {Quantum trajectory
  simulations of two-state behavior in an optical cavity containing one atom},\
  }\href {https://doi.org/10.1103/PhysRevA.46.R6801} {\bibfield  {journal}
  {\bibinfo  {journal} {Physical Review A}\ }\textbf {\bibinfo {volume} {46}},\
  \bibinfo {pages} {R6801} (\bibinfo {year} {1992})}\BibitemShut {NoStop}%
\bibitem [{\citenamefont {Dalibard}\ \emph {et~al.}(1992)\citenamefont
  {Dalibard}, \citenamefont {Castin},\ and\ \citenamefont
  {M\o{}lmer}}]{Dalibard1992}%
  \BibitemOpen
  \bibfield  {author} {\bibinfo {author} {\bibfnamefont {J.}~\bibnamefont
  {Dalibard}}, \bibinfo {author} {\bibfnamefont {Y.}~\bibnamefont {Castin}},\
  and\ \bibinfo {author} {\bibfnamefont {K.}~\bibnamefont {M\o{}lmer}},\
  }\bibfield  {title} {\bibinfo {title} {Wave-function approach to dissipative
  processes in quantum optics},\ }\href
  {https://doi.org/10.1103/PhysRevLett.68.580} {\bibfield  {journal} {\bibinfo
  {journal} {Physical Review Letters}\ }\textbf {\bibinfo {volume} {68}},\
  \bibinfo {pages} {580} (\bibinfo {year} {1992})}\BibitemShut {NoStop}%
\bibitem [{\citenamefont {M\o{}lmer}\ \emph {et~al.}(1993)\citenamefont
  {M\o{}lmer}, \citenamefont {Castin},\ and\ \citenamefont
  {Dalibard}}]{Molmer1993}%
  \BibitemOpen
  \bibfield  {author} {\bibinfo {author} {\bibfnamefont {K.}~\bibnamefont
  {M\o{}lmer}}, \bibinfo {author} {\bibfnamefont {Y.}~\bibnamefont {Castin}},\
  and\ \bibinfo {author} {\bibfnamefont {J.}~\bibnamefont {Dalibard}},\
  }\bibfield  {title} {\bibinfo {title} {Monte {Carlo} wave-function method in
  quantum optics},\ }\href {https://doi.org/10.1364/JOSAB.10.000524} {\bibfield
   {journal} {\bibinfo  {journal} {Journal of the Optical Society of America
  B}\ }\textbf {\bibinfo {volume} {10}},\ \bibinfo {pages} {524} (\bibinfo
  {year} {1993})}\BibitemShut {NoStop}%
\bibitem [{\citenamefont {Brun}(2002)}]{Brun2002}%
  \BibitemOpen
  \bibfield  {author} {\bibinfo {author} {\bibfnamefont {T.~A.}\ \bibnamefont
  {Brun}},\ }\bibfield  {title} {\bibinfo {title} {A simple model of quantum
  trajectories},\ }\href {https://doi.org/10.1119/1.1475328} {\bibfield
  {journal} {\bibinfo  {journal} {American Journal of Physics}\ }\textbf
  {\bibinfo {volume} {70}},\ \bibinfo {pages} {719} (\bibinfo {year}
  {2002})}\BibitemShut {NoStop}%
\bibitem [{\citenamefont {Yao}\ and\ \citenamefont {Hughes}(2009)}]{Yao2009}%
  \BibitemOpen
  \bibfield  {author} {\bibinfo {author} {\bibfnamefont {P.}~\bibnamefont
  {Yao}}\ and\ \bibinfo {author} {\bibfnamefont {S.}~\bibnamefont {Hughes}},\
  }\bibfield  {title} {\bibinfo {title} {Controlled cavity {QED} and
  single-photon emission using a photonic-crystal waveguide cavity system},\
  }\href {https://doi.org/10.1103/PhysRevB.80.165128} {\bibfield  {journal}
  {\bibinfo  {journal} {Physical Review B}\ }\textbf {\bibinfo {volume} {80}},\
  \bibinfo {pages} {165128} (\bibinfo {year} {2009})}\BibitemShut {NoStop}%
\bibitem [{\citenamefont {Gonzalez-Ballestero}\ \emph
  {et~al.}(2015)\citenamefont {Gonzalez-Ballestero}, \citenamefont
  {Gonzalez-Tudela}, \citenamefont {Garcia-Vidal},\ and\ \citenamefont
  {Moreno}}]{Ballestero2015}%
  \BibitemOpen
  \bibfield  {author} {\bibinfo {author} {\bibfnamefont {C.}~\bibnamefont
  {Gonzalez-Ballestero}}, \bibinfo {author} {\bibfnamefont {A.}~\bibnamefont
  {Gonzalez-Tudela}}, \bibinfo {author} {\bibfnamefont {F.~J.}\ \bibnamefont
  {Garcia-Vidal}},\ and\ \bibinfo {author} {\bibfnamefont {E.}~\bibnamefont
  {Moreno}},\ }\bibfield  {title} {\bibinfo {title} {Chiral route to
  spontaneous entanglement generation},\ }\href
  {https://doi.org/10.1103/PhysRevB.92.155304} {\bibfield  {journal} {\bibinfo
  {journal} {Physical Review B}\ }\textbf {\bibinfo {volume} {92}},\ \bibinfo
  {pages} {155304} (\bibinfo {year} {2015})}\BibitemShut {NoStop}%
\bibitem [{\citenamefont {le~Feber}\ \emph {et~al.}(2015)\citenamefont
  {le~Feber}, \citenamefont {Rotenberg},\ and\ \citenamefont
  {Kuipers}}]{LeFeber2015}%
  \BibitemOpen
  \bibfield  {author} {\bibinfo {author} {\bibfnamefont {B.}~\bibnamefont
  {le~Feber}}, \bibinfo {author} {\bibfnamefont {N.}~\bibnamefont
  {Rotenberg}},\ and\ \bibinfo {author} {\bibfnamefont {L.}~\bibnamefont
  {Kuipers}},\ }\bibfield  {title} {\bibinfo {title} {Nanophotonic control of
  circular dipole emission},\ }\href {https://doi.org/10.1038/ncomms7695}
  {\bibfield  {journal} {\bibinfo  {journal} {Nature Communications}\ }\textbf
  {\bibinfo {volume} {6}},\ \bibinfo {pages} {6695} (\bibinfo {year}
  {2015})}\BibitemShut {NoStop}%
\bibitem [{\citenamefont {Young}\ \emph {et~al.}(2015)\citenamefont {Young},
  \citenamefont {Thijssen}, \citenamefont {Beggs}, \citenamefont
  {Androvitsaneas}, \citenamefont {Kuipers}, \citenamefont {Rarity},
  \citenamefont {Hughes},\ and\ \citenamefont {Oulton}}]{Young2015}%
  \BibitemOpen
  \bibfield  {author} {\bibinfo {author} {\bibfnamefont {A.}~\bibnamefont
  {Young}}, \bibinfo {author} {\bibfnamefont {A.}~\bibnamefont {Thijssen}},
  \bibinfo {author} {\bibfnamefont {D.}~\bibnamefont {Beggs}}, \bibinfo
  {author} {\bibfnamefont {P.}~\bibnamefont {Androvitsaneas}}, \bibinfo
  {author} {\bibfnamefont {L.}~\bibnamefont {Kuipers}}, \bibinfo {author}
  {\bibfnamefont {J.}~\bibnamefont {Rarity}}, \bibinfo {author} {\bibfnamefont
  {S.}~\bibnamefont {Hughes}},\ and\ \bibinfo {author} {\bibfnamefont
  {R.}~\bibnamefont {Oulton}},\ }\bibfield  {title} {\bibinfo {title}
  {Polarization {Engineering} in {Photonic} {Crystal} {Waveguides} for
  {Spin}-{Photon} {Entanglers}},\ }\href
  {https://doi.org/10.1103/PhysRevLett.115.153901} {\bibfield  {journal}
  {\bibinfo  {journal} {Physical Review Letters}\ }\textbf {\bibinfo {volume}
  {115}},\ \bibinfo {pages} {153901} (\bibinfo {year} {2015})}\BibitemShut
  {NoStop}%
\bibitem [{\citenamefont {Söllner}\ \emph {et~al.}(2015)\citenamefont
  {Söllner}, \citenamefont {Mahmoodian}, \citenamefont {Hansen}, \citenamefont
  {Midolo}, \citenamefont {Javadi}, \citenamefont {Kiršanskė}, \citenamefont
  {Pregnolato}, \citenamefont {El-Ella}, \citenamefont {Lee}, \citenamefont
  {Song}, \citenamefont {Stobbe},\ and\ \citenamefont {Lodahl}}]{Sollner2015}%
  \BibitemOpen
  \bibfield  {author} {\bibinfo {author} {\bibfnamefont {I.}~\bibnamefont
  {Söllner}}, \bibinfo {author} {\bibfnamefont {S.}~\bibnamefont
  {Mahmoodian}}, \bibinfo {author} {\bibfnamefont {S.~L.}\ \bibnamefont
  {Hansen}}, \bibinfo {author} {\bibfnamefont {L.}~\bibnamefont {Midolo}},
  \bibinfo {author} {\bibfnamefont {A.}~\bibnamefont {Javadi}}, \bibinfo
  {author} {\bibfnamefont {G.}~\bibnamefont {Kiršanskė}}, \bibinfo {author}
  {\bibfnamefont {T.}~\bibnamefont {Pregnolato}}, \bibinfo {author}
  {\bibfnamefont {H.}~\bibnamefont {El-Ella}}, \bibinfo {author} {\bibfnamefont
  {E.~H.}\ \bibnamefont {Lee}}, \bibinfo {author} {\bibfnamefont {J.~D.}\
  \bibnamefont {Song}}, \bibinfo {author} {\bibfnamefont {S.}~\bibnamefont
  {Stobbe}},\ and\ \bibinfo {author} {\bibfnamefont {P.}~\bibnamefont
  {Lodahl}},\ }\bibfield  {title} {\bibinfo {title} {Deterministic
  photon–emitter coupling in chiral photonic circuits},\ }\href
  {https://doi.org/10.1038/nnano.2015.159} {\bibfield  {journal} {\bibinfo
  {journal} {Nature Nanotechnology}\ }\textbf {\bibinfo {volume} {10}},\
  \bibinfo {pages} {775} (\bibinfo {year} {2015})}\BibitemShut {NoStop}%
\bibitem [{\citenamefont {Coles}\ \emph {et~al.}(2016)\citenamefont {Coles},
  \citenamefont {Price}, \citenamefont {Dixon}, \citenamefont {Royall},
  \citenamefont {Clarke}, \citenamefont {Kok}, \citenamefont {Skolnick},
  \citenamefont {Fox},\ and\ \citenamefont {Makhonin}}]{Coles2016}%
  \BibitemOpen
  \bibfield  {author} {\bibinfo {author} {\bibfnamefont {R.~J.}\ \bibnamefont
  {Coles}}, \bibinfo {author} {\bibfnamefont {D.~M.}\ \bibnamefont {Price}},
  \bibinfo {author} {\bibfnamefont {J.~E.}\ \bibnamefont {Dixon}}, \bibinfo
  {author} {\bibfnamefont {B.}~\bibnamefont {Royall}}, \bibinfo {author}
  {\bibfnamefont {E.}~\bibnamefont {Clarke}}, \bibinfo {author} {\bibfnamefont
  {P.}~\bibnamefont {Kok}}, \bibinfo {author} {\bibfnamefont {M.~S.}\
  \bibnamefont {Skolnick}}, \bibinfo {author} {\bibfnamefont {A.~M.}\
  \bibnamefont {Fox}},\ and\ \bibinfo {author} {\bibfnamefont {M.~N.}\
  \bibnamefont {Makhonin}},\ }\bibfield  {title} {\bibinfo {title} {Chirality
  of nanophotonic waveguide with embedded quantum emitter for unidirectional
  spin transfer},\ }\href {https://doi.org/10.1038/ncomms11183} {\bibfield
  {journal} {\bibinfo  {journal} {Nature Communications}\ }\textbf {\bibinfo
  {volume} {7}},\ \bibinfo {pages} {11183} (\bibinfo {year}
  {2016})}\BibitemShut {NoStop}%
\bibitem [{\citenamefont {Scheucher}\ \emph {et~al.}(2016)\citenamefont
  {Scheucher}, \citenamefont {Hilico}, \citenamefont {Will}, \citenamefont
  {Volz},\ and\ \citenamefont {Rauschenbeutel}}]{Scheucher2016}%
  \BibitemOpen
  \bibfield  {author} {\bibinfo {author} {\bibfnamefont {M.}~\bibnamefont
  {Scheucher}}, \bibinfo {author} {\bibfnamefont {A.}~\bibnamefont {Hilico}},
  \bibinfo {author} {\bibfnamefont {E.}~\bibnamefont {Will}}, \bibinfo {author}
  {\bibfnamefont {J.}~\bibnamefont {Volz}},\ and\ \bibinfo {author}
  {\bibfnamefont {A.}~\bibnamefont {Rauschenbeutel}},\ }\bibfield  {title}
  {\bibinfo {title} {Quantum optical circulator controlled by a single chirally
  coupled atom},\ }\href {https://doi.org/10.1126/science.aaj2118} {\bibfield
  {journal} {\bibinfo  {journal} {Science}\ }\textbf {\bibinfo {volume}
  {354}},\ \bibinfo {pages} {1577} (\bibinfo {year} {2016})}\BibitemShut
  {NoStop}%
\bibitem [{\citenamefont {Lodahl}\ \emph {et~al.}(2017)\citenamefont {Lodahl},
  \citenamefont {Mahmoodian}, \citenamefont {Stobbe}, \citenamefont
  {Rauschenbeutel}, \citenamefont {Schneeweiss}, \citenamefont {Volz},
  \citenamefont {Pichler},\ and\ \citenamefont {Zoller}}]{Lodahl2017}%
  \BibitemOpen
  \bibfield  {author} {\bibinfo {author} {\bibfnamefont {P.}~\bibnamefont
  {Lodahl}}, \bibinfo {author} {\bibfnamefont {S.}~\bibnamefont {Mahmoodian}},
  \bibinfo {author} {\bibfnamefont {S.}~\bibnamefont {Stobbe}}, \bibinfo
  {author} {\bibfnamefont {A.}~\bibnamefont {Rauschenbeutel}}, \bibinfo
  {author} {\bibfnamefont {P.}~\bibnamefont {Schneeweiss}}, \bibinfo {author}
  {\bibfnamefont {J.}~\bibnamefont {Volz}}, \bibinfo {author} {\bibfnamefont
  {H.}~\bibnamefont {Pichler}},\ and\ \bibinfo {author} {\bibfnamefont
  {P.}~\bibnamefont {Zoller}},\ }\bibfield  {title} {\bibinfo {title} {Chiral
  quantum optics},\ }\href {https://doi.org/10.1038/nature21037} {\bibfield
  {journal} {\bibinfo  {journal} {Nature}\ }\textbf {\bibinfo {volume} {541}},\
  \bibinfo {pages} {473} (\bibinfo {year} {2017})}\BibitemShut {NoStop}%
\bibitem [{\citenamefont {Barik}\ \emph {et~al.}(2018)\citenamefont {Barik},
  \citenamefont {Karasahin}, \citenamefont {Flower}, \citenamefont {Cai},
  \citenamefont {Miyake}, \citenamefont {DeGottardi}, \citenamefont {Hafezi},\
  and\ \citenamefont {Waks}}]{Barik2018}%
  \BibitemOpen
  \bibfield  {author} {\bibinfo {author} {\bibfnamefont {S.}~\bibnamefont
  {Barik}}, \bibinfo {author} {\bibfnamefont {A.}~\bibnamefont {Karasahin}},
  \bibinfo {author} {\bibfnamefont {C.}~\bibnamefont {Flower}}, \bibinfo
  {author} {\bibfnamefont {T.}~\bibnamefont {Cai}}, \bibinfo {author}
  {\bibfnamefont {H.}~\bibnamefont {Miyake}}, \bibinfo {author} {\bibfnamefont
  {W.}~\bibnamefont {DeGottardi}}, \bibinfo {author} {\bibfnamefont
  {M.}~\bibnamefont {Hafezi}},\ and\ \bibinfo {author} {\bibfnamefont
  {E.}~\bibnamefont {Waks}},\ }\bibfield  {title} {\bibinfo {title} {A
  topological quantum optics interface},\ }\href
  {https://doi.org/10.1126/science.aaq0327} {\bibfield  {journal} {\bibinfo
  {journal} {Science}\ }\textbf {\bibinfo {volume} {359}},\ \bibinfo {pages}
  {666} (\bibinfo {year} {2018})}\BibitemShut {NoStop}%
\bibitem [{\citenamefont {Martin-Cano}\ \emph {et~al.}(2019)\citenamefont
  {Martin-Cano}, \citenamefont {Haakh},\ and\ \citenamefont
  {Rotenberg}}]{Martin-Cano2019}%
  \BibitemOpen
  \bibfield  {author} {\bibinfo {author} {\bibfnamefont {D.}~\bibnamefont
  {Martin-Cano}}, \bibinfo {author} {\bibfnamefont {H.~R.}\ \bibnamefont
  {Haakh}},\ and\ \bibinfo {author} {\bibfnamefont {N.}~\bibnamefont
  {Rotenberg}},\ }\bibfield  {title} {\bibinfo {title} {Chiral {Emission} into
  {Nanophotonic} {Resonators}},\ }\href
  {https://doi.org/10.1021/acsphotonics.8b01555} {\bibfield  {journal}
  {\bibinfo  {journal} {ACS Photonics}\ }\textbf {\bibinfo {volume} {6}},\
  \bibinfo {pages} {961} (\bibinfo {year} {2019})}\BibitemShut {NoStop}%
\bibitem [{\citenamefont {Mehrabad}\ \emph {et~al.}(2020)\citenamefont
  {Mehrabad}, \citenamefont {Foster}, \citenamefont {Dost}, \citenamefont
  {Clarke}, \citenamefont {Patil}, \citenamefont {Fox}, \citenamefont
  {Skolnick},\ and\ \citenamefont {Wilson}}]{Mehrabad2020}%
  \BibitemOpen
  \bibfield  {author} {\bibinfo {author} {\bibfnamefont {M.~J.}\ \bibnamefont
  {Mehrabad}}, \bibinfo {author} {\bibfnamefont {A.~P.}\ \bibnamefont
  {Foster}}, \bibinfo {author} {\bibfnamefont {R.}~\bibnamefont {Dost}},
  \bibinfo {author} {\bibfnamefont {E.}~\bibnamefont {Clarke}}, \bibinfo
  {author} {\bibfnamefont {P.~K.}\ \bibnamefont {Patil}}, \bibinfo {author}
  {\bibfnamefont {A.~M.}\ \bibnamefont {Fox}}, \bibinfo {author} {\bibfnamefont
  {M.~S.}\ \bibnamefont {Skolnick}},\ and\ \bibinfo {author} {\bibfnamefont
  {L.~R.}\ \bibnamefont {Wilson}},\ }\bibfield  {title} {\bibinfo {title}
  {Chiral topological photonics with an embedded quantum emitter},\ }\href
  {https://doi.org/10.1364/OPTICA.393035} {\bibfield  {journal} {\bibinfo
  {journal} {Optica}\ }\textbf {\bibinfo {volume} {7}},\ \bibinfo {pages}
  {1690} (\bibinfo {year} {2020})}\BibitemShut {NoStop}%
\bibitem [{\citenamefont {Hauff}\ \emph {et~al.}(2022)\citenamefont {Hauff},
  \citenamefont {Le~Jeannic}, \citenamefont {Lodahl}, \citenamefont {Hughes},\
  and\ \citenamefont {Rotenberg}}]{Hauff2021}%
  \BibitemOpen
  \bibfield  {author} {\bibinfo {author} {\bibfnamefont {N.~V.}\ \bibnamefont
  {Hauff}}, \bibinfo {author} {\bibfnamefont {H.}~\bibnamefont {Le~Jeannic}},
  \bibinfo {author} {\bibfnamefont {P.}~\bibnamefont {Lodahl}}, \bibinfo
  {author} {\bibfnamefont {S.}~\bibnamefont {Hughes}},\ and\ \bibinfo {author}
  {\bibfnamefont {N.}~\bibnamefont {Rotenberg}},\ }\bibfield  {title} {\bibinfo
  {title} {Chiral quantum optics in broken-symmetry and topological photonic
  crystal waveguides},\ }\href
  {https://doi.org/10.1103/PhysRevResearch.4.023082} {\bibfield  {journal}
  {\bibinfo  {journal} {Physical Review Research}\ }\textbf {\bibinfo {volume}
  {4}},\ \bibinfo {pages} {023082} (\bibinfo {year} {2022})}\BibitemShut
  {NoStop}%
\bibitem [{\citenamefont {Kabuss}\ \emph {et~al.}(2011)\citenamefont {Kabuss},
  \citenamefont {Carmele}, \citenamefont {Richter}, \citenamefont {Chow},\ and\
  \citenamefont {Knorr}}]{Kabuss2011}%
  \BibitemOpen
  \bibfield  {author} {\bibinfo {author} {\bibfnamefont {J.}~\bibnamefont
  {Kabuss}}, \bibinfo {author} {\bibfnamefont {A.}~\bibnamefont {Carmele}},
  \bibinfo {author} {\bibfnamefont {M.}~\bibnamefont {Richter}}, \bibinfo
  {author} {\bibfnamefont {W.~W.}\ \bibnamefont {Chow}},\ and\ \bibinfo
  {author} {\bibfnamefont {A.}~\bibnamefont {Knorr}},\ }\bibfield  {title}
  {\bibinfo {title} {Inductive equation of motion approach for a semiconductor
  {QD}-{QED}: {Coherence} induced control of photon statistics},\ }\href
  {https://doi.org/10.1002/pssb.201000851} {\bibfield  {journal} {\bibinfo
  {journal} {Physica Status Solidi B}\ }\textbf {\bibinfo {volume} {248}},\
  \bibinfo {pages} {872} (\bibinfo {year} {2011})}\BibitemShut {NoStop}%
\bibitem [{\citenamefont {Carmichael}(2008)}]{Carmichael2}%
  \BibitemOpen
  \bibfield  {author} {\bibinfo {author} {\bibfnamefont {H.~J.}\ \bibnamefont
  {Carmichael}},\ }\href@noop {} {\emph {\bibinfo {title} {Statistical
  {M}ethods in {Q}uantum {O}ptics 2: {N}on-{C}lassical {F}ields}}}\ (\bibinfo
  {publisher} {Springer-Verlag Berlin Heidelberg},\ \bibinfo {year}
  {2008})\BibitemShut {NoStop}%
\bibitem [{\citenamefont {Carmichael}(1999)}]{Carmichael1}%
  \BibitemOpen
  \bibfield  {author} {\bibinfo {author} {\bibfnamefont {H.~J.}\ \bibnamefont
  {Carmichael}},\ }\href@noop {} {\emph {\bibinfo {title} {Statistical
  {M}ethods in {Q}uantum {O}ptics 1: {M}aster {E}quations and {F}okker-{P}lanck
  {E}quations}}}\ (\bibinfo  {publisher} {Springer-Verlag Berlin Heidelberg},\
  \bibinfo {year} {1999})\BibitemShut {NoStop}%
\bibitem [{\citenamefont {Heitler}(1954)}]{Heitler1954}%
  \BibitemOpen
  \bibfield  {author} {\bibinfo {author} {\bibfnamefont {W.}~\bibnamefont
  {Heitler}},\ }\href@noop {} {\emph {\bibinfo {title} {The {Q}uantum {T}heory
  of {R}adiation}}}\ (\bibinfo  {publisher} {Clarendon Press},\ \bibinfo
  {address} {Oxford},\ \bibinfo {year} {1954})\BibitemShut {NoStop}%
\bibitem [{\citenamefont {Matthiesen}\ \emph {et~al.}(2012)\citenamefont
  {Matthiesen}, \citenamefont {Vamivakas},\ and\ \citenamefont
  {Atatüre}}]{Matthiesen2012}%
  \BibitemOpen
  \bibfield  {author} {\bibinfo {author} {\bibfnamefont {C.}~\bibnamefont
  {Matthiesen}}, \bibinfo {author} {\bibfnamefont {A.~N.}\ \bibnamefont
  {Vamivakas}},\ and\ \bibinfo {author} {\bibfnamefont {M.}~\bibnamefont
  {Atatüre}},\ }\bibfield  {title} {\bibinfo {title} {Subnatural {Linewidth}
  {Single} {Photons} from a {Quantum} {Dot}},\ }\href
  {https://doi.org/10.1103/PhysRevLett.108.093602} {\bibfield  {journal}
  {\bibinfo  {journal} {Physical Review Letters}\ }\textbf {\bibinfo {volume}
  {108}},\ \bibinfo {pages} {093602} (\bibinfo {year} {2012})}\BibitemShut
  {NoStop}%
\bibitem [{\citenamefont {Manga~Rao}\ and\ \citenamefont
  {Hughes}(2007)}]{Manga2007}%
  \BibitemOpen
  \bibfield  {author} {\bibinfo {author} {\bibfnamefont {V.~S.~C.}\
  \bibnamefont {Manga~Rao}}\ and\ \bibinfo {author} {\bibfnamefont
  {S.}~\bibnamefont {Hughes}},\ }\bibfield  {title} {\bibinfo {title} {Single
  quantum-dot {Purcell} factor and $\beta$ factor in a photonic crystal
  waveguide},\ }\href {https://doi.org/10.1103/PhysRevB.75.205437} {\bibfield
  {journal} {\bibinfo  {journal} {Physical Review B}\ }\textbf {\bibinfo
  {volume} {75}},\ \bibinfo {pages} {205437} (\bibinfo {year}
  {2007})}\BibitemShut {NoStop}%
\bibitem [{\citenamefont {Reimer}\ \emph {et~al.}(2016)\citenamefont {Reimer},
  \citenamefont {Bulgarini}, \citenamefont {Fognini}, \citenamefont {Heeres},
  \citenamefont {Witek}, \citenamefont {Versteegh}, \citenamefont {Rubino},
  \citenamefont {Braun}, \citenamefont {Kamp}, \citenamefont {H\"{o}fling},
  \citenamefont {Dalacu}, \citenamefont {Lapointe}, \citenamefont {Poole},\
  and\ \citenamefont {Zwiller}}]{Reimer2016}%
  \BibitemOpen
  \bibfield  {author} {\bibinfo {author} {\bibfnamefont {M.~E.}\ \bibnamefont
  {Reimer}}, \bibinfo {author} {\bibfnamefont {G.}~\bibnamefont {Bulgarini}},
  \bibinfo {author} {\bibfnamefont {A.}~\bibnamefont {Fognini}}, \bibinfo
  {author} {\bibfnamefont {R.~W.}\ \bibnamefont {Heeres}}, \bibinfo {author}
  {\bibfnamefont {B.~J.}\ \bibnamefont {Witek}}, \bibinfo {author}
  {\bibfnamefont {M.~A.~M.}\ \bibnamefont {Versteegh}}, \bibinfo {author}
  {\bibfnamefont {A.}~\bibnamefont {Rubino}}, \bibinfo {author} {\bibfnamefont
  {T.}~\bibnamefont {Braun}}, \bibinfo {author} {\bibfnamefont
  {M.}~\bibnamefont {Kamp}}, \bibinfo {author} {\bibfnamefont {S.}~\bibnamefont
  {H\"{o}fling}}, \bibinfo {author} {\bibfnamefont {D.}~\bibnamefont {Dalacu}},
  \bibinfo {author} {\bibfnamefont {J.}~\bibnamefont {Lapointe}}, \bibinfo
  {author} {\bibfnamefont {P.~J.}\ \bibnamefont {Poole}},\ and\ \bibinfo
  {author} {\bibfnamefont {V.}~\bibnamefont {Zwiller}},\ }\bibfield  {title}
  {\bibinfo {title} {Overcoming power broadening of the quantum dot emission in
  a pure wurtzite nanowire},\ }\href
  {https://doi.org/10.1103/PhysRevB.93.195316} {\bibfield  {journal} {\bibinfo
  {journal} {Physical Review B}\ }\textbf {\bibinfo {volume} {93}},\ \bibinfo
  {pages} {195316} (\bibinfo {year} {2016})}\BibitemShut {NoStop}%
\bibitem [{\citenamefont {Kuhlmann}\ \emph {et~al.}(2013)\citenamefont
  {Kuhlmann}, \citenamefont {Houel}, \citenamefont {Ludwig}, \citenamefont
  {Greuter}, \citenamefont {Reuter}, \citenamefont {Wieck}, \citenamefont
  {Poggio},\ and\ \citenamefont {Warburton}}]{War2013}%
  \BibitemOpen
  \bibfield  {author} {\bibinfo {author} {\bibfnamefont {A.~V.}\ \bibnamefont
  {Kuhlmann}}, \bibinfo {author} {\bibfnamefont {J.}~\bibnamefont {Houel}},
  \bibinfo {author} {\bibfnamefont {A.}~\bibnamefont {Ludwig}}, \bibinfo
  {author} {\bibfnamefont {L.}~\bibnamefont {Greuter}}, \bibinfo {author}
  {\bibfnamefont {D.}~\bibnamefont {Reuter}}, \bibinfo {author} {\bibfnamefont
  {A.~D.}\ \bibnamefont {Wieck}}, \bibinfo {author} {\bibfnamefont
  {M.}~\bibnamefont {Poggio}},\ and\ \bibinfo {author} {\bibfnamefont {R.~J.}\
  \bibnamefont {Warburton}},\ }\bibfield  {title} {\bibinfo {title} {Charge
  noise and spin noise in a semiconductor quantum device},\ }\href
  {https://doi.org/10.1038/nphys2688} {\bibfield  {journal} {\bibinfo
  {journal} {Nature Phys}\ }\textbf {\bibinfo {volume} {9}},\ \bibinfo {pages}
  {570} (\bibinfo {year} {2013})}\BibitemShut {NoStop}%
\bibitem [{\citenamefont {Somaschi}\ \emph {et~al.}(2016)\citenamefont
  {Somaschi}, \citenamefont {Giesz}, \citenamefont {Santis}, \citenamefont
  {Loredo}, \citenamefont {Almeida}, \citenamefont {Hornecker}, \citenamefont
  {Portalupi}, \citenamefont {Grange}, \citenamefont {Ant{\'{o}}n},
  \citenamefont {Demory}, \citenamefont {G{\'{o}}mez}, \citenamefont {Sagnes},
  \citenamefont {Lanzillotti-Kimura}, \citenamefont {Lema{\'{\i}}tre},
  \citenamefont {Auffeves}, \citenamefont {White}, \citenamefont {Lanco},\ and\
  \citenamefont {Senellart}}]{Som2016}%
  \BibitemOpen
  \bibfield  {author} {\bibinfo {author} {\bibfnamefont {N.}~\bibnamefont
  {Somaschi}}, \bibinfo {author} {\bibfnamefont {V.}~\bibnamefont {Giesz}},
  \bibinfo {author} {\bibfnamefont {L.~D.}\ \bibnamefont {Santis}}, \bibinfo
  {author} {\bibfnamefont {J.~C.}\ \bibnamefont {Loredo}}, \bibinfo {author}
  {\bibfnamefont {M.~P.}\ \bibnamefont {Almeida}}, \bibinfo {author}
  {\bibfnamefont {G.}~\bibnamefont {Hornecker}}, \bibinfo {author}
  {\bibfnamefont {S.~L.}\ \bibnamefont {Portalupi}}, \bibinfo {author}
  {\bibfnamefont {T.}~\bibnamefont {Grange}}, \bibinfo {author} {\bibfnamefont
  {C.}~\bibnamefont {Ant{\'{o}}n}}, \bibinfo {author} {\bibfnamefont
  {J.}~\bibnamefont {Demory}}, \bibinfo {author} {\bibfnamefont
  {C.}~\bibnamefont {G{\'{o}}mez}}, \bibinfo {author} {\bibfnamefont
  {I.}~\bibnamefont {Sagnes}}, \bibinfo {author} {\bibfnamefont {N.~D.}\
  \bibnamefont {Lanzillotti-Kimura}}, \bibinfo {author} {\bibfnamefont
  {A.}~\bibnamefont {Lema{\'{\i}}tre}}, \bibinfo {author} {\bibfnamefont
  {A.}~\bibnamefont {Auffeves}}, \bibinfo {author} {\bibfnamefont {A.~G.}\
  \bibnamefont {White}}, \bibinfo {author} {\bibfnamefont {L.}~\bibnamefont
  {Lanco}},\ and\ \bibinfo {author} {\bibfnamefont {P.}~\bibnamefont
  {Senellart}},\ }\bibfield  {title} {\bibinfo {title} {Near-optimal
  single-photon sources in the solid state},\ }\href
  {https://doi.org/10.1038/nphoton.2016.23} {\bibfield  {journal} {\bibinfo
  {journal} {Nature Photon}\ }\textbf {\bibinfo {volume} {10}},\ \bibinfo
  {pages} {340} (\bibinfo {year} {2016})}\BibitemShut {NoStop}%
\bibitem [{\citenamefont {Edwards}(1983)}]{Edwards1983}%
  \BibitemOpen
  \bibfield  {author} {\bibinfo {author} {\bibfnamefont {M.}~\bibnamefont
  {Edwards}},\ }\bibfield  {title} {\bibinfo {title} {Effect of adiabatic and
  near-adiabatic field turn-on on the resonance fluorescence spectrum of a
  two-level atom},\ }\href {https://doi.org/10.1088/0022-3700/16/5/011}
  {\bibfield  {journal} {\bibinfo  {journal} {Journal of Physics B: Atomic and
  Molecular Physics}\ }\textbf {\bibinfo {volume} {16}},\ \bibinfo {pages}
  {767} (\bibinfo {year} {1983})}\BibitemShut {NoStop}%
\bibitem [{\citenamefont {Ulhaq}\ \emph {et~al.}(2013)\citenamefont {Ulhaq},
  \citenamefont {Weiler}, \citenamefont {Roy}, \citenamefont {Ulrich},
  \citenamefont {Jetter}, \citenamefont {Hughes},\ and\ \citenamefont
  {Michler}}]{Ulhaq2013}%
  \BibitemOpen
  \bibfield  {author} {\bibinfo {author} {\bibfnamefont {A.}~\bibnamefont
  {Ulhaq}}, \bibinfo {author} {\bibfnamefont {S.}~\bibnamefont {Weiler}},
  \bibinfo {author} {\bibfnamefont {C.}~\bibnamefont {Roy}}, \bibinfo {author}
  {\bibfnamefont {S.~M.}\ \bibnamefont {Ulrich}}, \bibinfo {author}
  {\bibfnamefont {M.}~\bibnamefont {Jetter}}, \bibinfo {author} {\bibfnamefont
  {S.}~\bibnamefont {Hughes}},\ and\ \bibinfo {author} {\bibfnamefont
  {P.}~\bibnamefont {Michler}},\ }\bibfield  {title} {\bibinfo {title}
  {Detuning-dependent {Mollow} triplet of a coherently-driven single quantum
  dot},\ }\href {https://doi.org/10.1364/OE.21.004382} {\bibfield  {journal}
  {\bibinfo  {journal} {Optics Express}\ }\textbf {\bibinfo {volume} {21}},\
  \bibinfo {pages} {4382} (\bibinfo {year} {2013})}\BibitemShut {NoStop}%
\bibitem [{\citenamefont {Gustin}\ \emph {et~al.}(2018)\citenamefont {Gustin},
  \citenamefont {Manson},\ and\ \citenamefont {Hughes}}]{Gustin2018}%
  \BibitemOpen
  \bibfield  {author} {\bibinfo {author} {\bibfnamefont {C.}~\bibnamefont
  {Gustin}}, \bibinfo {author} {\bibfnamefont {R.}~\bibnamefont {Manson}},\
  and\ \bibinfo {author} {\bibfnamefont {S.}~\bibnamefont {Hughes}},\
  }\bibfield  {title} {\bibinfo {title} {Spectral asymmetries in the resonance
  fluorescence of two-level systems under pulsed excitation},\ }\href
  {https://doi.org/10.1364/OL.43.000779} {\bibfield  {journal} {\bibinfo
  {journal} {Optics Letters}\ }\textbf {\bibinfo {volume} {43}},\ \bibinfo
  {pages} {779} (\bibinfo {year} {2018})}\BibitemShut {NoStop}%
\bibitem [{\citenamefont {Mnaymneh}\ \emph {et~al.}(2019)\citenamefont
  {Mnaymneh}, \citenamefont {Dalacu}, \citenamefont {McKee}, \citenamefont
  {Lapointe}, \citenamefont {Haffouz}, \citenamefont {Weber}, \citenamefont
  {Northeast}, \citenamefont {Poole}, \citenamefont {Aers},\ and\ \citenamefont
  {Williams}}]{2019Dan}%
  \BibitemOpen
  \bibfield  {author} {\bibinfo {author} {\bibfnamefont {K.}~\bibnamefont
  {Mnaymneh}}, \bibinfo {author} {\bibfnamefont {D.}~\bibnamefont {Dalacu}},
  \bibinfo {author} {\bibfnamefont {J.}~\bibnamefont {McKee}}, \bibinfo
  {author} {\bibfnamefont {J.}~\bibnamefont {Lapointe}}, \bibinfo {author}
  {\bibfnamefont {S.}~\bibnamefont {Haffouz}}, \bibinfo {author} {\bibfnamefont
  {J.~F.}\ \bibnamefont {Weber}}, \bibinfo {author} {\bibfnamefont {D.~B.}\
  \bibnamefont {Northeast}}, \bibinfo {author} {\bibfnamefont {P.~J.}\
  \bibnamefont {Poole}}, \bibinfo {author} {\bibfnamefont {G.~C.}\ \bibnamefont
  {Aers}},\ and\ \bibinfo {author} {\bibfnamefont {R.~L.}\ \bibnamefont
  {Williams}},\ }\bibfield  {title} {\bibinfo {title} {On-chip integration of
  single photon sources via evanescent coupling of tapered nanowires to {SiN}
  waveguides},\ }\href {https://doi.org/10.1002/qute.201900021} {\bibfield
  {journal} {\bibinfo  {journal} {Advanced Quantum Technologies}\ }\textbf
  {\bibinfo {volume} {3}},\ \bibinfo {pages} {1900021} (\bibinfo {year}
  {2019})}\BibitemShut {NoStop}%
\bibitem [{\citenamefont {Kannan}\ \emph {et~al.}(2020)\citenamefont {Kannan},
  \citenamefont {Ruckriegel}, \citenamefont {Campbell}, \citenamefont {Kockum},
  \citenamefont {Braum\"{u}ller}, \citenamefont {Kim}, \citenamefont
  {Kjaergaard}, \citenamefont {Krantz}, \citenamefont {Melville}, \citenamefont
  {Niedzielski}, \citenamefont {Veps\"{a}l\"{a}inen}, \citenamefont {Winik},
  \citenamefont {Yoder}, \citenamefont {Nori}, \citenamefont {Orlando},
  \citenamefont {Gustavsson},\ and\ \citenamefont {Oliver}}]{2020nori}%
  \BibitemOpen
  \bibfield  {author} {\bibinfo {author} {\bibfnamefont {B.}~\bibnamefont
  {Kannan}}, \bibinfo {author} {\bibfnamefont {M.~J.}\ \bibnamefont
  {Ruckriegel}}, \bibinfo {author} {\bibfnamefont {D.~L.}\ \bibnamefont
  {Campbell}}, \bibinfo {author} {\bibfnamefont {A.~F.}\ \bibnamefont
  {Kockum}}, \bibinfo {author} {\bibfnamefont {J.}~\bibnamefont
  {Braum\"{u}ller}}, \bibinfo {author} {\bibfnamefont {D.~K.}\ \bibnamefont
  {Kim}}, \bibinfo {author} {\bibfnamefont {M.}~\bibnamefont {Kjaergaard}},
  \bibinfo {author} {\bibfnamefont {P.}~\bibnamefont {Krantz}}, \bibinfo
  {author} {\bibfnamefont {A.}~\bibnamefont {Melville}}, \bibinfo {author}
  {\bibfnamefont {B.~M.}\ \bibnamefont {Niedzielski}}, \bibinfo {author}
  {\bibfnamefont {A.}~\bibnamefont {Veps\"{a}l\"{a}inen}}, \bibinfo {author}
  {\bibfnamefont {R.}~\bibnamefont {Winik}}, \bibinfo {author} {\bibfnamefont
  {J.~L.}\ \bibnamefont {Yoder}}, \bibinfo {author} {\bibfnamefont
  {F.}~\bibnamefont {Nori}}, \bibinfo {author} {\bibfnamefont {T.~P.}\
  \bibnamefont {Orlando}}, \bibinfo {author} {\bibfnamefont {S.}~\bibnamefont
  {Gustavsson}},\ and\ \bibinfo {author} {\bibfnamefont {W.~D.}\ \bibnamefont
  {Oliver}},\ }\bibfield  {title} {\bibinfo {title} {Waveguide quantum
  electrodynamics with superconducting artificial giant atoms},\ }\href
  {https://doi.org/10.1038/s41586-020-2529-9} {\bibfield  {journal} {\bibinfo
  {journal} {Nature}\ }\textbf {\bibinfo {volume} {583}},\ \bibinfo {pages}
  {775} (\bibinfo {year} {2020})}\BibitemShut {NoStop}%
\bibitem [{\citenamefont {Blais}\ \emph {et~al.}(2020)\citenamefont {Blais},
  \citenamefont {Girvin},\ and\ \citenamefont {Oliver}}]{2020blais}%
  \BibitemOpen
  \bibfield  {author} {\bibinfo {author} {\bibfnamefont {A.}~\bibnamefont
  {Blais}}, \bibinfo {author} {\bibfnamefont {S.~M.}\ \bibnamefont {Girvin}},\
  and\ \bibinfo {author} {\bibfnamefont {W.~D.}\ \bibnamefont {Oliver}},\
  }\bibfield  {title} {\bibinfo {title} {Quantum information processing and
  quantum optics with circuit quantum electrodynamics},\ }\href
  {https://doi.org/10.1038/s41567-020-0806-z} {\bibfield  {journal} {\bibinfo
  {journal} {Nature Physics}\ }\textbf {\bibinfo {volume} {16}},\ \bibinfo
  {pages} {247} (\bibinfo {year} {2020})}\BibitemShut {NoStop}%
\bibitem [{\citenamefont {Mirhosseini}\ \emph {et~al.}(2019)\citenamefont
  {Mirhosseini}, \citenamefont {Kim}, \citenamefont {Zhang}, \citenamefont
  {Sipahigil}, \citenamefont {Dieterle}, \citenamefont {Keller}, \citenamefont
  {Asenjo-Garcia}, \citenamefont {Chang},\ and\ \citenamefont
  {Painter}}]{Mirhosseini2019}%
  \BibitemOpen
  \bibfield  {author} {\bibinfo {author} {\bibfnamefont {M.}~\bibnamefont
  {Mirhosseini}}, \bibinfo {author} {\bibfnamefont {E.}~\bibnamefont {Kim}},
  \bibinfo {author} {\bibfnamefont {X.}~\bibnamefont {Zhang}}, \bibinfo
  {author} {\bibfnamefont {A.}~\bibnamefont {Sipahigil}}, \bibinfo {author}
  {\bibfnamefont {P.~B.}\ \bibnamefont {Dieterle}}, \bibinfo {author}
  {\bibfnamefont {A.~J.}\ \bibnamefont {Keller}}, \bibinfo {author}
  {\bibfnamefont {A.}~\bibnamefont {Asenjo-Garcia}}, \bibinfo {author}
  {\bibfnamefont {D.~E.}\ \bibnamefont {Chang}},\ and\ \bibinfo {author}
  {\bibfnamefont {O.}~\bibnamefont {Painter}},\ }\bibfield  {title} {\bibinfo
  {title} {Cavity quantum electrodynamics with atom-like mirrors},\ }\href
  {https://doi.org/10.1038/s41586-019-1196-1} {\bibfield  {journal} {\bibinfo
  {journal} {Nature}\ }\textbf {\bibinfo {volume} {569}},\ \bibinfo {pages}
  {692} (\bibinfo {year} {2019})}\BibitemShut {NoStop}%
\bibitem [{\citenamefont {Reimer}\ \emph {et~al.}(2010)\citenamefont {Reimer},
  \citenamefont {Dalacu}, \citenamefont {Poole},\ and\ \citenamefont
  {Williams}}]{Reimer2010}%
  \BibitemOpen
  \bibfield  {author} {\bibinfo {author} {\bibfnamefont {M.~E.}\ \bibnamefont
  {Reimer}}, \bibinfo {author} {\bibfnamefont {D.}~\bibnamefont {Dalacu}},
  \bibinfo {author} {\bibfnamefont {P.~J.}\ \bibnamefont {Poole}},\ and\
  \bibinfo {author} {\bibfnamefont {R.~L.}\ \bibnamefont {Williams}},\
  }\bibfield  {title} {\bibinfo {title} {Biexciton binding energy control in
  site-selected quantum dots},\ }\href
  {https://doi.org/10.1088/1742-6596/210/1/012019} {\bibfield  {journal}
  {\bibinfo  {journal} {Journal of Physics: Conference Series}\ }\textbf
  {\bibinfo {volume} {210}},\ \bibinfo {pages} {012019} (\bibinfo {year}
  {2010})}\BibitemShut {NoStop}%
\end{thebibliography}%

\appendix

\section{Deriving the Interaction Hamiltonian}
\label{sec:App1}

The free field of the waveguide is described by
\begin{equation}
    \mathcal{E}(x) = \mathcal{E}_-(x) + \mathcal{E}_+(x),
\end{equation}
where $\mathcal{E}_-(x)$ and $\mathcal{E}_+(x)$ are the left and right propagating fields respectively, with the explicit form
\begin{equation}
    \mathcal{E}_{\pm}(x) = \mp \frac{1}{\sqrt{2\pi}} \int_{\infty}^{\infty} e^{\pm i\omega x/c}b(\omega) d\omega,
    \label{E_form}
\end{equation}
where $c$ is the group velocity of the waveguide mode with frequency $\omega$. Due to the reflection from the mirror, we can write $\mathcal{E}_+(0)$ in terms of $\mathcal{E}_-(0)$ by including the phase change picked up from the round trip,
\begin{equation}
    \mathcal{E}_+(0) = e^{i(\reedit{\phi_M} + \omega \tau)} \mathcal{E}_-(0),
    \label{E_relate}
\end{equation}
where $x = 0$ is the location of the mirror. Then the interaction between the TLS and this free field at $x = 0$ is
\begin{equation}
    H_{\rm I} = \reedit{\sqrt{\frac{\gamma}{2}}} \sigma^+ \frac{\mathcal{E}_-(0)}{2} + \reedit{\sqrt{\frac{\gamma}{2}}} \sigma^+ \frac{\mathcal{E}_+(0)}{2} + {\rm H.c.}
    \label{preHint}
\end{equation}

Next, substituting Eqs.~\eqref{E_form} and~\eqref{E_relate} into Eq.~\eqref{preHint}, we get our interaction Hamiltonian
\begin{align}
    H_{\rm I} ={} & \int_{-\infty}^{\infty} d\omega \left[ \left( \reedit{\sqrt{\frac{\gamma}{4\pi}}} \sigma^+ b(\omega) \right. \right. \\
    & {}+ \left. \left. \reedit{\sqrt{\frac{\gamma}{4\pi}}} e^{i(\reedit{\phi_M} + \omega \tau)} \sigma^+ b(\omega) \right) + \rm H.c. \right]. \nonumber
\end{align}\\

\section{Deriving the Evolution of the Waveguide Operators}
\label{sec:App2}

Generally, the free evolution of the waveguide is described by the unitary operator $U_{\rm W} (t) = e^{-i H_{\rm W} t}$ and the evolution of each discrete bin is $U_{\rm W}^{\dagger}(t) B_{n} U_{\rm W} (t)$. Over one time step these exponentials can be expanded to first order in $\Delta t$ to be
\begin{equation}
    U_{\rm W} (\Delta t) = e^{-i H_{\rm W} \Delta t} = 1 - i H_{\rm W} \Delta t,
\end{equation}
which can be substituted into the evolution of $B_n$ to get
\begin{equation}
    U_{\rm W}^{\dagger}(\Delta t) B_{n} U_{\rm W} (\Delta t) = (1 + i H_{\rm W}^{\dagger} \Delta t) B_{n} (1 - i H_{\rm W} \Delta t).
\end{equation}
Then expressing this in terms of the $b_k$ operators and expanding to first order in $\Delta t$, we get
\begin{align}
    U_{\rm W}^{\dagger}(\Delta t) B_{n} U_{\rm W} (\Delta t) & {}= \frac{1}{\sqrt{N}} \sum_{k=0}^{N-1} b_{k} e^{i \reedit{(\omega_k - \omega_0)} n \Delta t} \\
    & \hspace{-3cm} {}+ \frac{i \Delta t}{\sqrt{N}} \sum_{k,k'=0}^{N-1} \reedit{(\omega_k - \omega_{\rm L})} e^{i \reedit{(\omega_{k'} - \omega_0)} n \Delta t} \left( b_k^{\dagger} b_k b_{k'} - b_{k'} b_k^{\dagger} b_k \right). \nonumber
\end{align}
Noting that the commutation relation for $b_k$ is $[b_k,b^{\dagger}_{k'}] = \delta_{k,k'}$ this reduces to
\begin{align}
    U_{\rm W}^{\dagger}(\Delta t) B_{n} U_{\rm W} (\Delta t) & {}= \frac{1}{\sqrt{N}} \sum_{k=0}^{N-1} b_{k} e^{i \reedit{(\omega_k - \omega_0)} n \Delta t} \\
    & \hspace{-1.55cm} {}- \frac{i \Delta t}{\sqrt{N}} \sum_{k,k'=0}^{N-1} \reedit{(\omega_k - \omega_{\rm L})} \delta_{k,k'} b_k e^{i \reedit{(\omega_{k'} - \omega_0)} n \Delta t}, \nonumber \\
    & \hspace{-1.55cm} {}= \frac{1}{\sqrt{N}} \sum_{k=0}^{N-1} b_{k} e^{i \reedit{(\omega_k - \omega_0)} n \Delta t} \left( 1 - i \Delta t \reedit{(\omega_k - \omega_{\rm L})} \right), \nonumber \\
    & \hspace{-1.55cm} {}= \frac{\reedit{e^{-i(\omega_0 - \omega_{\rm L})\Delta t}}}{\sqrt{N}} \sum_{k=0}^{N-1} b_{k} e^{i \reedit{(\omega_{k'} - \omega_0)} (n-1) \Delta t}, \nonumber \\
    & \hspace{-1.55cm} {}= \reedit{e^{-i \delta \Delta t}} B_{n-1}, \nonumber
\end{align}
where we have used that 
\begin{equation}
    (1 - i \Delta t \reedit{(\omega_k - \omega_{\rm L})}) = e^{-i \reedit{(\omega_k - \omega_{\rm L})} \Delta t}
\end{equation} to first order in $\Delta t$.

\end{document}